%% file: main.tex
\documentclass[a4paper,fleqn]{cas-sc}

\usepackage[numbers,sort&compress]{natbib}

\usepackage{amsthm}
\usepackage{mathrsfs}%
\usepackage{xcolor}
\usepackage{float}
\definecolor{color1}{HTML}{303a2b}
\definecolor{color2}{HTML}{00798c}
\definecolor{color3}{HTML}{d1495b}
\definecolor{color4}{HTML}{eddb53}
\usepackage[separate-uncertainty=true]{siunitx}  
\usepackage{xfrac}
\usepackage{mathrsfs}
\usepackage{dcolumn}
\usepackage{bold-extra}

\usepackage{tikz}
\usetikzlibrary{arrows.meta,positioning,fit,calc,trees,shapes,bending,shapes.geometric,backgrounds,fadings}

\usepackage{pgfplots}

\usepgfplotslibrary{patchplots}
\usepgfplotslibrary{fillbetween}
\pgfplotsset{%
    layers/standard/.define layer set={%
        background,axis background,axis grid,axis ticks,axis lines,axis tick labels,pre main,main,axis descriptions,axis foreground%
    }{
        grid style={/pgfplots/on layer=axis grid},%
        tick style={/pgfplots/on layer=axis ticks},%
        axis line style={/pgfplots/on layer=axis lines},%
        label style={/pgfplots/on layer=axis descriptions},%
        legend style={/pgfplots/on layer=axis descriptions},%
        title style={/pgfplots/on layer=axis descriptions},%
        colorbar style={/pgfplots/on layer=axis descriptions},%
        ticklabel style={/pgfplots/on layer=axis tick labels},%
        axis background@ style={/pgfplots/on layer=axis background},%
        3d box foreground style={/pgfplots/on layer=axis foreground},%
    },
}

\usepackage{caption}
\usepackage{subcaption}
\usepackage{placeins}

\newcommand{\code}[1]{\texttt{\detokenize{#1}}}

\newcommand{\posi}{x} 
\newcommand{\pos}{\mathbf{\posi}} 
\newcommand{\veli}{u} 
\newcommand{\vel}{\mathbf{\veli}} 
\newcommand{\straintens}{\mathbb{S}}
\newcommand{\nonlin}{\mathbf{H}} 

\newcommand{\sposi}{X} 
\newcommand{\com}{G} 
\newcommand{\spos}{\mathbf{\sposi}} 
\newcommand{\sveli}{V} 
\newcommand{\svel}{\mathbf{\sveli}} 
\newcommand{\sacc}{\mathbf{A}} 
\newcommand{\quati}{m} 
\newcommand{\quatv}{\mathbf{\quati}} 
\newcommand{\quat}{\underline{\quati}} 
\renewcommand{\fint}{\mathbf{f}^\text{I}} 
\newcommand{\fhid}{\mathbf{f}^\text{H}} 
\newcommand{\Fhid}{\mathbf{F}^\text{H}} 
\newcommand{\fext}{\mathbf{f}^\text{E}} 
\newcommand{\Fext}{\mathbf{F}^\text{E}} 
\newcommand{\fib}{\mathbf{f}^\text{IB}} 
\newcommand{\Mext}{\mathbf{M}^\text{E}} 
\newcommand{\Mhid}{\mathbf{M}^\text{H}} 
\newcommand{\momin}{\boldsymbol{I}} 
\newcommand{\velang}{\boldsymbol{\omega}} 
\newcommand{\viscstress}{\boldsymbol{\tau}} 
\newcommand{\lagpos}{\mathbf{X}} 

\DeclareSIUnit\dyne{dyne}

\newcommand{\Fluid}{\mathbf{q}}
\newcommand{\Solid}{\mathbf{z}}

\newcommand{\pdv}[2][]{\frac{\partial #1}{\partial #2}}
\newcommand{\pdvv}[2][]{\frac{\partial^2 #1}{\partial { #2 }^2}}

\ExplSyntaxOn
\RenewDocumentEnvironment { Abstract } { o }
{
  \group_begin:
  \IfNoValueTF { #1 } { } { \tex_gdef:D \abstractname { #1 } }
  \parindent \c_zero_dim
  \vbox_set:Nw \l_tmpa_box
    \hsize = .65\textwidth \linewidth = \hsize \parindent \c_zero_dim
    \noindent \abstractname \par
    \skip_vertical:n { -4pt }
    \noindent \rule{.65\textwidth}{.2pt}\par \footnotesize
    \ignorespaces \everypar { \parindent=1.5em }
}
{
  \par \vbox_set_end:
  \noindent
  \hbox_to_wd:nn { \textwidth }
  {
    \hbox_to_wd:nn { .35\textwidth } { \box_use_drop:N \g_stm_key_box \hss }
    \vbox_top:n { \vbox_unpack_drop:N \l_tmpa_box } \hss
  }
  \par
  \group_end:
}
\ExplSyntaxOff

\ExplSyntaxOn
\RenewDocumentCommand \dashrule { O{.4pt} m m }
   {
     \group_begin: \color{black!50}
     \skip_vertical:n { #2 }
     \noindent \rule { \linewidth } { #1}
     \group_end: \skip_vertical:n { #3 }
   }
\ExplSyntaxOff

\begin{document}
\let\WriteBookmarks\relax
\def\floatpagepagefraction{1}
\def\textpagefraction{.001}

\shorttitle{IB-Flows: an open-source multi-GPU immersed boundary FSI code}

\shortauthors{G. Vagnoli et~al.}

\title[mode = title]{IB-Flows: an open-source multi-GPU immersed boundary code for fluid--structure interaction}


\author[1]{Giovanni Vagnoli}
\fnmark[$\dagger$]
\credit{}

\author[1,2]{Martino Andrea Scarpolini}
\fnmark[$\dagger$]
\credit{}

\author[3]{Fabio Guglietta}
\fnmark[$\ddagger$]
\credit{}

\author[4]{Joshua Romero}
\credit{}

\author[4]{Massimiliano Fatica}
\credit{}

\author[1,5]{Roberto Verzicco}
\credit{}

\author[1,2]{Francesco Viola}[orcid=0000-0003-1303-5934]
\cormark[1]
\ead{francesco.viola@gssi.it}
\credit{}

\affiliation[1]{organization={Gran Sasso Science Institute (GSSI)},
            city={L'Aquila},
            country={Italy}}
\affiliation[2]{organization={INFN--Laboratori Nazionali del Gran Sasso},
            city={Assergi},
            country={Italy}}
\affiliation[3]{organization={Department of Physics and INFN, University of Rome, Tor Vergata},
            country={Italy}}
\affiliation[4]{organization={NVIDIA Corporation},
            city={Santa Clara},
            state={CA},
            country={USA}}
\affiliation[5]{organization={Physics of Fluids Group, University of Twente},
            city={Enschede},
            country={The Netherlands}}

\cortext[1]{Corresponding author}

\newcommand\gugliettanote{Currently at INAIL Research, Monte Porzio Catone, Italy.}
\newcommand\equalnote{These authors contributed equally to this work.}
\let\origprintfnotes\printfnotes
\renewcommand\printfnotes{%
  \origprintfnotes
  \def\thefootnote{$\dagger$}%
  \footnotetext{\rule{0pt}{3ex}\equalnote}%
  \def\thefootnote{$\ddagger$}%
  \footnotetext{\gugliettanote}%
}

\begin{abstract}
\relax 
We present IB-Flows, an open-source, multi-GPU solver for the direct numerical simulation of incompressible fluid--structure interaction (FSI) problems. The code couples a second-order finite-difference fractional-step Navier--Stokes solver on a staggered Cartesian grid with two immersed boundary methods: a Lagrangian method based on moving least-squares (MLS) interpolation and an Eulerian sharp-interface method. Rigid bodies are advanced with a quaternion-based Newton--Euler solver, while deformable surfaces embedded in three-dimensional flows are described by a structural solver based on the interaction potentials; a predictor--corrector scheme provides either loose or strong fluid--structure coupling. Non-Newtonian fluids with shear-thinning or shear-thickening rheology are handled through a strain-rate-dependent viscosity, and the same framework accommodates subgrid-scale eddy-viscosity models for large-eddy simulations of turbulent flows. The solver is written in CUDA Fortran with MPI domain decomposition, so that the fluid, interpolation and structural kernels are all executed on the GPUs, while the distributed transposes required by the Poisson and implicit solvers are further accelerated by the cuDecomp library through GPU-aware communication. Strong- and weak-scaling tests show near-ideal intra-node scaling and good multi-node efficiency. The solver is validated against a set of benchmark problems spanning wall-bounded turbulence, rigid-body and deformable-body fluid--structure interaction, and a biomedical application. IB-Flows is intended as a reproducible reference implementation for immersed boundary FSI simulations on modern GPU clusters and as a transparent platform to be used for tackling multiphysics problems.
\end{abstract}



\maketitle
\vspace{-0.5cm}
\section*{Program summary}
\noindent%
\textit{Program Title:} IB-Flows\\
\textit{CPC Library link to program files:} (to be added by Technical Editor)\\
\textit{Developer's repository link:} \url{https://gitlab.com/gssi-fluids/ib-flows}\\
\textit{Licensing provisions:} BSD 3-clause\\
\textit{Programming language:} CUDA Fortran, MPI\\
\textit{Nature of problem:} Direct numerical simulation of incompressible flows interacting with rigid or deformable immersed bodies (fluid--structure interaction), including non-Newtonian rheology and large-eddy-simulation closures.\\
\textit{Solution method:} Second-order finite-difference fractional-step Navier--Stokes solver on a staggered Cartesian grid with an FFT-based Poisson solver; Lagrangian moving-least-squares and Eulerian sharp-interface immersed boundary methods; quaternion-based Newton--Euler rigid-body solver and interaction-potential structural model for thin deformable surfaces; loose or strong predictor--corrector fluid--structure coupling; MPI slab decomposition with CUDA Fortran kernels and optional cuDecomp transposes.\\
\textit{Additional comments including restrictions and unusual features:} Requires the NVIDIA HPC SDK, an MPI library, HDF5, cuFFT, cuRAND and NCCL, and runs on NVIDIA GPUs only. Periodic boundary conditions and uniform grid spacing are required in the $x_1$ and $x_2$ directions; the computational domain is decomposed into slabs along $x_3$ only.%

\section{Introduction}
\label{sec:intro}
\noindent Open-source computational fluid dynamics (CFD) software is now widespread, and robust implementations of many well-established numerical algorithms are readily available. Examples include general-purpose CFD and multiphysics suites such as OpenFOAM~\cite{weller1998tensorial} and Neko~\cite{jansson2021neko}, high-order finite-element and spectral-element libraries such as Nektar++~\cite{cantwell2015nektar}, MFEM~\cite{anderson2019mfem}, deal.II~\cite{arndt2019dealii}, FEniCS~\cite{logg2012automated} and FreeFEM~\cite{hecht2012freefem}, and compact research solvers designed around specific numerical methods, such as WaterLily.jl~\cite{weymouth2025waterlily} and STREAmS~\cite{bernardini2021streams}. In contrast, open-source fluid--structure interaction (FSI) frameworks remain comparatively scarce, especially when the target application is large-scale direct numerical simulation (DNS) on modern GPU-accelerated high-performance computing (HPC) systems. FSI codes must combine a fluid solver, a structural solver, interface force and motion transfer, time-coupling algorithms, and parallel data movement.\\
\indent Several commercial packages with FSI capabilities are currently available and benefit from a long history of validation and extensive user communities, including LS-DYNA~\cite{lstc2024lsdyna}, Simcenter STAR-CCM+~\cite{siemens2025starccm}, Abaqus~\cite{dassault2024abaqus} and COMSOL Multiphysics~\cite{comsol2025manual}. These tools can address complex multiphysics problems using monolithic or partitioned coupling strategies. Their use in academic method development, however, is often limited by significant barriers: closed-source implementations reduce transparency and reproducibility, licensing costs restrict access to large-scale runs, and specialized or unconventional algorithms can be difficult to implement through user subroutines alone. These limitations are particularly relevant when the scientific goal is not only to solve a specific engineering case, but also to inspect, modify, and extend the numerical method itself.\\
\indent Open-source finite-element and multiphysics libraries, such as FEniCS~\cite{logg2012automated}, MFEM~\cite{anderson2019mfem}, deal.II~\cite{arndt2019dealii} and FreeFEM~\cite{hecht2012freefem}, provide an important foundation for developing FSI applications. Similarly, coupling libraries such as preCICE \cite{chourdakis2021precice} can connect independent fluid and structural solvers in a flexible partitioned workflow. These approaches are valuable because they encourage modularity and reuse mature single-physics components. Nevertheless, for large-scale DNS they do not remove the need for a tightly designed parallel implementation. Interface interpolation, immersed boundary forcing, structural updates and communication patterns must all be coordinated with the target architecture. In practice, this often requires a dedicated solver in which the numerical method and the HPC implementation are developed together.\\
\indent A major distinction among FSI methods is the treatment of the moving interface. Arbitrary Lagrangian--Eulerian (ALE) formulations use body-fitted meshes and can represent stresses and boundary conditions accurately at conforming fluid--solid interfaces. They are widely used in biomedical and engineering software, including OpenFOAM-based solid/FSI toolboxes~\cite{cardiff2018solids4foam}, SimVascular~\cite{updegrove2017simvascular}, CRIMSON~\cite{arthurs2021crimson} and lifex-cfd~\cite{africa2023lifexcfd}. Their main cost is that the fluid mesh must deform with the structure and may require interpolation and remeshing. These operations become increasingly demanding for large displacements, multiple bodies, contact, complex deformable kinematics and long time integrations. They also complicate GPU acceleration because the mesh connectivity, load balance and data-access patterns evolve during the simulation.\\
\indent Immersed and unfitted methods follow a different philosophy. The fluid is solved on a background Eulerian grid, while the solid is represented by a Lagrangian surface or volume that exchanges forces and kinematic information with the Eulerian field. This avoids body-fitted remeshing and makes it comparatively simple to introduce bodies of arbitrary shape, prescribe or solve their motion, and handle configurations in which the body trajectory is not known in advance. These properties have motivated several open-source immersed boundary or immersed-interface solvers, including IBAMR~\cite{bhalla2013unified}, PetIBM~\cite{chuang2018petibm}, cuIBM~\cite{layton2011cuibm}, WaterLily.jl~\cite{weymouth2025waterlily} and OpenIFEM~\cite{cheng2019openifem}. The same simplicity also makes immersed formulations attractive as research platforms for testing new interpolation, forcing, load-reconstruction and coupling strategies.

\subsection*{Available open-source FSI codes}
\noindent A representative list of open-source FSI and moving-boundary CFD codes is summarized in Table~\ref{tab:fsi_codes}. The table is not intended to be exhaustive: for example, meshless multiphysics libraries such as SPHinXsys \cite{zhang2021sphinxsys}, educational immersed boundary implementations such as IB2d \cite{battista2017ib2d}, and solver-coupling ecosystems based on preCICE address important parts of the open-source FSI landscape. The focus here is instead on established or methodologically close solvers that are useful for positioning the present code.\\
\begin{table}
\centering
\small
\begin{tabular}{lccccl}
\hline
\textbf{Name} & \textbf{Language} & \textbf{Method} & \textbf{MPI} & \textbf{GPU} & \textbf{Citation} \\
\hline
OpenIFEM & C++ & IFEM & yes & no & \citet{cheng2019openifem} \\
solids4Foam & C++ & FV/ALE & yes & no & \citet{cardiff2018solids4foam} \\
PetIBM & C++ & IBM-FDM & yes & yes & \citet{chuang2018petibm} \\
cuIBM & CUDA-C++ & IBM-FDM & no & yes & \citet{layton2011cuibm} \\
CRIMSON & Fortran/C++ & FEM/ALE & yes & no & \citet{arthurs2021crimson} \\
SimVascular & C++ & FEM/ALE & yes & no & \citet{updegrove2017simvascular} \\
lifex-cfd & C++ & FEM/ALE & yes & no & \citet{africa2023lifexcfd} \\
SU2 & C++ & FVM/FEM & yes & no & \citet{economon2016su2} \\
IBAMR & C++ & IBM-FDM/FEM & yes & no & \citet{bhalla2013unified} \\
WaterLily.jl & Julia & BDIM-FDM & no & yes & \citet{weymouth2025waterlily} \\
\textbf{IB-Flows} & CUDA-Fortran & IBM-FDM & yes & yes & (present code) \\
\hline
\end{tabular}
\caption{Representative open-source FSI and moving-boundary CFD codes. Abbreviations are: finite-difference method (FDM), finite-volume method (FVM), finite-element method (FEM), immersed boundary method (IBM), immersed finite element method (IFEM), boundary data immersion method (BDIM) and arbitrary Lagrangian--Eulerian formulation (ALE). The GPU column indicates reported GPU execution of the main flow solver or linear-algebra kernels.}
\label{tab:fsi_codes}
\end{table}
\indent The codes in Table~\ref{tab:fsi_codes} cover different objectives. SimVascular, CRIMSON and lifex-cfd are oriented toward cardiovascular workflows, including image-based geometry preparation, physiological boundary conditions, and finite-element discretizations. solids4Foam and SU2 provide flexible finite-volume or multiphysics environments for engineering applications on unstructured or body-fitted meshes. IBAMR is a mature immersed boundary framework with adaptive mesh refinement and a substantial record of applications in biological and bio-inspired flows. PetIBM and cuIBM emphasize reproducible immersed boundary research workflows on Cartesian grids, with PetIBM targeting distributed-memory systems and cuIBM providing a compact single-GPU implementation. WaterLily.jl is a recent example of a concise, differentiable, backend-agnostic Julia solver for flows around dynamic immersed bodies. Long-standing sharp-interface immersed boundary developments by Mittal and co-workers, including recent GPU-resident implementations such as ViCar3D \cite{mittal2026gpu,mittal2008sharp_IB_method,mittal2011sharp_IB_method}, further demonstrate the performance potential of GPU-resident immersed boundary algorithms, although they do not fill the open-source niche addressed here.\\
\indent The software presented in this work is designed specifically for large-scale DNS of incompressible FSI problems on GPU clusters. It combines a finite-difference method for the Navier--Stokes equations with two immersed boundary methods, a Lagrangian one based on moving least-squares (MLS) interpolation and an Eulerian sharp-interface one, while in both cases the hydrodynamic loads are computed with MLS normal probes. The code supports rigid and  deformable bodies embedded and interacting with three-dimensional flows.  The main contribution of the present release is the integration of this numerical approach into an open-source CUDA--MPI implementation intended for production-scale simulations. The (Cartesian) Eulerian grid keeps the flow solver, interpolation stencils and immersed boundary operations algorithmically simple and naturally parallel. At the same time, the Lagrangian representation of the bodies makes the setup of moving rigid or deformable geometries substantially simpler than in body-fitted approaches: no conforming fluid mesh has to be generated around the structure, and no mesh-deformation algorithm is required when the body undergoes large or unpredictable motion. This makes the code useful not only as an application solver for turbulent FSI, particulate flows, flapping structures and biomedical valves, but also as a transparent platform for testing new immersed boundary, force-reconstruction and coupling algorithms under realistic HPC constraints.\\
\indent The remainder of the paper describes the mathematical formulation, time-coupling strategy and GPU/MPI implementation; documents validation and demonstration cases spanning turbulent channel flow, rigid-body FSI, deformable-body FSI and a mechanical-valve application; and reports scalability on modern multi-GPU systems. The intent is to provide a reproducible open-source reference implementation for high-resolution immersed boundary FSI simulations.

\section{Material and Methods} \label{sec:mm}

\begin{figure*}
  \centering
  \resizebox{\textwidth}{!}{\input{setup.tex}}
  \caption{Overview of the immersed boundary computational setup of the solver. (a) The immersed body (solid) is embedded in a  Cartesian domain where the fluid equations are solved. (b) The fluid is discretized on a fixed Eulerian grid, while the surface of the body is discretized by a Lagrangian grid of triangles that does not conform to the fluid mesh. (c) Zoom on the triangulated surface, showing the triangle vertices (black dots) and the Lagrangian markers (orange squares) located at the triangle centroids, where the no-slip condition is enforced in the Lagrangian approach. (d) Two-dimensional sketch of the moving-least-squares (MLS) support domain of the Lagrangian IBM around a selected Lagrangian marker $\mathbf{X}_l$ (red square): the $3\times3\times3$ Eulerian nodes $\mathbf{x}_k$ (black dots) enclosed by the red dashed box are used to interpolate the fluid velocity onto the marker and to spread back the immersed boundary forcing; $\Delta x_1$ and $\Delta x_2$ denote the Eulerian grid spacings.}
  \label{fig:setup}
\end{figure*}
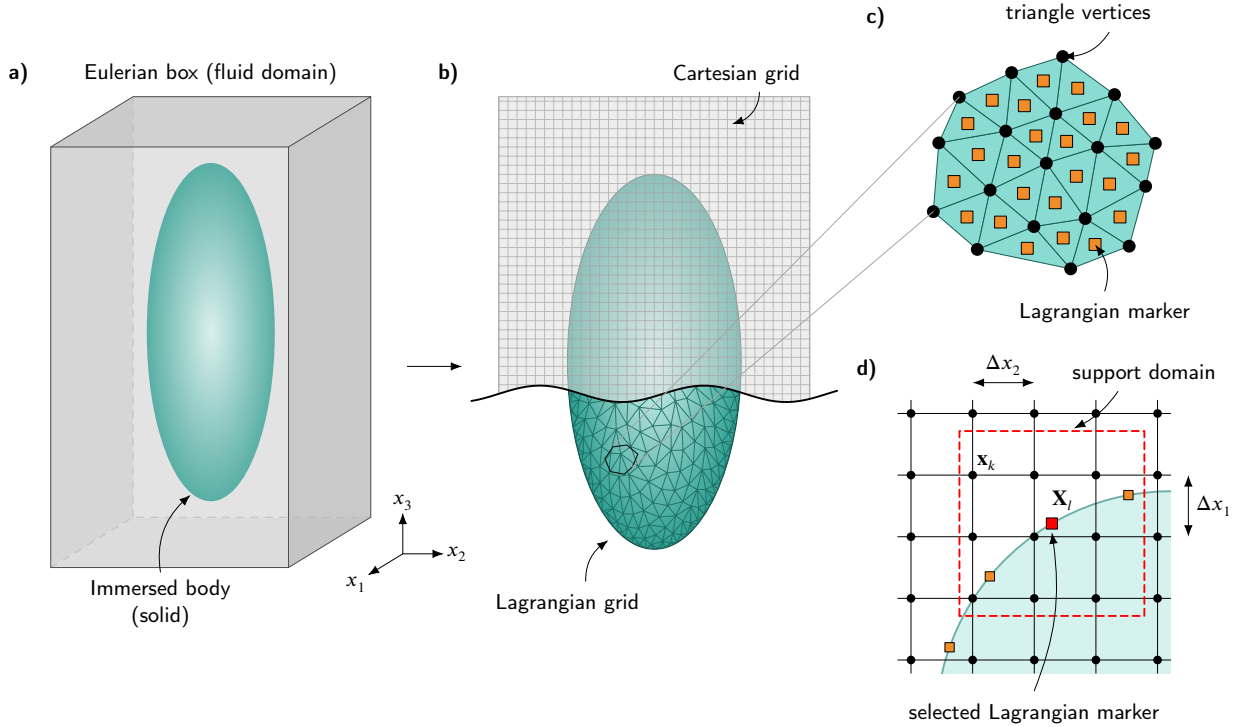

\subsection{Fluid Mechanics}\label{sssec:fluidMech}
\noindent The incompressible fluid of density $\rho_f$ and kinematic viscosity $\nu$ is described by the velocity $\vel(\pos,t)$ and pressure $p(\pos,t)$ fields which are governed by the incompressible Navier-Stokes equations:
\begin{subequations}\label{eq:ns}
    \begin{align}
    \rho_f \left(\frac{\partial \vel}{\partial t} + \vel\cdot\nabla\vel\right) &= -\nabla p + \nabla\cdot \viscstress + \fib \, , \label{eq:nsmom} \\
    \nabla\cdot\vel &= 0 \, . \label{eq:nscont}
    \end{align}
\end{subequations}
The viscous stress tensor $\viscstress$ for incompressible fluids can be written in a general form as $\viscstress = 2\rho_f \nu(\straintens)\straintens$, where $\straintens=(\nabla \vel + \nabla \vel^T )/2$ is the strain rate tensor. For a Newtonian fluid, $\nu$ does not depend on the strain rate and therefore $\viscstress$ is linearly related to $\straintens$. In contrast, other rheological behaviours are possible if the kinematic viscosity depends on $\straintens$. 
\\ \indent The solver allows for a strain-dependent, non-constant viscosity through the flag \code{visc_model}, which enables the solution of the Navier--Stokes equations with a generalized viscosity law.
By setting \code{visc_model = carreau_yasuda}, a non-Newtonian shear-thinning fluid (e.g. blood) is prescribed using the Carreau-Yasuda model \cite{carreau1972rheological,yasuda1981shear,lupi2025impact}
\begin{equation}
    \nu(\straintens) = \nu(S)= \nu_{\infty} + (\nu_0 - \nu_{\infty}) [1 + (\lambda S)^a]^{(n-1)/a},
\end{equation}
where $S = |\straintens|$ is the norm of the strain rate tensor, $\nu_0 = \nu(S\rightarrow 0)$ whereas $\nu_{\infty} =  \nu(S\rightarrow \infty)$, and the parameters $\{\lambda,a,n\}$ depend on the haematocrit of blood.
The same constitutive form of the viscous stress tensor is adopted in large-eddy simulations (LES). In this case, the velocity and pressure fields are understood as spatially filtered quantities, and the kinematic viscosity is written as $\nu(\straintens) = \nu_m + \nu_t(\straintens)$, where $\nu_m$ denotes the molecular (constant) viscosity and $\nu_t$ is the subgrid-scale, or turbulent, eddy viscosity that models the effects of the unknown turbulent stress tensor. 
In the code, turbulent viscosity can be modeled using the Smagorinsky model (\code{visc_model = smago}), $\nu_t = C \Delta ^2 S$, where $C$ is the Smagorinsky constant and $\Delta$ is the local grid size.
In the presence of walls, the Smagorinsky model can fail to predict zones of laminar flow near solid boundaries, thus two subgrid-scale models allowing for solid walls are also included.
The first model is proposed by Vreman~\cite{vremanEddyviscositySubgridscaleModel2004} (\code{visc_model = vreman})
\begin{equation}
    \nu_t = c \sqrt{B_{\beta}/(\nabla{\mathbf{u}}:\nabla{\mathbf{u}})}, \quad B_{\beta} = \beta_{11}\beta_{22} - \beta_{12}^2 + \beta_{11}\beta_{33} - \beta_{13}^2 + \beta_{22}\beta_{33} - \beta_{23}^2,
\end{equation}
where $c \approx 2.5C^2$, $\beta_{ij} = \sum_{m = 1}^3\Delta_m^2 \partial u_i /\partial x_m  \partial u_j /\partial x_m$ and the symbol : denotes the tensor contraction $A:B = \sum_{ij}A_{ij} B_{ij}$.
The second model is the WALE model~\cite{nicoud1999wale} (\code{visc_model = wale})
\begin{equation}
    \nu_t = (C_w \Delta)^2 \frac{\mathcal{S}_d^{3/2}}{(\straintens:\straintens)^{5/2} + \mathcal{S}_d^{5/4}},\quad \mathcal{S}_d = \frac{1}{6}((\straintens:\straintens)^{2} + (\mathbb{W}:\mathbb{W})^{2}) + \frac{2}{3} (\straintens:\straintens)(\mathbb{W}:\mathbb{W}) + 2 I_4,
\end{equation}
where $C_w$ is a constant, $\mathbb{W}=(\nabla \vel - \nabla \vel^T)/2$ is the rotation rate tensor and $I_4 = \sum_{i,j,k,l}\straintens_{ik}\straintens_{kj} \mathbb{W}_{jl}\mathbb{W}_{li}$.

\indent The fluid equations~\eqref{eq:ns} are solved in dimensionless form using the characteristic length $L$, velocity $U$ and density $\rho_f$ as reference quantities, which give rise to the Reynolds number $Re=UL/\nu$. These equations are advanced in time using a fractional step method \cite{kimApplicationFractionalstepMethod1985,van2015pencil}.\\
\indent The momentum equation~\eqref{eq:nsmom} is advanced from time $t_{n}$ to $t_{n+1}$ using several possible integration schemes, both single and multi-step (chosen using the \code{nrk} input variable): explicit Euler (\code{nrk}=1, \code{fluid_eulerian_time_int = true}), second-order Adams--Bashforth (\code{nrk}=1, \code{fluid_eulerian_time_int = false}), or low-storage Runge--Kutta schemes with two or three substeps (\code{nrk}=2 or 3, respectively). In the Runge--Kutta cases, the nonlinear terms are advanced with coefficients $(\gamma,\rho)=(1/2,0),(1,-1/2)$ for the second-order scheme and $(\gamma,\rho)=(8/15,0),(5/12,-17/60),(3/4,-5/12)$ for the third-order one, while $\alpha=\gamma+\rho$ multiplies the pressure gradient. The viscous terms can either be advanced explicitly, using the same discretisation as the nonlinear terms, or treated implicitly with a Crank--Nicolson scheme. In the explicit case, the viscous terms are included in $\nonlin$ and advanced with the same $\gamma,\rho$ weights as the nonlinear terms. In the implicit case, the Crank--Nicolson viscous terms are multiplied by $\alpha$. The input variable \code{implicit_NS} selects between the two treatments: if \code{false}, the viscous terms are advanced explicitly, while if \code{true}, they are treated implicitly using Crank--Nicolson. 
In the following we describe the latter case in which viscous terms are treated implicitly and the integration in time is performed using Adams-Bashforth/Crank-Nicolson: given $\vel _n$ and $p_n$, velocity and pressure fields at time $t_n$, and $\Delta t$ the time step, the provisional non-solenoidal velocity field $\hat{\vel}$ satisfies
\begin{equation}\label{eq:nstime_discr}
\frac{\hat{\vel}-\vel_n}{\Delta t} = -\alpha \nabla p_n + \gamma \nonlin_n + \rho \nonlin_{n-1} + \frac{\alpha}{2 Re}\nabla^2(\hat{\vel} + \vel_n)
\end{equation}
where $\nonlin_n$ ($\nonlin_{n-1}$) incorporates the non-linear terms at the time $t_n$ ($t_{n-1})$ and each possible volume force, while $\gamma=3/2$, $\rho=-1/2$ and $\alpha = \gamma + \rho = 1$ are the coefficients of the Adams–Bashforth/Crank-Nicolson time advancement scheme \cite{deTullio2016MLS}.\\
\indent The equations are discretised in space using a second-order accurate centred finite-difference (FD) scheme on a Cartesian staggered grid, hereafter denoted as Eulerian grid \cite{verzicco1996finite}. $N_1$, $N_2$ and $N_3$ points are used in each Cartesian direction ($x_i$ for $i=1,2,3$, with unit vectors $\hat{\pos}_i$), and the corresponding grid size in the three spatial directions of the $k$-th Eulerian cell is $\Delta x_1^k$, $\Delta x_2^k$ and $\Delta x_3^k$. Periodic boundary conditions are applied in the $x_1$ and $x_2$ directions, while along $x_3$ different boundary conditions can be imposed: Dirichlet, Neumann or the radiative condition \cite{Verzicco_Orlandi_Eisenga_Heijst_Carnevale_1996} where each velocity component is advected using a constant outward velocity $\mathbf{c}=c\,\hat{\pos}_3$
\begin{equation}\label{eq:radiative}
\frac{\partial \vel}{\partial t} + c\frac{\partial \vel}{\partial x_3} = 0\,.
\end{equation}
\indent Regarding velocity initial conditions, these are set by default to zero everywhere. In special cases such as the turbulent channel test case, the solution can be initialized to Poiseuille-like solutions in order to speed up the statistically stationary solution.\\
\indent The no-slip condition on the immersed surfaces can be enforced with either of the two immersed boundary methods (IBMs) available in the code: a Lagrangian IBM based on moving least-squares (MLS) interpolation and an Eulerian sharp-interface IBM. Each method is activated by its own input flag, \code{mlsIB = true} for the Lagrangian MLS IBM and \code{sharpIB = true} for the Eulerian sharp-interface IBM. These flags are specified separately for each immersed body, so that within the same simulation one body can be treated with the Lagrangian MLS IBM and another one with the Eulerian sharp-interface IBM. Both methods yield a forcing field $\fib$ that corrects the provisional velocity (equation~\eqref{eq:mlsForcing}), and in both cases the hydrodynamic loads acting on the immersed bodies are computed with the MLS normal-probe approach described in Section~\ref{sec:IBM_interpolation}.

\paragraph{Lagrangian MLS IBM.}
In the Lagrangian IBM, the no-slip condition is imposed at the Lagrangian markers uniformly distributed on the immersed surfaces. These surfaces are discretized using triangular elements (Figure~\ref{fig:setup}), with their centroids serving as Lagrangian marker nodes. A 3D support domain consisting of $N_e=3\times 3\times 3 =27$ Eulerian nodes is created around each Lagrangian marker whose position is $\lagpos$. The fluid velocity at the marker location $\hat{\veli}_i(\lagpos)$ is computed through interpolation using the velocity at the $N_e$ support Eulerian points:
\begin{equation}\label{eq:interp}
\hat{\veli}_i(\lagpos) = \sum_{k=1}^{N_e} \phi^k(\lagpos)\hat{\veli}_i(\pos_k)
\end{equation}
where $\phi^k(\lagpos)$ are the transfer operators which depend on the shape functions used for the interpolation. Generally, this interpolated velocity $\hat{\vel}(\lagpos)$ does not match the corresponding Lagrangian marker velocity $\svel(\lagpos)$ (coming from the structural solver), and their difference is therefore used to compute a source term $\fib(\lagpos) = [\svel(\lagpos)-\hat{\vel}(\lagpos)]/\Delta t$, which is then transferred back to the Eulerian grid points as a distributed forcing $\fib$ by an equation similar to~\eqref{eq:interp}. This procedure is repeated for all Lagrangian markers, and the resulting forcing field is used to update the provisional velocity
\begin{equation}\label{eq:mlsForcing}
  \vel^* = \hat{\vel} + \Delta t \,\fib \, .
\end{equation}

\paragraph{Eulerian sharp-interface IBM.}
In the Eulerian IBM, the forcing is computed directly at Eulerian forcing nodes using the sharp-interface method introduced in \cite{vagnoli2026fastconsistentsharpinterfaceimmersed}, which applies to stationary, moving and deformable bodies of arbitrary thickness. The same triangular discretization of MLS-IBM is required, and, at each time step, a tagging procedure examines the Eulerian nodes surrounding each triangular element and classifies a node as a forcing node when the immersed interface separates it from at least one of its six neighbors on the Eulerian lattice. 
The underlying identification of surface-intersecting Eulerian cells relies on a triangle-local ray-tracing method, which is based on the Möller--Trumbore intersection algorithm.
For each forcing node, the corresponding closest point on the wall is determined, and a probe is extended along the local wall-normal direction to locate an external point. The velocity at this external point is reconstructed from the surrounding Eulerian nodes through trilinear interpolation. The velocity at the forcing node is then obtained by linear interpolation between the wall velocity and the external velocity. This procedure is applied independently on both sides of the wet surface, providing a sharp representation of both zero- and finite-thickness structures on the fluid grid. The forcing $\fib(\pos_k)=[\svel(\pos_k)-\hat{\vel}(\pos_k)]/\Delta t$, where $\svel(\pos_k)$ is the target velocity reconstructed at the forcing node $\pos_k$, is then applied through equation~\eqref{eq:mlsForcing}. Compared with the Lagrangian IBM, this approach substantially reduces the spurious transpiration across solid surfaces \cite{vagnoli2026fastconsistentsharpinterfaceimmersed}; for moving bodies, it must be combined with the Runge--Kutta time integrators (including explicit Euler).\\
\indent Since $\vel^*$ satisfies the no-slip boundary conditions but it is still a non-solenoidal field, it is projected onto a divergence-free space by a correction
\begin{equation}\label{eq:velsolen}
  \vel_{n+1}=\vel^*-\alpha\Delta t \, \nabla \varphi \, ,
\end{equation}
where the scalar field $\varphi$ comes from the elliptic equation
\begin{equation}\label{eq:elliptic}
\nabla^2\varphi = \frac{\nabla \cdot \vel^* - q}{\alpha\Delta t},
\end{equation}
where the quantity $q$ denotes a spatial distribution of mass sources/sinks, which is nonzero when using the sharp Eulerian IBM and is required to ensure numerical consistency in the presence of solid walls \cite{vagnoli2026fastconsistentsharpinterfaceimmersed}. The updated pressure is given by
\begin{equation}\label{eq:pressure}
  p_{n+1} = p_n + \varphi - \frac{\alpha \Delta t }{2 Re} \,\nabla^2\varphi \, .
\end{equation}
The last term accounts for the implicit treatment of the viscous terms and is therefore included only when \code{implicit_NS = true}; with explicit viscous terms the pressure is simply updated as $p_{n+1}=p_n+\varphi$.\\
\indent Boundary conditions for the momentum and projection steps are imposed consistently with the fractional-step formulation. The two homogeneous directions, $x_1$ and $x_2$ are periodic for the velocity, pressure and pressure-correction fields; this periodicity is also the basis of the horizontal FFTs used in the Poisson solver. In the $x_3$ direction, the default configuration prescribes the velocity at the lower and upper boundaries, corresponding to no-slip walls or to user-defined boundary values. The Poisson problem for $\varphi$ then uses the complementary homogeneous Neumann condition, so that the projection does not modify the imposed normal velocity; the arbitrary constant of $\varphi$ is removed by fixing a reference value. For open-flow configurations, activated through \code{radiative_cond}, the lower boundary is treated as a prescribed inflow profile, while the upper boundary is advanced with the radiative outflow condition of equation~\eqref{eq:radiative}; the outlet normal velocity is additionally corrected to enforce global mass conservation. The integer value of \code{radiative_cond} selects the inlet profile (uniform, Hagen--Poiseuille or hyperbolic-tangent; see Table~\ref{tab:input_var_names}). Finally, when \code{neumann_velocity = true}, the upper boundary condition is swapped: a zero-normal-gradient condition is imposed on the velocity and the pressure-correction equation uses the homogeneus Dirichlet condition at the outlet.\\
\indent It should be noted that the projection step $\vel^*\to\vel_{n+1}$ of equation~\eqref{eq:velsolen}, enforcing the divergence-free condition for the velocity, slightly perturbs the field $\vel^*$, which satisfies the IB condition imposed on equation~\eqref{eq:mlsForcing}. Since reducing the residual mass flux through the immersed surfaces is extremely important in some applications (especially when large pressure gradients exist across IBs), the code also implements additional methods to deal with better interpolation and pressure correction. For the Lagrangian IBM, a first degree of improvement can be achieved by iterating steps (\ref{eq:interp}--\ref{eq:pressure})
to obtain an updated velocity field $\vel_{n+1}$ complying, at the same time, both with the solenoidal- and no-slip boundary condition up to a certain tolerance (setting the input variable $\code{iterMLSdiv}>1$). To improve the MLS interpolation by itself, instead of iterating the elliptic solver, another possibility is to iterate equations~\eqref{eq:interp} and~\eqref{eq:mlsForcing} alone, which is still effective and much faster (which can be done setting $\code{iterMLS}>1$).\\
\indent Two further options are available to reduce the boundary error without resorting to costly implicit formulations. The first, specific to the Lagrangian IBM, is the MLS-$\kappa$ correction \cite{vagnoli2025kappa} (enabled with the flag \code{kappa_correction}). Since the interpolation and spreading operators are not reciprocal and the support domains of neighbouring markers overlap, the explicit forcing $\fib(\lagpos_l)$ obtained from equation~\eqref{eq:mlsForcing} does not enforce the no-slip condition exactly. By formulating the determination of the IB forces as a linear system coupling all the Lagrangian markers and applying an approximate factorisation of its matrix, one obtains a local, fully explicit correction $\fib(\lagpos_l)\to\kappa(\lagpos_l)\fib(\lagpos_l)$, in which the coefficient $\kappa(\lagpos_l)$ is computed for each marker and each velocity component from the MLS shape functions of the neighbouring markers and from the local value of the explicit forcing itself. The correction therefore accounts both for the spreading error and for the overlap of the support domains, adapts dynamically to the instantaneous flow, requires no linear solve and adds a negligible cost, while it substantially reduces the slip and transpiration velocities at the immersed surfaces and improves the convergence properties of the MLS forcing. The second option, also specific to the Lagrangian IBM, is the pressure-decoupling algorithm \cite{yildiran2024pressure} (flag \code{activate_prbc}), which addresses the perturbation introduced by the projection step. Following a ghost-fluid-like treatment, the homogeneous Neumann condition for the pseudo-pressure is explicitly enforced on the immersed surface by adding to the right-hand side of the momentum equation~\eqref{eq:nstime_discr} 
a pressure jump term, evaluated from the pressure field at the previous time step through the normal probes described in Section~\ref{sec:IBM_interpolation}, while the coefficient matrix of the Poisson equation is left unchanged. As a result, the solutions inside and outside the body are decoupled, the residual mass flux through the immersed surface is drastically reduced, and the FFT-based direct solver of the code can still be used. This treatment is not required by the Eulerian IBM, whose sharp formulation decouples the two sides of the immersed boundary by construction.\\
\indent  Table~\ref{tab:input_var_names} illustrates a selection of the major input variables of the code which will be described in this work.\\
\begin{table}
\centering
\begin{tabular}{r|p{0.47\textwidth}p{0.2\textwidth}}
\hline
\textbf{code variable name} & \textbf{description} & \textbf{possibilities}\\
\hline
\code{nrk} & fluid integration scheme steps & $\{1,2,3\}$\\
\code{fluid_eulerian_time_int} & use Euler scheme when \code{nrk}\,=\,1 & \{\code{true,false}\}\\
\code{implicit_NS} & Use semi-implicit viscous NS solver & \{\code{true,false}\}\\
\code{visc_model} & viscosity model: Newtonian, Carreau--Yasuda non-Newtonian, or LES subgrid-scale models (Smagorinsky, Vreman, WALE) & \{\code{none}, \code{carreau_yasuda}, \code{smago}, \code{vreman}, \code{wale}\}\\[2pt]
\code{radiative_cond} & set open-flow configuration along $x_3$: prescribed inflow at the lower boundary and radiative outflow condition (equation~\ref{eq:radiative}) at the upper one. The value selects the inlet profile: 0 = no inflow/outflow (both $x_3$ boundaries are solid walls, as in the channel flow of Test~\#1), 1 = Hagen--Poiseuille profile, 2 = hyperbolic-tangent profile, any other value = uniform inflow & \{0,1,2,\ldots\}\\
\code{neumann_velocity} & zero-normal-gradient velocity condition (with Dirichlet pressure correction) at the upper boundary & \{\code{true,false}\}\\
  \code{mlsIB} & activate the Lagrangian MLS IBM & \{\code{true,false}\}\\
\code{hdx} & Lagrangian adaptive mesh refinement ratio (triangle edge /  grid spacing) & $>0$\\
\code{iterMLSdiv} & iterations of the MLS forcing and projection steps (equations~\ref{eq:interp}--\ref{eq:pressure}) & $\ge1$\\
\code{iterMLS} & iterations of the MLS forcing step alone (equations~\ref{eq:interp}--\ref{eq:mlsForcing}) & $\ge1$ \\
\code{kappa_correction} & MLS-kappa IB correction (Lagrangian IBM only) & \{\code{true,false}\}\\
\code{activate_prbc} & pressure decoupling (Lagrangian IBM only) & \{\code{true,false}\}\\
  \code{sharpIB} & activate the Eulerian sharp-interface IBM & \{\code{true,false}\}\\
\code{solid_scheme} & time integration scheme of the structural solver (\texttt{h2}: second order, \texttt{h4}: Hamming fourth order) & \{\texttt{"h2"},\texttt{"h4"},...\} \\
\code{fsi_tol} & FSI coupling tolerance (0: loose coupling, $>0$: strong coupling) & $[0,1)$ \\
\code{hydroloads} & activate the computation of the hydrodynamic loads exerted by the fluid on the immersed bodies & \{\code{true,false}\}\\
\code{wet_pos} & Hydrodynamic loads computation along positive normal probe $\mathbf{n}^+$ & \{\code{true,false}\}\\
\code{wet_neg} & Hydrodynamic loads computation along negative normal probe $\mathbf{n}^-$ & \{\code{true,false}\}\\
\code{taycorrp} & Taylor reconstruction of the pressure loads (0: first, 1: second, 2: enhanced second order) & $\{0,1,2\}$ \\
\code{taycorrv} & Taylor reconstruction of the viscous loads (0: first, 1: second, 2: enhanced second order) & $\{0,1,2\}$ \\
\code{lvlhalo}&halo/ghost layer thickness for the fluid domain decomposition ($\lceil \max_k\left(h^k/\Delta^k x_3\right)+|s|\rceil$, Section~\ref{sssec:fluid-parallelization})&$\ge1$\\
\hline
\end{tabular}
\caption{Possible input variable names and descriptions for the fluid and structural solvers, as well as IB method.}
\label{tab:input_var_names}
\end{table}
\indent The elliptic equation~\eqref{eq:elliptic} is solved on the Cartesian grid using fast Fourier transforms in the periodic directions \cite{kimApplicationFractionalstepMethod1985}. This requires a uniform grid spacing in the $x_1$ and $x_2$ directions ($\Delta x_1^k = \Delta x_1$ and $\Delta x_2^k = \Delta x_2$), whereas a non-uniform grid can be used in $x_3$.

\subsection{Structural mechanics \label{ssec:struc}}
The motion of immersed solid bodies of uniform mass density $\rho_s$ is obtained by solving the structural mechanics problem that governs the evolution of their configuration in time. Let $\spos(t)$ denote the position of a material point of the solid surface, $\svel=\dot{\spos}$ its velocity and $\sacc=\ddot{\spos}$ its acceleration. The structural dynamics, made non-dimensional using the characteristic length $L$, velocity $U$, and density $\rho_f$ (the same ones of the fluid), results from the balance between inertia, internal structural forces $\fint$, hydrodynamic loads $\fhid$ provided by the fluid solver, and possible external forces $\fext$. Depending on the nature of the solid, the configuration update reduces either to rigid body dynamics, where the relative distance between material points is preserved, or to deformable body dynamics, where internal elastic forces depend on the local deformation of the structure. While rigid bodies can be either empty or filled, deformable bodies must be two-dimensional thin structures.

\subsubsection{Rigid bodies}
For rigid bodies no deformation occurs, $\fint=\mathbf 0$ and only global force and moment balances are required. The linear dynamics of the center of mass position $\spos_\com$ and velocity $\svel_\com$ is solved in the inertial frame of reference of the fluid, while the rotational one is solved in a body-fixed reference frame aligned with the principal axes of inertia, with $\velang$ the angular velocity of the body. The body orientation (i.e. the orientation of the body-fixed frame with respect to the inertial one) is described through quaternions \cite{wieSpaceVehicleDynamics2008} which are a convenient way of representing rotations using four-dimensional vectors $\quat=(\quati_0,\quatv)$. The rotation matrix $R=R(\quat)$ to transform vectors between the inertial and body-fixed frames can be easily obtained from $\quat$. In this configuration the equations to be numerically integrated are the following:
\begin{subequations}\label{eq:struc_rigid_dynamics}
  \begin{align}
    m\dot{\svel}_\com &= \Fext + \Fhid, \\
    \dot{\spos}_\com &= \svel_\com, \\
    \momin \dot{\velang} &= -\velang \times \momin \velang + \Mext + \Mhid, \\
    \dot{\quatv} &= \sfrac12 \, (\quati_0\velang - \velang \times \quatv), \\
    \dot{\quati}_0 &= -\sfrac12 \, (\velang \cdot \quatv),
  \end{align}
\end{subequations}
where $m$ is the total mass of the body, $\momin$ its inertia tensor evaluated on the body-fixed frame, and $\Fext$, $\Fhid$, $\Mext$ and $\Mhid$ are the external and hydrodynamic forces and moments integrated over each body surface (computed with respect to its centre of mass). $\momin$ is computed once during initialization and, thanks to the body-fixed frame of reference, remains constant during the simulation. It is calculated as surface integral (subroutine \code{compute_inertia_tensor_and_initial_quaternions}) from $\rho_s$, i.e. the (uniform) density of the solid and its geometrical properties for both hollow (thin-shell) and solid rigid bodies. \\
\indent For problems involving unbounded trajectories, such as freely falling or rising bodies (see Section~\ref{ssec:rising}), the code can solve the Navier--Stokes equations in a translating, non-rotating, non-inertial reference frame whose origin is attached to the body center. This formulation avoids the need for vertically elongated computational domains to capture the entire body trajectory. Instead, it enables the use of smaller domains, hence substantially reducing the computational cost. Simulations in the non-inertial frame can be enabled by setting the flag \code{non_inertial_eulerian_rs = true}. In this case, the body acceleration is included in the fluid equations as a spatially uniform inertial forcing term, $-\dot{\svel}_\com$. Accordingly, an inflow velocity $\vel=-\svel_\com$ is prescribed at the bottom boundary (uniform inflow profile), while a radiative condition, given in equation~\eqref{eq:radiative} and including the body acceleration, is imposed at the top boundary. Within this framework, the body occupies a fixed region of the fluid domain, while retaining the ability to rotate about its center through its rotational degrees of freedom.

\subsubsection{Deformable bodies}
For deformable bodies, the configuration is described by the positions $\spos_i$ of the mesh vertices forming the triangulated surface (see Figure~\ref{fig:setup}). The mass of each triangular element $j$, computed as the product of its surface area $A_j$, thickness $s$ and density $\rho_s$, is distributed equally among its three vertices. This results in the total mass of the structure lumped at each vertex $i$ as:
\begin{equation}
  m_i = \frac13\sum_{j\in T_i}\rho_s s A_j
\end{equation}
where the sum runs over $T_i$ which contains each triangle $j$ that shares vertex $i$, $s$ is the thickness of the structure and $A_j$ is the surface area of triangle $j$.\\
\indent In this case, the internal forces $\fint_i$ arise from elastic deformation and depend on the local displacement field. The structural model is based on a spring-network formulation using an interaction potential approach \cite{fedosov2010systematic,hammer2011mass,gelder1998elastic_membranes}. Each edge of the triangulated surface is associated with an elastic spring simulating the in-plane resistance with associated potential $W^\text{el}=\sfrac12 k_e (\ell-\ell_0)^2$, where $\ell$ and $\ell_0$ are the current and rest edge lengths. To compute the spring constant $k_e$ for a given spring, we use an equation proposed by van Gelder \cite{gelder1998elastic_membranes} for calculating spring constants throughout an unstructured triangular mesh in order to produce uniform membrane behavior according to a given Young modulus $E$: $k_e=Es(A_1+A_2)/\ell^2$, 
where $A_1$ and $A_2$ are the surface areas of the two triangles adjacent to the given edge.\\
\indent The resistance against bending is modeled through an additional angular spring on the angle $\theta$ between adjacent triangles sharing an edge $W^b=k_b [1-\cos(\theta-\theta_0)]$, where $\theta_0$ is the angle at rest configuration. The bending constant is equal to $k_b = 2B/\sqrt{3}$, with $B = Es^3/ [12(1 - \nu^2_m)]$ the bending modulus of a planar structure and $\nu_m$ the Poisson ratio of the material \cite{deTullio2016MLS}.\\
\indent The nodal forces generated by each potential can be computed explicitly by calculating the gradient with respect to node displacements, as detailed in \cite{deTullio2016MLS}. In this code, spring constants and every sub-component of the total internal force  $\fint$  are computed inside \code{internalforce} subroutine, which uses a CUF kernel for the in-plane and the \code{compute_fbxyz_kernel} kernel for the out-of-plane nodal forces. This approach can be extended to hyperelastic materials, as described in \cite{hammer2011mass,deTullio2016MLS,viola2023high}, and is implemented in the code as well for the Fung-type materials \cite{fung1993biomechanics}.\\
\indent The motion of each vertex is governed by the balance between inertia, the hydrodynamic force $\fhid_i$ provided by the fluid solver (see equation~\eqref{eq:traction} later and related discussion), the internal elastic force $\fint_i$, and any prescribed external force $\fext_i$:
\begin{equation}\label{eq:struc_def_dynamics}
m_i\ddot{\spos}_i = \fint_i + \fhid_i + \fext_i .
\end{equation}
\indent 
The integration in time of both rigid and deformable bodies is performed after the structural and hydrodynamic loads have been computed (more details in Section~\ref{sec:FSI-coupling}). The accelerations of vertices (for deformable bodies) or of the centre of mass (for rigid bodies) are computed explicitly.  These accelerations ($\sacc_{n}$) are used in the solid time-integration scheme with a second-order Adams--Bashforth predictor for the velocity and a trapezoidal, or Crank--Nicolson, update for the position:
\begin{subequations}
  \begin{align}
    \svel_{n+1} &= \svel_n + \Delta t \left(\frac32\sacc_n - \frac12\sacc_{n-1} \right), \\
    \spos_{n+1} &= \spos_n + \frac{\Delta t}{2} \left(\svel_{n+1} + \svel_{n}\right) .
  \end{align}
\end{subequations}
For rigid bodies, the rotational variables, namely angular velocity $\velang$ and quaternion $\quat$, are advanced with a three-stage Runge--Kutta integration of the angular momentum and quaternion equations. The quaternion is normalized after each discrete update to preserve unit length.

\subsection{MLS interpolation and hydrodynamic loads}\label{sec:IBM_interpolation}
\noindent Equation~\eqref{eq:interp} is based on the moving least square (MLS) interpolation \cite{deTullio2016MLS,spandan_2017_potential,spandan2018fast,viola2023high}, which is used by the Lagrangian IBM. It uses a linear basis function with an exponential weight function centered at the location of the Lagrangian marker. Regardless of the IBM selected (Lagrangian or Eulerian), MLS interpolation is also used to compute the hydrodynamic loads transferred to the structural solver. These loads are evaluated at the same Lagrangian markers located on the immersed body surface, i.e. on each triangle barycentre. Since Eulerian quantities cannot be evaluated across the immersed surface (because the stencil would mix values from both sides of the surface), the evaluation is based on the normal probe approach introduced in \cite{vanellabalaras2009,deTullio2016MLS}: for each marker $l$, a probe in the wall normal direction $\mathbf{n}^l$ is cast from the center of the marker $\spos^l$ to the bulk of the fluid $\spos^{l_e}$ at distance $h$:
\begin{equation}\label{eq:force_probe}
  \spos^{l_e} = \spos^l + h \mathbf{n}^l \,.
\end{equation}
In general, the value of $h$ is proportional to the local Eulerian grid spacing, thus $h$ can vary over the surface of the body depending on the size of the local cell (see Section~\ref{sssec:fluid-parallelization}).
To extrapolate the interpolated quantities back onto the original Lagrangian marker, the pressure and the velocity gradients are expanded in Taylor series, as detailed in \cite{vagnoli2026taycorr}, from $\spos^{l_e}$ backwards to $\spos^l$, leading to the following expressions:
\begin{subequations}\label{eq:p_gradu_taylor}
  \begin{align}
    p(\spos^l) &= p (\spos^{l_e}) - \pdv[p]{n} \bigg|_{\spos^{l_e}}\!\!\! h + \frac{1}{2} \pdvv[p]{n}\bigg|_{\spos^{l_e}} \!\!\! h^2,\\
    \nabla{\vel}\bigg|_{\spos^l} & = \nabla{\vel}\bigg|_{\spos^{l_e}} - \pdv{n} ( \nabla\vel)\bigg|_{\spos^{l_e}}\!\!\!h + \frac12 \pdvv{n}\left(\nabla{\vel}\right)\bigg|_{\spos^{l_e}}\!\!\!h^2.
  \end{align}
\end{subequations}
The coefficients of the Taylor series on the point $\mathbf{X}^{l_e}$ are evaluated through a second order MLS interpolation of the flow quantities defined on the Eulerian grid. The traction $\mathbf{t}$ on each marker is given by:
\begin{equation}\label{eq:traction}
  \mathbf{t}(\spos^l) = -p(\spos^l)\mathbf{n}_l + \frac{1}{Re}\left( \nabla\vel(\spos^l) + \nabla\vel^\text{T}(\spos^l) \right)\cdot \mathbf{n}_l\,,
\end{equation}
leading to the force on each triangle $\fhid(\spos^l)=\mathbf{t}(\spos^l)A_l$, being $A_l$ the area of the $l$-th triangle. The latter forces are then transferred from the barycentre $\fhid(\spos^l)$ to the vertex of each triangle $\fhid_i$, where the structural dynamics is actually solved for deformable structures (equation~\eqref{eq:struc_def_dynamics}).
Finally, the total force and torque around the point $\spos_p$ is given by:
\begin{equation}\label{eq:hid_force_tot}
  \Fhid = \sum_{l=1}^{N_t}\fhid(\spos^l), \quad \Mhid = \sum_{l=1}^{N_t}(\spos^l -\spos_p)\times\fhid(\spos^l)\,.
\end{equation}
\indent The hydrodynamic loads computation, starting from equation~\eqref{eq:force_probe} up to equation~\eqref{eq:hid_force_tot}, can be performed using the positive $\mathbf{n}^+=\mathbf{n}$ and/or negative $\mathbf{n}^-=-\mathbf{n}$ normal directions of the immersed boundary. The choice depends on the specific case. For a solid represented as a closed surface, for example, only the surface outward normal direction should be considered. For a thin tissue wet on both sides (as for the flag test case in Section~\ref{ssec:tc:flag}), instead, both directions should be considered. The input variables \code{wet_pos} and \code{wet_neg} for each immersed body control whether the hydrodynamic load computation (along each immersed body surface triangles) is performed along the positive or negative normal probes. When only one of the \code{wet_pos} or \code{wet_neg} variables is set to \code{false} the code interprets the body as a solid with a defined volume, whereas when both variables are set to \code{true}, instead, the code interprets the body as a thin surface. The positive normal direction is dictated by the triangle nodes order and is thus a property of each immersed surface given as input to the code. The Taylor expansions in equation~\eqref{eq:p_gradu_taylor} can be truncated to first, second  order or enhanced second order accuracy  \cite{vagnoli2026taycorr} using the input variables \code{taycorrp} and \code{taycorrv} equal to 0, 1 or 2, respectively. In the case of non-Newtonian fluid model, the local viscosity appears in the traction expression (equation~\eqref{eq:traction}).\\
\indent To achieve a robust and accurate two-way transfer of information between Lagrangian and Eulerian grids when using the Lagrangian MLS IBM, some geometrical constraints have to be satisfied. A working compromise is typically to have a surface triangulation with equilateral triangles having edges of size $\simeq 0.7$ the local grid spacing; this choice does not penalize the computational overhead while ensuring a sufficiently dense spatial distribution of Lagrangian markers. However, satisfying this constraint can become difficult in two cases: first, if the Eulerian grid is refined and, second, when the immersed body deforms. Moreover, this geometrical constraint is driven by the fluid solver, but the same Lagrangian mesh that is used for enforcing the interfacial boundary condition is used for the structural solver. This implies that a refinement in the IBM mesh automatically implies a refinement in the structural solver mesh which may lead to unnecessary computational nodes for discretising the deformation governing equations and small time steps to track the dynamics of small structural elements.\\
\indent These problems are prevented by using an adaptive Lagrangian mesh refinement procedure where the initial triangular mesh is automatically subdivided into virtual subtriangles (the "tiles") until each one gets smaller than the local Eulerian grid size, thus avoiding “holes” in the interfacial boundary condition. In this way, the initial mesh which is loaded into the solver is the coarsest one and is calibrated to be used with the structural solver. For the Lagrangian MLS IBM, instead, each Lagrangian marker is dynamically obtained by considering the tiles, i.e. the subdivisions of the initial mesh triangles, such that the correct Lagrangian-to-Eulerian grid ratio is maintained throughout the whole simulation \cite{spandan2018fast,viola2023high}. The adaptive Lagrangian refinement is controlled in the code through the input variable \code{hdx}, which regulates the ratio between triangle edge length and the Eulerian grid spacing. This variable is set to $0.7$ by default.
For the Eulerian IBM, tiles are not required, since the no-slip boundary condition is imposed directly on the Eulerian grid. Therefore, in this case the initial geometry provided to the code must be properly calibrated to accurately represent the continuous geometry and to ensure an accurate description of the body dynamics by the structural solver.

\subsection{FSI coupling}\label{sec:FSI-coupling}
The structural and fluid problems are not independent. The structural update requires the hydrodynamic loads generated by the flow, while the Navier--Stokes solver requires the instantaneous immersed geometry and interface velocity imposed by the structure. Equations~\eqref{eq:ns} and~\eqref{eq:struc_rigid_dynamics} (for rigid bodies, or \eqref{eq:struc_def_dynamics} for deformable tissues) must therefore be advanced in a coupled manner.\\
\indent FSI algorithms are commonly classified as monolithic or partitioned. Monolithic methods solve the fluid and structural unknowns as a single system, which gives tight coupling but at high computational and implementation cost. The present code follows a partitioned strategy: the fluid and solid solvers remain distinct modules, each with its own discretization and time-integration method, and exchange interface data at well-defined stages of the time step. We denote the solid state by $\Solid=(\spos,\svel)$, or by $\Solid=(\spos_\com,\svel_\com,\quat,\velang)$ for rigid bodies, and the fluid state by $\Fluid=(\vel,p)$. In compact form, the structural solver is written as $\dot{\Solid}=\mathcal{Z}(\Solid,\Fluid)$, because the structural equations depend on the hydrodynamic force $\fhid$. Conversely, the fluid solver is written as $\dot{\Fluid}=\mathcal{Q}(\Fluid,\Solid)$, because the immersed boundary forcing $\fib$ depends on the current structural configuration.\\
\indent The default strategy is loose coupling, obtained by setting \code{fsi_tol = 0}. In this case only the predictor branch in Figure~\ref{fig:coupling} is executed. The solid is first advanced from $t_n$ to $t_{n+1}$ using the loads available from $\Fluid_n$, giving the predicted state $\Solid_{n+1}^{p}$. The fluid solver is then advanced using this updated structural configuration, giving $\Fluid_{n+1}^{p}$. Since no corrector loop is used, these predicted values are accepted as the solution at the new time level: $\Solid_{n+1}=\Solid_{n+1}^{p}$ and $\Fluid_{n+1}=\Fluid_{n+1}^{p}$. This staggered update is inexpensive and is the standard operating mode of the code, although stability may require a smaller time step in strongly coupled regimes.
\begin{figure*}
  \centering
  \input{loose_strong.tex}
  \caption{Predictor--corrector fluid--structure coupling strategy. Solid and fluid states are denoted by $\Solid$ and $\Fluid$, respectively. In the default loose-coupling mode, only the predictor is used and the predicted states are accepted at $t_{n+1}$. If strong coupling is enabled, the predicted pair $(\Solid_{n+1}^{p},\Fluid_{n+1}^{p})$ initializes the corrector loop, where solid and fluid updates are repeated until the prescribed coupling tolerance is reached.}
  \label{fig:coupling}
\end{figure*}
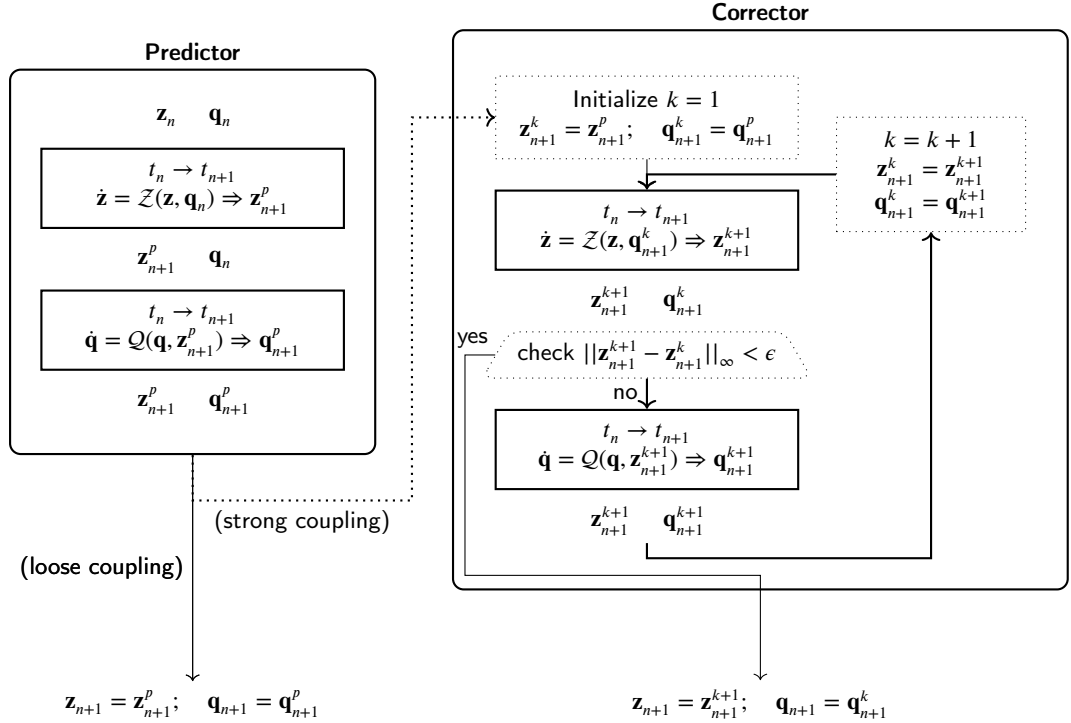\\
\indent Strong coupling is activated by setting \code{fsi_tol > 0}. The predictor still provides the initial estimate at $t_{n+1}$, but the solution is then refined through the corrector branch shown in Figure~\ref{fig:coupling}. At each corrector iteration the solid is advanced using the most recent fluid state $\Fluid_{n+1}^{k}$, the convergence of the structural state (whose updated configuration is $\Solid_{n+1}^{k+1}$) is checked, and, if the tolerance is not met, the fluid is advanced again using the updated structure. This loop recomputes the hydrodynamic loads within the same time step (immediately after each fluid update) and provides a more stable estimate of the coupled force balance. The iteration stops when the selected residual, here represented as $|| \Solid_{n+1}^{k+1} - \Solid_{n+1}^{k} ||_\infty$, falls below the tolerance \code{fsi_tol}.\\
\indent The structural time-integration order is selected with \code{solid_scheme}. The option \code{solid_scheme = h2} denotes the second-order scheme and is the default choice, particularly for loose coupling. The option \code{solid_scheme = h4} denotes the Hamming fourth-order scheme \cite{hamming1959stable,deTullio2016MLS}. It is typically used together with strong coupling, where the corrector iterations exploit its improved convergence properties, as shown in \cite{deTullio2016MLS}. Conversely, \code{h4} is not recommended for loose coupling because, without the within-step correction of the FSI loads, it can reduce the stability margin of the staggered update. The appropriate choice between loose and strong coupling depends on the physical parameters of the problem. When structural inertia dominates fluid inertia, loose coupling is often stable and efficient, possibly with a reduced time step \cite{viola2020fsei,verzicco_2022_EFM}; stronger added-mass effects usually require the iterative corrector.

\subsection{Implementation and domain decomposition}
\noindent The discretized FSI problem demands substantial computational resources and is advanced in time with a hybrid MPI--CUDA implementation. Both the fluid and structural computations are parallelized at two levels: first, the Eulerian and Lagrangian domains are decomposed employing distributed memory parallelism (using MPI) and, secondly, the computation on each subdomain is accelerated by running it in parallel on a Graphics Processing Unit (GPU). The GPU porting was mainly performed and described in \cite{viola2022fsei_gpu} and is based on CUDA Fortran \cite{ruetsch2024cuda}. In CUDA, the code executed on the GPU is written as kernels, i.e. subroutines executed concurrently by a grid of threads organized in blocks. Blocks are scheduled independently on the streaming multiprocessors of the GPU, and threads within the same block can share fast on-chip memory. The majority of regular loops are offloaded to the GPU through CUF kernels directives which preserve the structure of the original Fortran loop. This is done simply adding CUDA directive decorators before Fortran do loops. The latter appear as a comment to the compiler if GPU code generation is disabled (similar to the OpenMP directives). Instead, in performance-critical or irregular operations, computation is offloaded to the GPU implementing dedicated GPU subroutine (CUDA kernels). The latter are used, for example, when several threads must cooperate on the same Lagrangian marker. CUDA-enabled GPUs thus provide thousands of processor cores which allow to run tens of thousands of threads concurrently resulting in an effective speed-up of the operations over large computational grids as is the present case. The whole fluid and structural computations are performed on the GPUs, whereas the CPUs are only used to stage the data needed during the communication phases and to handle the input/output operations. Large output operations, such as three-dimensional flow fields, are written with parallel MPI I/O in VTK format, while lightweight diagnostics are post-processed online and written as text files.\\
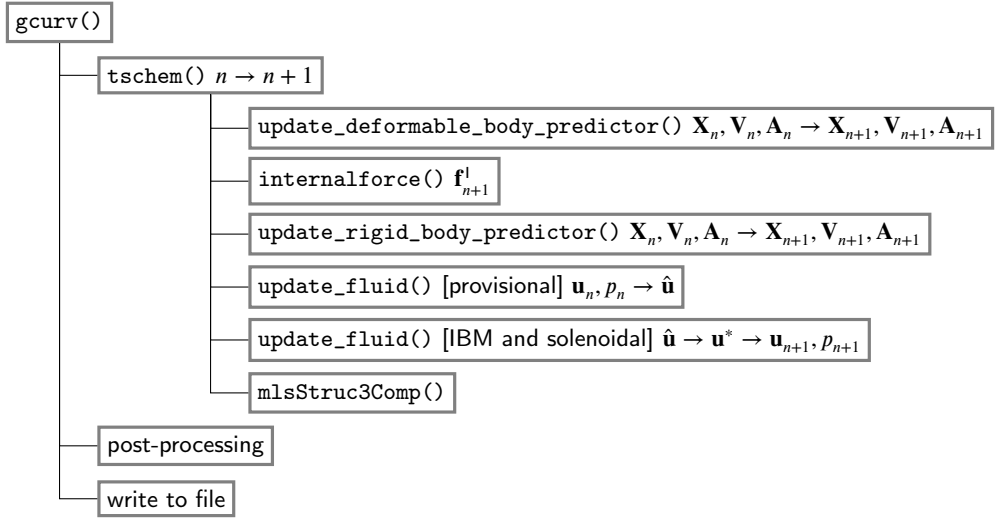
\begin{figure}
  \centering
  \tikzstyle{every node}=[draw=gray, very thick,anchor=west]
  \begin{tikzpicture}[%
    grow via three points={one child at (0.5,-0.7) and %
    two children at (0.5,-0.7) and (0.5,-1.4)},%
    edge from parent path={(\tikzparentnode.south)%
    |- (\tikzchildnode.west)}]
    \node {\code{gcurv()}}
        child { node {\code{tschem()} $n \to n+1$}
           child {node {\code{update_deformable_body_predictor()} $\spos_n, \svel_n, \sacc_n \to \spos_{n+1}, \svel_{n+1}, \sacc_{n+1} $}}
           child {node {\code{internalforce()} $\fint_{n+1}$}}
           child {node {\code{update_rigid_body_predictor()}  $\spos_n, \svel_n, \sacc_n \to \spos_{n+1}, \svel_{n+1}, \sacc_{n+1} $}}
           child {node {\code{update_fluid()} [provisional] $\vel_n, p_n \to \hat{\vel}$}}
           child {node {\code{update_fluid()} [IBM and solenoidal] $\hat{\vel} \to \vel^* \to \vel_{n+1}, p_{n+1}$}}
           child {node {\code{mlsStruc3Comp()}}}
        }
        child [missing] {}
        child [missing] {}
        child [missing] {}
        child [missing] {}
        child [missing] {}
        child [missing] {}
        child { node {post-processing}}
        child { node {write to file}};
  \end{tikzpicture}
  \caption{Main subroutine sequence for one default loose-coupling time step. Optional modules and the strong-coupling corrector loop are omitted for clarity.}
  \label{fig:stacktrace}
\end{figure}
\indent The main time-step sequence is shown in Figure~\ref{fig:stacktrace}. The driver routine \code{gcurv()} calls \code{tschem()}, which first advances the solid predictor. For deformable bodies this includes the evaluation of the internal elastic forces. The fluid is then advanced by \code{update_fluid()} using the fractional-step method: first, the provisional velocity is computed (equation~\ref{eq:nstime_discr}), then the immersed boundary forcing is applied (equation~\ref{eq:mlsForcing}) with either the Lagrangian MLS or the Eulerian sharp-interface IBM and, lastly, the projected velocity and pressure fields are obtained by enforcing mass conservation (equations~\ref{eq:velsolen}--\ref{eq:pressure}). The final operation in the loose-coupling time step is the hydrodynamic-load evaluation, where pressure and velocity-gradient information from the updated fluid state are interpolated to the immersed surface.
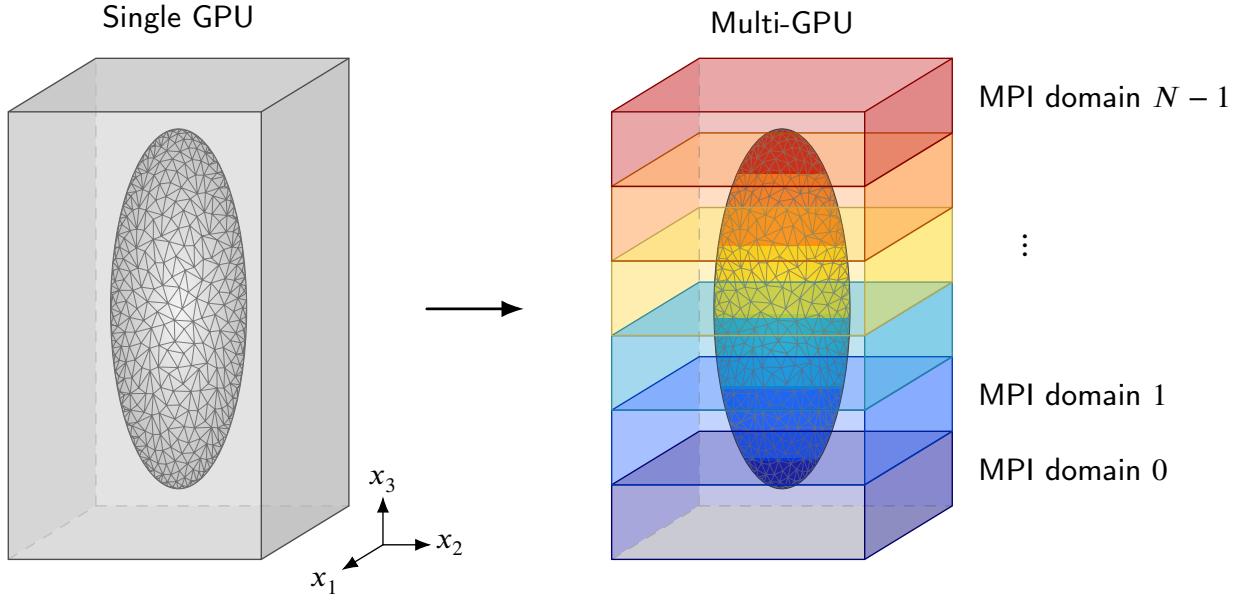
\begin{figure*}
  \centering
  \resizebox{\textwidth}{!}{\input{decomposition.tex}}
  \caption{Eulerian and Lagrangian domain decompositions. The Eulerian fluid arrays are distributed over slab subdomains, while Lagrangian markers are assigned to the MPI ranks whose Eulerian slabs contain the corresponding interpolation or probe stencils. This local Lagrangian ownership is used both for immersed boundary forcing $\fib$ and for hydrodynamic-load evaluation $\fhid$.}
  \label{fig:decomp}
\end{figure*}

\subsubsection{Fluid domain\label{sssec:fluid-parallelization}}
The Navier--Stokes solver uses a one-dimensional MPI decomposition of the Cartesian domain into slabs along the $x_3$ direction \cite{spandan_2017_potential,spandan2018fast}, as sketched in Figure~\ref{fig:decomp}. Eulerian arrays are divided between each slab subdomain, assigned to each MPI rank, storing only the owned range $x_{3,start}:x_{3,end}$ (the rank-local indices \code{kstart} and \code{kend} mark the first and last owned indices of each array, which are allocated respecting this convention in order to mimic the indexing of a global array). Arrays that require stencil operations (like finite difference derivatives) own halo layers of size \code{lvlhalo}, corresponding to a $x_3$-slab spanning \code{kstart-lvlhalo:kend+lvlhalo} to allow easy communication with surrounding subdomains. 
\\ \indent A single halo layer (\code{lvlhalo=1}) is sufficient for the flow equations, whose second-order finite-difference stencils only involve the immediate neighbors of each cell. The evaluation of the hydrodynamic loads is more demanding: the probes of equation~\eqref{eq:force_probe} extend away from the immersed surface by a distance $h$, and the MLS support used to reconstruct pressure and velocity gradients at the probe tip may lie several cells outside the slab owned by the rank.
The halo thickness must therefore be increased, depending on the length of the probe used for the evaluation of hydrodynamic loads.
The length of the probe can vary locally due to the grid stretching, thus the halo thickness is set to $\code{lvlhalo}=\lceil \max_k \left(h^k/\Delta x^k_3\right)+|s|\rceil$, where $h^k$ is the local probe length, $x^k_3$ is the local grid spacing in the $x_3$ direction, $|s|$ is the size of the interpolation stencil and $\lceil\cdot\rceil$ denotes the ceiling function.
When the centred scheme method is used to evaluate the hydrodynamic loads (up to third order accuracy, which reduces to an enhanced second order accuracy when applied to a second order flow solver as in this case, see \cite{vagnoli2026taycorr}), the interpolation stencil is $s=\pm 2$ with respect to the cell containing the probe tip. Consequently, the probe length $h^k$ is taken equal to $2.5$ times the local cell diagonal, to avoid interpolation nodes lying on the opposite side of the immersed interface. The halo layer should thus contain both the probe tip and its corresponding interpolation stencil: on a uniform and isotropic grid  $h=2.5\sqrt{3}\,\Delta x$ and, consequently, $\code{lvlhalo}=7$.
Alternatively, if the hydrodynamic loads are evaluated through the off-centred scheme (up to second order accuracy \cite{vagnoli2026taycorr}) the interpolation stencil is $s=\pm 1$ with respect to the cell containing the probe tip, and a shorter probe equal to $1.5$ time the local cell diagonal can be used. Hence, on a uniform and isotropic grid $h=1.5\sqrt{3}\,\Delta x$ and $\code{lvlhalo}=4$.
These values guarantee that the probe stencil is always contained in the local slab for bodies of arbitrary orientation with respect to the Eulerian grid; for fixed bodies with a known orientation, shorter probes, and hence thinner halos, may be considered.\\
\indent The fluid update follows the fractional-step algorithm described in Section~\ref{sssec:fluidMech}. In the provisional step, the explicit terms are assembled and the viscous terms are advanced either explicitly or through the factored implicit solver. In the latter case, the implicit operator is split into direction-wise tridiagonal solves using Thomas' algorithm (with a Sherman--Morrison perturbation in the two periodic dimensions), avoiding the assembly of a global sparse matrix.\\
\indent The projection step requires the solution of a Poisson equation~\eqref{eq:elliptic}. The solver applies batched two-dimensional real-to-complex fast Fourier transforms (cuFFT) on the local $x_1$--$x_2$ planes, diagonalizing the periodic directions and leaving independent tridiagonal systems along $x_3$ for each horizontal Fourier mode. Since the domain is stored as $x_3$ slabs, the transformed coefficients are transposed before the line solves so that each MPI rank owns complete $x_3$ columns for a subset of modes. After the GPU-accelerated tridiagonal solve, the transpose is reversed and inverse cuFFT transforms return the correction field to physical space.\\
\indent The same transpose interface supports two communication backends. In the default version, GPU buffers are packed and exchanged through MPI all-to-all communication. When the code is compiled with \code{USE_CUDECOMP} compilation flag, these redistributions are performed by the cuDecomp library \cite{romero2022distributed}. The library is also used for the implicit viscous solves when \code{implicit_NS=true}, and for the halo exchanges of the Eulerian fields, which otherwise rely on non-blocking MPI point-to-point communication. Independently of the backend, the additive halo exchange of the immersed boundary force-spreading arrays and the broadcast of the inflow/outflow boundary planes are performed with NCCL collectives, so that these operations remain GPU-resident.

\subsubsection{IBM: Lagrangian and Eulerian methods}\label{sssec:IBM-parallelization}
\noindent The immersed boundary implementation follows the parallelization strategy of \cite{spandan_2017_potential,spandan2018fast}. The Lagrangian geometry is stored in global arrays, while the actual work is assigned locally. At each time step, the code determines the Eulerian cell indices associated with the Lagrangian markers or normal probes and builds rank-local lists of faces or tiles whose interpolation stencils lie inside the local slab. This task-based ownership avoids processing every Lagrangian marker on every rank.\\
\indent MLS interpolation appears in two distinct parts of the algorithm. During the fluid update, the Lagrangian immersed boundary module interpolates the provisional Eulerian velocity to the Lagrangian markers, computes the no-slip correction, and spreads the resulting force density back to the Eulerian grid. This operation is performed on the marker or tile list owned by the current MPI rank. Although we use CUF kernels extensively, these subroutines are implemented as explicit GPU kernels since the $4\times4$ MLS system of equations that has to be solved for each Lagrangian marker to compute the MLS weights is better handled using multiple threads (namely 16) concurrently \cite{viola2022fsei_gpu}.\\
\indent During hydrodynamic-load evaluation, the same MLS framework is used in the opposite direction: pressure and velocity-gradient information are interpolated from the Eulerian grid to the normal probes defined in equation~\eqref{eq:force_probe}, and the reconstructed traction is integrated over the immersed surface. Pressure and viscous contributions are accumulated on triangle barycentres, transferred to vertices for deformable structures, and summed into total forces and torques for rigid bodies. Since each rank owns only part of the Lagrangian work, global hydrodynamic forces, torques, and nodal load arrays are assembled with MPI reductions.\\
\indent The same parallelization strategy is adopted for the Eulerian IBM. At each time step, MPI ownership is determined by constructing a rank-local list of triangles with at least one vertex inside the local slab. For these triangles, the surrounding Eulerian grid nodes are examined to identify the forcing nodes, where the prescribed target velocity is imposed. Both operations are performed within a single explicit GPU kernel.
The sink/source term in equation~\eqref{eq:elliptic} is evaluated directly on the Eulerian grid using a CUF kernel directive.

\subsubsection{Structural solver}\label{sssec:struc-parallelization}
Structural data are stored as Lagrangian arrays containing the current vertex positions, velocities, accelerations, connectivity, and material parameters. The predictor kernels update the nodal state of deformable bodies and the centre-of-mass and quaternion state of rigid bodies. \\
\indent The \code{internalforce} routine evaluates internal elastic loads $\fint_i$ (equation~\eqref{eq:struc_def_dynamics}) partitioning vertex, edge, and face ranges across MPI ranks. On the GPU, the in-plane (edge) contributions are computed with CUF kernel directives, whereas the bending (out-of-plane) contributions are evaluated by the dedicated CUDA kernel \code{compute_fbxyz_kernel}; in both cases the nodal forces are accumulated with atomic operations, since several edges and faces share the same vertex. The partial force arrays are finally summed over MPI ranks, giving a consistent internal-force vector on all MPI ranks. Since these structural operations do not require Eulerian field values, their distribution is based only on the Lagrangian mesh rather than on the spatial position of the vertices within the Eulerian slab decomposition. This makes their distribution more balanced, thus reaching a better scaling performance.

\section{Scaling performance}
\begin{figure*}
  \centering
  \includegraphics[width=0.99\textwidth]{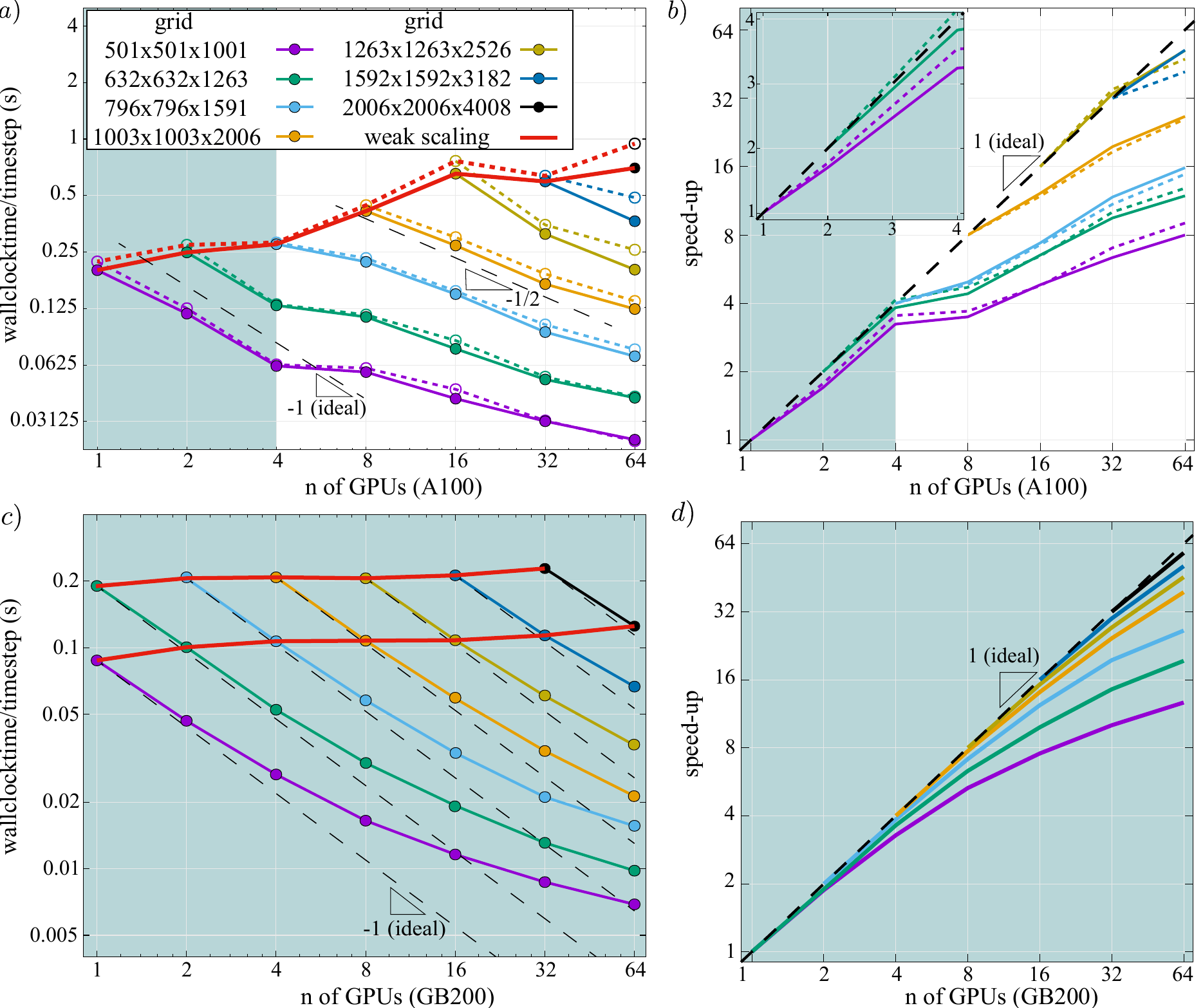}
  \caption{Strong and weak scalability of the code. (a) Wall-clock time per time step as a function of the number of GPUs, $n$, on the Leonardo Booster partition (NVIDIA A100 GPUs, four per node): the gray shaded area highlights the intra-node scaling, i.e. runs on 1--4 GPUs connected through NVLink within the same node; colors denote the Eulerian grid resolutions listed in the legend; empty symbols and dotted lines correspond to runs without \code{cuDecomp} (baseline MPI redistribution path), while filled symbols and solid lines correspond to runs with \code{cuDecomp} \cite{romero2022distributed}; the thick red dotted and solid lines identify the weak-scaling sequences of the two backends. The thin black dashed lines have slope $-1$ (ideal linear scaling) and $-1/2$, as indicated by the reference triangles. (b) Corresponding speed-up: for each grid, the wall-clock time is normalized to the run on the smallest number of GPUs on which the grid fits, and the curves are rescaled so as to start on the ideal dashed line of unit slope; the gray shaded area again marks the intra-node (NVLink) region, which is magnified in the inset. (c) Wall-clock time per time step on the NVIDIA GB200 NVL72 system (GPU devices connected through NVLink), using the \code{cuDecomp} backend: each colored curve shows the strong scaling of one grid, the thin dashed lines indicate the ideal slope $-1$, and the two thick red lines connect the weak-scaling sequences at a fixed load of approximately $0.25\times10^9$ (lower) and $0.5\times10^9$ (upper) grid points per GPU. (d) Corresponding speed-up on the GB200 NVL72 system, normalized as in panel (b).}
  \label{fig:scalability}
\end{figure*}
To assess the computational efficiency and scalability of the solver, a performance analysis was conducted on the Leonardo Booster partition at CINECA. Each compute node hosts four NVIDIA A100 GPUs with 64 GB of memory. Runs using up to four GPUs are therefore confined to a single node and benefit from the intra-node NVLink interconnect, whereas larger runs span multiple nodes and communicate through Leonardo's InfiniBand fabric. A second set of runs was performed on an NVIDIA GB200 NVL72 system, a rack-scale platform in which 72 Blackwell GPUs (paired with Grace CPUs) are connected in a single NVLink domain, so that all the runs up to 64 GPUs benefit from the NVLink interconnect and no inter-node network is traversed.\\
\indent The benchmark problem is a rigid sphere immersed in a fluid domain and subjected to gravity. The sphere surface is initially discretized with approximately 20000 triangular elements, while the surrounding fluid is represented on a structured Eulerian grid. The smallest grid, $N_1 = N_2 = 501$, $N_3=1001$, fills approximately 90\% of the memory of one A100 GPU. Larger cases are obtained by refining the Eulerian grid up to $N_1 = N_2 = 2006$, $N_3=4008$, corresponding to approximately 16 billion grid points. For weak scaling, the grid size is increased with the number of GPUs so that the workload per device remains approximately constant. The Lagrangian mesh is refined consistently with the Eulerian spacing using the dynamic refinement strategy of Section~\ref{sec:IBM_interpolation}, with \code{hdx}=0.7.\\
\indent Figure~\ref{fig:scalability}a reports the wall-clock time per time step as a function of the number of GPUs, $n$, on Leonardo. Open symbols show strong-scaling results obtained with the default MPI redistribution backend for the Poisson and implicit-solver transposes. The intra-node results are close to ideal linear scaling, as indicated by the black dashed reference line (with slope $-1$ in the log--log plot), because communication remains local to the node. When the runs extend beyond one node ($n>4$), the slope decreases as global transpositions and halo exchanges become limited by inter-node communication latency and bandwidth; the black dashed line (slope $-1/2$) is included as a visual guide for this multi-node regime, and the largest grids, which only fit on 16 or more GPUs, follow the same trend.
Filled symbols show the corresponding runs using \code{cuDecomp} for the distributed transposes. The scaling trend is similar to the baseline backend, but the absolute wall-clock time is reduced because the transpose operations are performed through an optimized GPU-aware communication layer. The benefit is most visible in the larger multi-GPU cases, where the Poisson and implicit-solver redistributions represent a significant fraction of the time step, with reductions of the wall-clock time of up to about 20--30\% at 32--64 GPUs.
The thick red dotted (baseline) and solid (\code{cuDecomp}) curves highlight the weak-scaling sequences. The wall-clock time remains nearly constant within a single node and increases more noticeably only when the run spans multiple nodes, reaching below one second per time step at the largest grid and GPU count considered here (about $0.7$~s with \code{cuDecomp} for the 16-billion-point grid on 64 GPUs).
\\ \indent
Figure~\ref{fig:scalability}b shows the corresponding speed-up: for each grid, the wall-clock time is normalized to the run on the smallest number of GPUs on which the grid fits, and the curves are rescaled so as to start on the ideal dashed line of unit slope. Within a single node, i.e. for 1--4 GPUs connected through NVLink (gray shaded area, magnified in the inset), the measured speed-up closely follows the ideal one. Beyond the node boundary the curves bend as inter-node communication comes into play, and the departure from the ideal line is largest for the smallest grids, whose local workload per GPU is not sufficient to hide the communication overhead. Conversely, the larger the grid, the longer the speed-up remains close to ideal, with the largest grids following the ideal line up to 32 GPUs. The baseline (dotted lines) and \code{cuDecomp} (solid lines) backends exhibit a similar parallel efficiency, confirming that the benefit of \code{cuDecomp} lies mainly in the reduction of the absolute wall-clock time rather than in a change of the scaling slope.
\\ \indent
Figure~\ref{fig:scalability}c reports the same analysis on the GB200 NVL72 system, using the \code{cuDecomp} backend for all the grids. Since all the GPUs belong to the same NVLink domain, the sharp change of slope observed on Leonardo when crossing the node boundary is absent, and the wall-clock time decreases monotonically up to 64 GPUs for every grid. A gradual departure from the ideal slope $-1$ (thin dashed guides) is nevertheless visible already at 4--8 GPUs for the smallest grids and slowly accumulates as the GPU count grows. This is a strong-scaling (surface-to-volume) effect rather than a network bottleneck: with the one-dimensional slab decomposition, the $x_3$ extent of the subdomain owned by each GPU shrinks as $N_3/n$, so the halo layers and the transposed buffers become progressively larger relative to the local workload; at 64 GPUs, each GPU of the $501\times501\times1001$ case owns a slab of only about 16 $x_1$--$x_2$ planes, to be compared with a halo of several planes on each side. The absolute performance improves considerably with respect to the A100 system: the $501\times501\times1001$ case requires about $0.09$~s per time step on a single GB200 GPU, roughly half of the A100 time, and about $7$~ms on 64 GPUs, i.e. four times faster than the corresponding Leonardo run. Moreover, owing to the larger on-board memory of the Blackwell GPUs, the $632\times632\times1263$ grid, with twice the points of the smallest case, also fits on a single device. The two thick red lines connect the weak-scaling sequences at a fixed load of approximately $0.25\times10^9$ and $0.5\times10^9$ grid points per GPU: the wall-clock time per time step remains nearly constant across the whole range, increasing by less than about 40\% (15\%) for the lower (upper) sequence, while the problem size grows by a factor of 64 (32). In particular, the largest grid, with about 16 billion points, is advanced in about $0.13$~s per time step on 64 GPUs, more than five times faster than on 64 A100 GPUs.
\\ \indent
Figure~\ref{fig:scalability}d shows the corresponding speed-up on the GB200 NVL72 system, normalized as in panel (b). The curves depart from the ideal line smoothly and progressively later as the grid size increases: the smallest grid attains a speed-up of about 13 at 64 GPUs, to be compared with about 8 on Leonardo, while the largest grid remains within about 15\% of the ideal value. These results indicate that the communication pattern of the solver, dominated by the transposes of the Poisson and implicit solvers, maps efficiently onto NVLink-connected multi-GPU systems, and that the code is ready to exploit the latest generation of GPU architectures without modifications.
\\ \indent 
The solver thus exhibits near-linear scaling within a four-A100 NVLink node. Across Leonardo nodes, transpose and halo communication reduce strong- and weak-scaling efficiency, although cuDecomp lowers absolute time per step by approximately 20–30$\%$ at 32–64 GPUs. On the GB200 NVL72 system, weak-scaling time increases by approximately 15–40$\%$, depending on workload, while strong scaling eventually becomes limited by the decreasing local slab depth.

\section{Test cases}

\noindent In this section, we present a collection of well-established benchmark problems in fluid dynamics and fluid–structure interaction that are commonly used in the literature. These test cases are employed to assess the accuracy, numerical stability, and robustness of the proposed solver, and to validate its ability to reproduce reference solutions across a range of flow regimes and coupling scenarios. Table~\ref{tab:tests} summarizes all test cases and reports the corresponding grid resolution, spatial and temporal discretization parameters ($\Delta x$ and $\Delta t$), Reynolds ($Re$) or Galilei ($Ga$) number, solid-to-fluid density ratio and number of GPUs used. In addition, the table includes the wall-clock time required to advance a single time step for each case, providing an overview of the computational performance of the code.

\begin{table}
\centering
\begin{tabular}{l|c|c|c|c|c|c|c}
test name & grid size & $\Delta x$ & $\Delta t$ & $Re$ or $Ga$ & $\rho_s /\rho_f$ & n. GPU & WCpt [s]\\
\hline
1 turbulent channel & $384\times384\times192$ & - & - & 3173 & - & 4 & 0.04 \\
2 rising/falling sphere & $301\times301\times401$ & 0.02 & $5\cdot10^{-5}$ & 204.45 & $0.1$--$1.2$ & 4 & 0.05 \\
3 sedimentation of oblate spheroid & $201\times201\times301$ & 0.02 & $5\cdot10^{-5}$ & 150 & 2.14 & 4 & 0.03 \\
4 flapping flag & $200\times250\times250$ & 0.02 & $1\cdot10^{-4}$ & 200 & 100 & 2 & 0.032 \\
5 aortic bi-leaflet valve & $400\times400\times1200$ & 0.005 & $5\cdot10^{-5}$ & 6996 & 2 & 4 & 0.11 \\
\end{tabular}
\caption{Summary of the details of each test case employed. WCpt stands for "wall clock time per timestep" and is reported in seconds.}
\label{tab:tests}
\end{table}

\subsection{Test \#1: Turbulent channel flow}
The first test case considers the fully developed turbulent flow of an incompressible Newtonian fluid between two parallel planar walls. This configuration is a standard benchmark for wall-bounded turbulence and is used here to validate the fluid solver in the absence of immersed structures.\\
\indent The flow is driven by a time-variable volume forcing term added to the momentum equation in the streamwise direction. The forcing term is computed in such a way that a constant inlet bulk velocity $U$ is achieved. At statistical equilibrium, this forcing is balanced by the wall shear stress $\tau_w=\rho_f\nu(\partial u_1/\partial x_3)_w$. The simulation is carried out in bulk units, i.e. lengths are normalized by the channel half-height $\delta$ and velocities by the bulk velocity $U$, so that the Reynolds number is $Re=U\delta/\nu=3173$. Using the friction velocity $u_\tau=(\langle \tau_w\rangle/\rho_f)^{1/2}$, defined by the mean wall shear stress, in place of the bulk velocity gives the friction Reynolds number $Re_\tau=u_\tau\delta/\nu$. The bulk Reynolds corresponds to a $Re_\tau=200$ and the results  are thus reported in friction (or wall) units, with $u_i^+=u_i/u_\tau$ and the wall-normal distance from the nearest wall is scaled by $\delta^+=\nu/u_\tau$.

\begin{figure*}
  \centering
  \includegraphics[width=\textwidth]{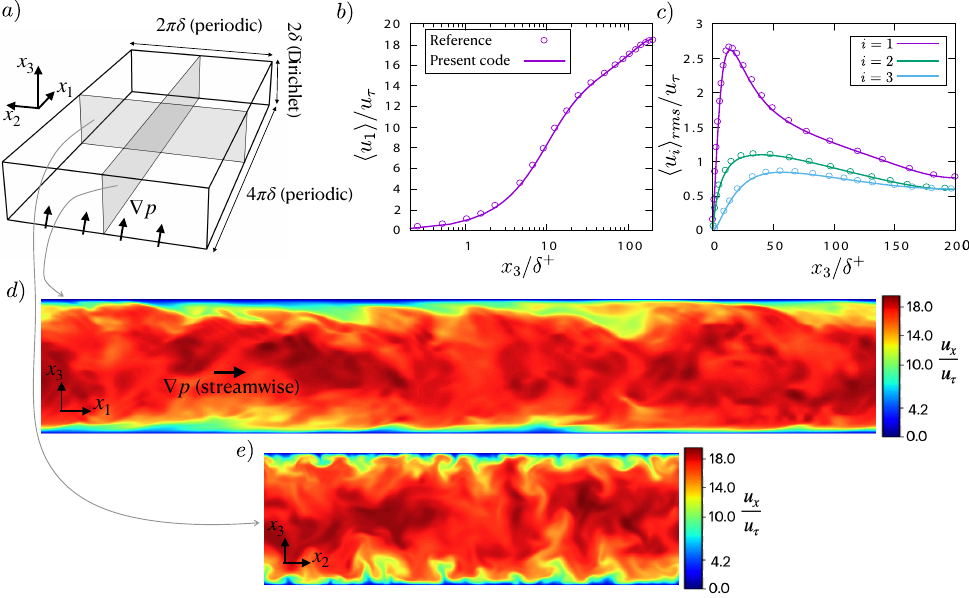}
  \caption{Turbulent channel. Panel (a) shows the computational domain and boundary conditions. Panel (b) reports the space-time averaged streamwise velocity profile $\langle u_1 \rangle$ in the wall-normal direction $x_3$, while panel (c) reports the root-mean-square velocity fluctuations $\langle u_i \rangle_\text{rms}$ in all three directions $x_i$ (reference data from \cite{quadrio2016does_forcing_affects}). Panels (d) and (e) show instantaneous streamwise-velocity fields on streamwise and spanwise slices, whose locations are indicated in panel (a).}
  \label{fig:channel}
\end{figure*}

As shown in Figure~\ref{fig:channel}a, the streamwise, spanwise and wall-normal directions correspond to the Cartesian coordinates $x_1$, $x_2$ and $x_3$, respectively. The two walls are separated by a distance $2\delta$. Periodic boundary conditions are imposed in the homogeneous streamwise and spanwise directions, while no-slip conditions are imposed at the walls. The computational domain has dimensions $4\pi \delta \times 2\pi \delta \times 2\delta$ and is discretized on a grid of $384 \times 384 \times 192$ points. Uniform spacing is used in $x_1$ and $x_2$, whereas a clamped Chebyshev-distributed grid clusters points near the walls in the $x_3$ direction. The resulting $\Delta x_3$ grid spacing ranges from a minimum of $0.002\delta$ near the wall to a maximum of $0.015\delta$ at the center of the channel.\\
\indent The flow is initialized with a Reichardt wall-law mean streamwise profile, symmetric about the channel centerline, and a finite-amplitude three-dimensional sinusoidal perturbation is added to accelerate the development of turbulence.
The flow is evolved for $T_0=1350 \,\delta/U$ time units at a variable timestep such that a constant CFL of 0.1 is achieved until a statistically stationary state is reached, as indicated by stationary wall shear stress and mean velocity profiles. Statistics are then collected over an additional interval $T=1350\,\delta/U$ time units. The reported profiles are obtained by averaging in time and over the two homogeneous directions. Denoting the streamwise and spanwise lengths by $L_1=4\pi\delta$ and $L_2=2\pi\delta$, the mean and root-mean-square (RMS) profiles are computed as

\begin{equation}
\langle u_i \rangle (x_3) =
\frac{1}{TL_1L_2}
\int_{T_0}^{T_0+T}\int_0^{L_1}\int_0^{L_2}
u_i(x_1,x_2,x_3,t)\,dx_2\,dx_1\,dt ,
\end{equation}
\begin{equation}
\langle u_i \rangle_{\mathrm{rms}}(x_3) =
\left(
\frac{1}{TL_1L_2}
\int_{T_0}^{T_0+T}\int_0^{L_1}\int_0^{L_2}
\left( u_i(x_1,x_2,x_3,t) - \langle u_i \rangle(x_3) \right)^2
\,dx_2\,dx_1\,dt
\right)^{1/2}.
\end{equation}
\indent Figure~\ref{fig:channel}b shows that the mean streamwise velocity agrees closely with the reference data of~\cite{quadrio2016does_forcing_affects}, including the near-wall region and the outer-layer profile. The RMS fluctuations in Figure~\ref{fig:channel}c reproduce the expected anisotropy of turbulent channel flow: the streamwise component exhibits the largest near-wall peak, followed by the spanwise and wall-normal components, and all three profiles remain in good agreement with the reference solution across the channel half-height. The instantaneous fields in Figure~\ref{fig:channel}d,e show the corresponding turbulent organization, with elongated streamwise velocity structures and strong wall-normal variations near the walls.

\subsection{Test \#2: Freely falling/rising sphere in a quiescent fluid \label{ssec:rising}}
\noindent This test considers the vertical motion of a single spherical particle in an otherwise quiescent fluid. The configuration is a classical validation problem for particle-resolved FSI solvers \cite{ten2002sphere,yang_and_stern2015}, because it probes both the accuracy of the hydrodynamic force evaluation and the stability of the rigid-body coupling when the particle-to-fluid density ratio is small.\\
\begin{figure}
  \centering
  \includegraphics[width=\textwidth]{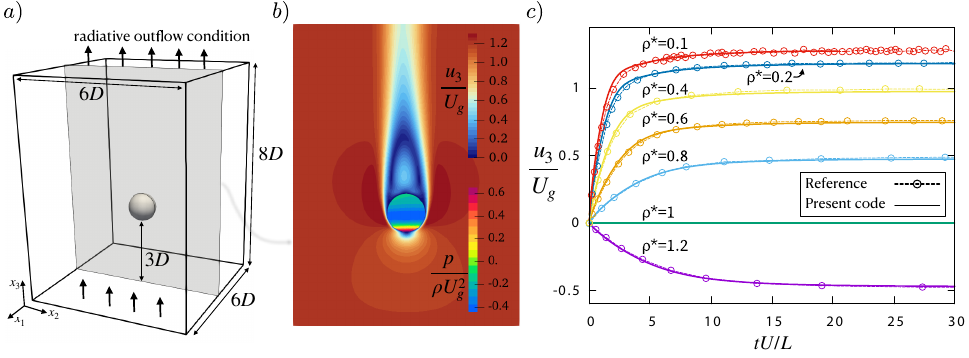}
  \caption{Freely rising/falling sphere in a quiescent fluid. Panel (a) shows the computational domain and boundary conditions in the translating reference frame. Panel (b) shows representative contours of vertical velocity and pressure at the terminal state.
For different density ratios $\rho^*$, panel (c) compares the time evolution of the particle velocity obtained with the present code against the reference data \cite{yang_and_stern2015}.}
  \label{fig:rising}
\end{figure}
\indent The particle has diameter $D$ and density ratio $\rho^*=\rho_s/\rho_f$, where $\rho_s$ and $\rho_f$ are the body and fluid densities, respectively. Gravity is directed along the negative $x_3$ direction. The reference velocity is the gravitational velocity $U_g=\sqrt{gD}$, with $g=|\mathbf{g}|$, and the Galilei number is
\begin{equation}
Ga = \frac{U_g D}{\nu} = 204.45.
\end{equation}
This value matches the benchmark of Yang and Stern~\cite{yang_and_stern2015}, who report $Re=367.4$ based on the terminal velocity and $Fr=U_\infty/\sqrt{gD}=1.797$, giving $Ga=Re/Fr$. Length, velocity, time, and density are nondimensionalized by $D$, $U_g$, $D/U_g$, and $\rho_f$, respectively. The density ratio is varied in the range $\rho^*=0.1$--$1.2$, spanning rising particles ($\rho^*<1$), the neutrally buoyant limit ($\rho^*=1$), and a settling particle ($\rho^*>1$).\\
\indent The non-inertial frame approach described in Section~\ref{ssec:struc} is used for this simulation, and the computational domain has size $6D\times6D\times8D$ and is discretized with uniform grid spacing $\Delta x=0.02D$. Periodic boundary conditions are imposed in the $x_1$ and $x_2$ directions.\\
\indent The rigid-body dynamics follows equations~\eqref{eq:struc_rigid_dynamics}. In FSI simulations, the maximum admissible time-step size is frequently governed by the stability requirements of the structural solver rather than by those of the fluid integration scheme. Accordingly, in the FSI cases presented below, the time step is held constant at a value dictated by structural stability, while ensuring that the corresponding CFL number remains within the bounds required for stable fluid integration.
In particular, here the constant time step $\Delta t=5\times10^{-5} D/U_g$ is used. The no-slip condition is imposed with the Lagrangian MLS IBM. Hydrodynamic loads are evaluated with a probe length $h=2.5\sqrt{3}\,\Delta x$ and second-order Taylor reconstructions for pressure and velocity gradients. Unless otherwise stated, the predictor loose-coupling strategy is used; the most demanding very-light case, $\rho^*=0.1$, is run with the strong-coupling option to suppress added-mass instabilities.\\
\indent Figure~\ref{fig:rising}c reports the time evolution of the vertical particle velocity $u_3/U_g$. The sign of the terminal velocity changes at $\rho^*=1$, as expected: particles lighter than the fluid rise, the neutrally buoyant particle remains at rest, and the heavier particle settles. The present results closely follow the reference curves throughout the acceleration phase and at terminal velocity for all density ratios. The agreement is particularly relevant for the low-density cases, where explicit or loosely coupled partitioned FSI algorithms are most sensitive to added-mass effects. The stable results down to $\rho^*=0.2$ with loose coupling, and the successful $\rho^*=0.1$ computation with strong coupling, demonstrate that the rigid-body update and hydrodynamic-load reconstruction remain robust over a wide density-ratio range. Figure~\ref{fig:rising}b further shows the steady wake and pressure distribution around the body in the translating frame, confirming that the non-inertial formulation captures the expected terminal state while keeping the sphere fixed inside the computational box.

\subsection{Test \#3: Sedimentation of an oblate spheroid}
\noindent The third test case considers the sedimentation of an oblate spheroid in a viscous fluid initially at rest. The body is rigid and freely moves under gravity, so that both the center-of-mass motion and the angular dynamics are determined by the hydrodynamic loads. Compared with the spherical-particle case of Section~\ref{ssec:rising}, this benchmark is more demanding because the wake, the particle orientation, and the lateral force are coupled. It therefore provides a useful validation of the torque evaluation and of the rigid-body FSI update.\\
\begin{figure}
  \centering
  \includegraphics[width=\textwidth]{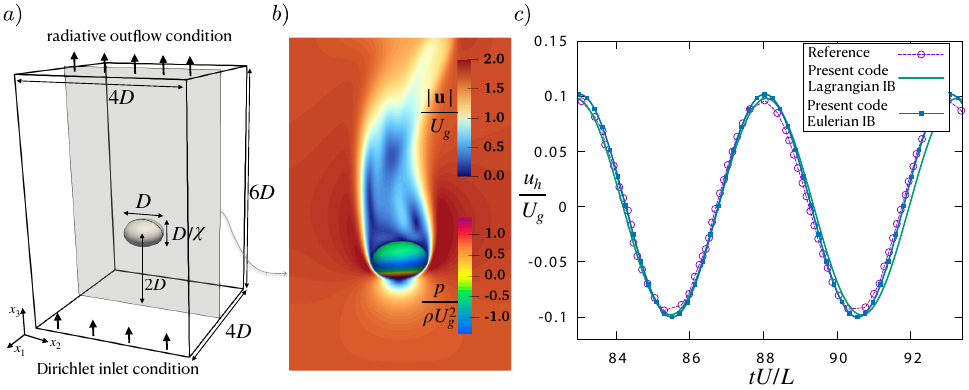}
  \caption{Sedimentation of an oblate spheroid. Panel (a) shows the computational domain and boundary conditions in the translating reference frame. Panel (b) shows representative contours of velocity magnitude and pressure during the terminal periodic regime. Panel (c) compares the signed horizontal particle velocity obtained with the Lagrangian MLS (green curve) and Eulerian sharp-interface (blue curve) IB methods available in the code against the reference data \cite{moricheSingleOblateSpheroid2021}.}
  \label{fig:osp}
\end{figure}
\indent The spheroid is obtained by revolving an ellipse around its minor axis. Its equatorial diameter is $D$, its polar diameter is $D/\chi$, and $\chi$ is the aspect ratio. The particle-to-fluid density ratio is $\rho^*=\rho_s/\rho_f$. Following the convention used in the sedimenting-spheroid benchmarks \cite{zhouPathInstabilitiesOblate2017,moricheSingleOblateSpheroid2021}, the reference velocity is based on the particle volume $V_b=\pi D^3/(6\chi)$,
\begin{equation}
U_g = \left(\frac{\pi |\rho^*-1| g D}{6\chi}\right)^{1/2},
\end{equation}
and the Galilei number is $Ga=U_gD/\nu$, with $\nu$ the kinematic viscosity. In the present validation we use $Ga=150$, $\rho^*=2.14$, and $\chi=1.5$. These parameters correspond to the vertical-periodic regime, in which the mean trajectory remains vertical but the particle develops a sustained oscillatory horizontal velocity and a periodic rotation.\\
\indent Lengths and velocities are nondimensionalized by $D$ and $U_g$, respectively. The computational box and boundary conditions are shown in Figure~\ref{fig:osp}a. Periodic boundary conditions are imposed in the horizontal directions, while the lower and upper boundaries use, respectively, an inflow condition (uniform inflow profile) and the radiative outflow condition of equation~\eqref{eq:radiative}. As for the freely rising/falling sphere, the computation is performed using the non-inertial approach described in Section~\ref{ssec:struc}.\\
The domain is $4D\times4D\times6D$ wide and is discretized with a spatially uniform grid with $\Delta x = 0.02D$. The simulation is advanced with a timestep $\Delta t=5\times10^{-5}D/U_g$.\\
\indent Hydrodynamic loads are evaluated with the same normal-probe reconstruction used in the previous rigid-body test, along the positive surface-normal direction only, with a probe length $h=2.5\sqrt{3}\,\Delta x$ and second-order Taylor corrections for both pressure and velocity-gradient reconstruction. These choices are consistent with the load-reconstruction framework assessed in \cite{vagnoli2026taycorr}.\\
\indent Figure~\ref{fig:osp}b shows representative contours of velocity magnitude and pressure after the transient. The wake is no longer axisymmetric and follows the instantaneous body orientation, as expected in the vertical-periodic regime. Figure~\ref{fig:osp}c compares the signed horizontal velocity $u_h$ with the reference data~\cite{moricheSingleOblateSpheroid2021}. The simulation reproduces the periodic response reported in the benchmark and in the MLS-load reconstruction study \cite{vagnoli2026taycorr}: the oscillation remains centered around zero, its amplitude is of order $10^{-1}U_g$, and its period agrees well with the reference over multiple cycles. This agreement indicates that the solver captures the exchange between rotational and translational motion that controls the lateral force, rather than only the mean settling speed.\\
\indent This test case is repeated using the Eulerian sharp-interface IBM detailed in Section~\ref{sssec:fluidMech} to enforce the no-slip condition on the solid walls. The resulting oscillation of $u_h$ is reported in Figure~\ref{fig:osp}c. As for the Lagrangian MLS IBM, the solver accurately reproduces the correct rotational and translational dynamics of the object, while exhibiting reduced transpiration \cite{vagnoli2026fastconsistentsharpinterfaceimmersed}.

\subsection{Test \#4: Three–dimensional flapping flag in a uniform free stream}\label{ssec:tc:flag}
\begin{figure*}
  \centering
  \includegraphics[width=\textwidth]{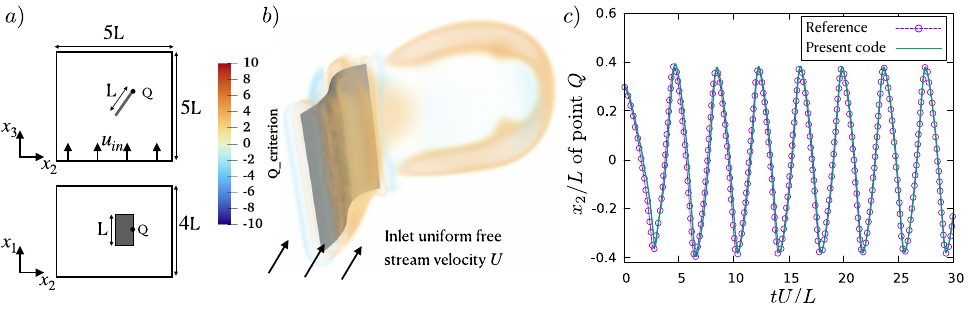}
  \caption{Three-dimensional flapping flag in a uniform free stream. Panel (a) shows the computational domain and the initially tilted flag. Panel (b) shows a representative instantaneous $Q$-criterion field during the periodic flapping regime. Panel (c) compares the transverse displacement of the trailing-edge midpoint $Q$ with the reference data of Huang et al.~\cite{huang2010flag}.}
  \label{fig:flag}
\end{figure*}

\noindent This test is the first validation case involving a deformable structure. It considers the interaction between a uniform free stream and a thin square flag clamped along its leading edge, a standard benchmark for three-dimensional immersed boundary FSI solvers \cite{huang2010flag,deTullio2016MLS}. The case is particularly useful because the long-time response is not a static equilibrium: the flag reaches a sustained flapping regime generated by the coupled action of internal elastic forces, inertia, and vortex shedding.\\
\indent The reference velocity and length are the inlet velocity $U$ and the flag side length $L$, respectively. The Reynolds number is $Re=UL/\nu=200$. The flag thickness is $s=0.01L$ and the density ratio is $\rho_s/\rho_f=100$. The structure is modeled as a thin deformable surface with the spring-mass network described in Section~\ref{ssec:struc} and in \cite{deTullio2016MLS}. A large in-plane stiffness, $E=10^3\,\rho_f U^2L/s$, is prescribed to keep the sheet nearly inextensible, while the bending stiffness is $B=10^{-4}\,\rho_f U^2L^3$.\\
\indent As described in Figure~\ref{fig:flag}a, the computational domain has size $4L\times5L\times5L$ and is discretized by $200\times250\times250$ grid points, corresponding to a uniform spacing $\Delta x=0.02L$. The time step is $\Delta t=10^{-4}L/U$. The inflow is imposed along the $x_3$ direction (uniform inflow profile), periodic boundary conditions are used in the transverse directions, and the outflow uses the radiative condition of equation~\eqref{eq:radiative}. The flag is initially planar, with the leading edge located two length units from the inlet and tilted by $\theta_0=0.1\pi$ with respect to the $x_1$--$x_3$ plane to trigger the flapping instability. The no-slip condition is imposed with the Lagrangian MLS IBM. Since the flag is a thin open surface wet on both sides, hydrodynamic loads are evaluated on both normal directions using a probe length $h=2.5\sqrt{3}\,\Delta x$ and second-order Taylor reconstruction of pressure and velocity gradients.\\
\indent Figure~\ref{fig:flag}b illustrates the fully developed regime. The flag rolls into a three-dimensional flapping shape and the wake organizes into an alternating pattern synchronized with the trailing-edge motion. The quantitative comparison is shown in Figure~\ref{fig:flag}c, where the transverse coordinate $x_2(t)$ of the trailing-edge midpoint $Q$ is plotted against the reference solution of Huang et al.~\cite{huang2010flag}. After the initial transient, the signal converges to a periodic limit cycle. The present computation reproduces both the peak-to-peak amplitude and the oscillation frequency with very high accuracy over several consecutive periods, with no visible drift in phase. This agreement is consistent with the results reported in the MLS hydrodynamic-load reconstruction study \cite{vagnoli2026taycorr} and confirms that the coupled solver captures the elastic-inertial balance controlling the flag response, as well as the hydrodynamic forcing generated by the surrounding vortical flow. It is worth noting that, owing to the large density ratio $\rho_s/\rho_f=100$, the added-mass contribution is weak and the default loose-coupling strategy (\code{fsi_tol = 0}) is sufficient to obtain a stable limit cycle, without resorting to the strong-coupling corrector loop of Figure~\ref{fig:coupling}. The whole simulation is run on 2 GPUs and requires about $0.032$~s of wall-clock time per time step (Table~\ref{tab:tests}), so that several flapping periods are advanced within a few hours.

\subsection{Test \#5: Bileaflet aortic mechanical valve}
\noindent The final test case considers the pulsatile non-Newtonian flow through a bileaflet mechanical prosthetic aortic valve. This configuration is more complex than the previous validation cases because the moving immersed bodies, the confined aortic-root geometry, the pulsatile inflow, the non-Newtonian Carreau-Yasuda rheology, and the transitional downstream flow all interact. For this exact combination of geometry, waveform, and leaflet model, there is not a single consolidated benchmark data set that is as widely used as the channel, particle-settling, or flag cases. The purpose of the test is therefore to demonstrate that the solver can handle a realistic biomedical FSI configuration and reproduce the expected qualitative valve dynamics.\\
\begin{figure*}
  \centering
  \includegraphics[width=\textwidth]{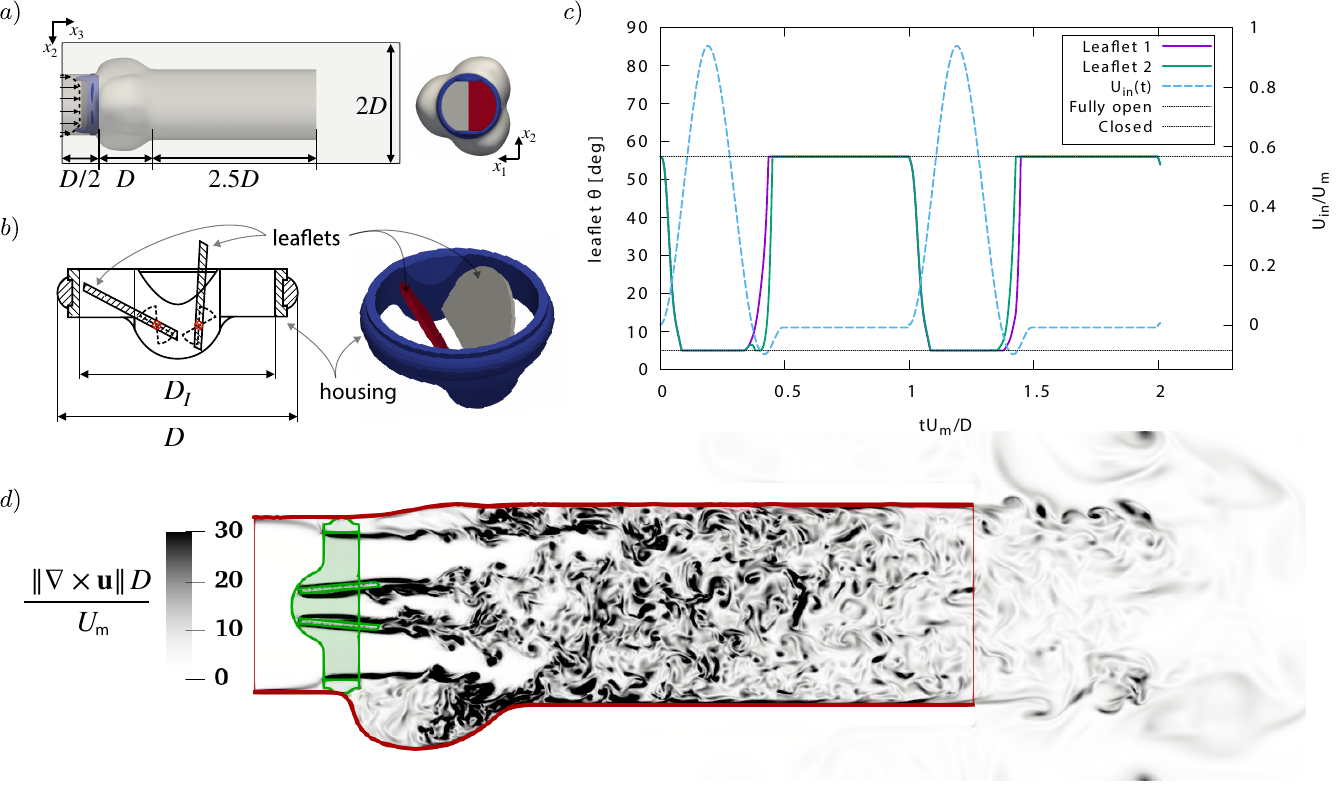}
  \caption{Bileaflet mechanical aortic valve in pulsatile flow. Panel (a) shows the idealized aortic-root domain and the computational setup. Panel (b) shows the valve housing, leaflets, and angular constraints. Panel (c) reports the prescribed flow-rate waveform and the leaflet opening angle $\theta$ over two cardiac cycles. Panel (d) shows the vorticity magnitude $\lVert\nabla\times\vel\rVert$ during systolic flow.}
  \label{fig:mech}
\end{figure*}
\indent Figure~\ref{fig:mech}a shows the computational domain. The valve is mounted in an idealized straight aortic tract with three sinus-like expansions downstream of the inlet section. The aortic diameter is $D=21~\unit{\mm}$, which is also used as the reference length. The computational box has nondimensional size $2\times2\times6$, with the aortic axis aligned with $x_3$, and is discretized on a uniform Cartesian grid of $400\times400\times1200$ points, resulting in $\Delta x=0.005D$. The peak inlet velocity is $U_m=1.1~\unit{\meter/\second}$, giving $Re=U_mD/\nu=6996$. The dimensionless cardiac period is $T=39.29$, corresponding to 80 bpm, and the time step is $\Delta t=5\times10^{-5}D/U_m$.\\
\indent Figure~\ref{fig:mech}b details the prosthesis. The geometry is CAD-reconstructed from a commercial bileaflet mechanical valve, 
with the microscopic hinge mechanism not explicitly resolved. The housing is fixed, while the two leaflets are treated as rigid bodies of density $\rho_s=2\rho_f$. Their motion follows equation~\eqref{eq:struc_rigid_dynamics}, constrained to one rotational degree of freedom about the hinge axis. At each time step the hydrodynamic torque is projected onto this axis and the leaflet angle is limited between the closed and fully open configurations. When a limiting angle is reached, an inelastic impact condition is applied by setting the angular momentum to zero. 
The maximum opening angle with respect to the stream-wise direction ($\theta=0$) is $\theta=5^\circ$; this leaves a small angle with the streamwise direction and favours closure when the pressure gradient reverses (as well as the flow rate) as in \cite{de2009dns_mech_valve}.
The fully closed configuration of the chosen prosthesis corresponds to $\theta=70^\circ$. In our setup, however, the actual fully closed position is limited to an angle of $\theta=56^\circ$: a small gap of $14^\circ$ is always left between the leaflets and the housing in order to allow some backflow when the valve is closed and to avoid very high values of the velocity in the quasi-closed configuration where cavitation might become an issue, as in \cite{de2009dns_mech_valve}.\\
\indent The valve is mounted at a $90^\circ$ axial orientation following the nomenclature described in G\"ulan and Holzner~\cite{gulan2018valve_axial_orientation}.\\
\indent A pulsatile inflow condition is prescribed at the inlet section along $x_3$,
\begin{equation}
  U_\text{in}(t,r) =
  U(t)\frac{\tanh\left[20(1-2r/D)\right]}{\tanh(20)} ,
\end{equation}
where $r=\sqrt{x_1^2+x_2^2}$ and $U(t)$ is selected so that the integral over the inlet section gives the imposed flow rate (hyperbolic-tangent inflow profile, \code{radiative_cond = 2}), as outlined by the dashed line in Figure~\ref{fig:mech}c. The waveform shown in Figure~\ref{fig:mech}c includes forward systolic flow, a short regurgitation phase, and a small diastolic backflow used to keep the valve closed until the next cycle. The average flow rate is approximately $4~\unit{\liter/\min}$. The outflow uses the radiative condition of equation~\eqref{eq:radiative}. To reduce outlet reflections and instabilities at this Reynolds number, a sponge layer is added in the final part of the domain (controlled by the \code{isponge}, \code{sponge_zstart}, \code{sponge_zend} and \code{sponge_strength} input variables) by smoothly increasing the viscosity from $\nu=U_m D/Re$ to $5U_m D/Re$ over $5.0D<x_3<5.25D$. The no-slip condition on the housing and leaflets is imposed with the Lagrangian MLS IBM. Hydrodynamic loads are evaluated with the normal-probe method using $h=2.5\sqrt{3}\,\Delta x$ and second-order Taylor reconstructions of pressure and velocity gradients.\\
\indent The leaflet kinematics in Figure~\ref{fig:mech}c show rapid opening during systolic acceleration, followed by a plateau at the maximum opening angle. The two leaflets remain nearly synchronized during opening, indicating that the axial asymmetry of the configuration, with one leaflet  positioned between two Valsalva sinuses and the other leaflet aligned with one sinus (as evidenced in Figure~\ref{fig:mech}d), has only a limited effect on the opening dynamics. During flow deceleration and reversal, the leaflets close sharply and small cycle-to-cycle differences appear during the initial phase of closure. This is consistent with previous DNS observations for bileaflet mechanical valves, where opening was found to be comparatively insensitive to the precise accelerating part of the inlet waveform, whereas closure was strongly affected by the flow-rate descent and by the amount of regurgitation \cite{deVita2016non-newtonian,deTullio2012mechHemolAorticValveProstheses}. In the present simulation, vortical structures generated during closure persist into the following cycle and slightly perturb the instantaneous pressure and torque acting on the leaflets; this makes the closing phase more sensitive to inflow and outflow boundary conditions than the opening phase. Moreover, the absence of post-impact rebounds during the closing phase may be partly due to the inelastic impact condition used at the angular stops.\\
\indent Figure~\ref{fig:mech}d shows the vorticity magnitude during systolic flow. The three jets formed by the two leaflet gaps and the central orifice break down downstream of the valve and generate a strongly three-dimensional vortical wake. High-vorticity layers are shed from the leaflet tips and housing edges, then interact with the confined downstream flow. The field illustrates why this case is a demanding test for the solver: the rigid-body motion must remain stable while the immersed boundary forcing resolves thin shear layers, vortex shedding, and the transition from organized jets near the valve to a more complex wake farther downstream.

\section{Discussion}\label{sec:disc}
\noindent This work presents an open-source solver for incompressible fluid--structure interaction problems on GPU-accelerated distributed-memory systems. The code combines a finite-difference fractional-step Navier--Stokes solver with Lagrangian (MLS) and Eulerian (sharp-interface) immersed boundary methods, rigid-body dynamics and a lightweight structural model for deformable surfaces. Its objective is to provide a transparent and scalable research code for high-resolution DNS of FSI problems.\\
\indent The main numerical advantage of the immersed boundary formulation is that moving and deforming bodies are represented on a Lagrangian mesh embedded in a fixed Cartesian grid. This avoids body-fitted mesh deformation and remeshing, which are often limiting steps in ALE approaches for large displacements or long time integrations \cite{cheng2019openifem}. The no-slip condition can be imposed either with the Lagrangian MLS IBM or with the Eulerian sharp-interface IBM, which substantially reduces spurious transpiration, while in both cases the same MLS normal-probe framework reconstructs the hydrodynamic loads from Eulerian pressure and velocity-gradient fields. This gives a consistent route from the fluid solver to structural forces and torques, while keeping the algorithm relatively simple to implement and parallelize.\\
\indent A distinctive feature of the present release is its HPC-oriented design. MPI domain decomposition and CUDA execution are part of the core implementation, allowing the solver to run efficiently on multi-GPU workstations and modern GPU clusters. This places the code between compact immersed boundary solvers such as cuIBM, PetIBM and WaterLily.jl \cite{layton2011cuibm,chuang2018petibm,weymouth2025waterlily}, and larger commercial multiphysics environments that are broader in scope but generally less specialized for GPU-resident DNS. The default loose-coupling strategy remains inexpensive and effective for many inertia-dominated configurations, while the optional corrector loop and higher-order structural scheme provide a path for more demanding strongly coupled cases.\\
\indent The test cases reported in this paper cover the main operating regimes of the software: a turbulent channel for the fluid solver, freely settling or rising rigid bodies, an oblate spheroid with coupled translation and rotation, a deformable flapping flag and a pulsatile mechanical-valve configuration. Together, these examples show that the code can reproduce established benchmark behavior while retaining scalability on large grids. The present implementation is intentionally specialized: it does not aim to replace general finite-element structural solvers or full image-to-simulation cardiovascular pipelines. Its strength is instead the ability to set up complex moving-boundary problems with modest geometric preprocessing and to run them at DNS resolution on contemporary GPU-based HPC systems.
The current release also has some structural limitations. First, the pressure Poisson equation is solved with fast Fourier transforms in the $x_1$ and $x_2$ directions, which requires these directions to be periodic and discretized with uniform grid spacing; grid stretching and non-periodic boundary conditions are therefore restricted to the $x_3$ direction. Second, the Eulerian domain is decomposed into slabs along $x_3$ only. This one-dimensional decomposition limits strong scaling at large GPU counts, because the number of $x_1$--$x_2$ planes owned by each GPU decreases as $N_3/n$ and the halo and transpose communication becomes a growing fraction of the local workload (Figure~\ref{fig:scalability}c,d).

\section*{Code availability}
IB-Flows is released as open-source software under the BSD-3-Clause licence and is available at {\url{https://gitlab.com/gssi-fluids}}, together with the input files of the test cases presented in this paper and the documentation of the input variables. The code is written in CUDA Fortran and requires the NVIDIA HPC SDK compilers, an MPI library, HDF5, and the cuFFT, cuRAND, NVTX and NCCL libraries (all shipped with the HPC SDK). NCCL is used for the GPU-to-GPU exchange of the halo layers of the immersed boundary force-spreading arrays and for the broadcast of the inflow/outflow boundary planes, whereas HDF5 is used only for the optional restart files (\code{dump_h5 = true}); the three-dimensional fields are written in VTK format through MPI I/O, as described in Section~\ref{sssec:fluid-parallelization}. The cuDecomp library (used for the distributed transposes and halo exchanges, Section~\ref{sssec:fluid-parallelization}) is an optional dependency, as is OpenVDB, whose volumetric output is still under development in the present release.

\section*{CRediT authorship contribution statement}
\textbf{Giovanni Vagnoli:} Conceptualization, Methodology, Software, Validation,  Formal analysis, Data curation, Writing -- original draft.
\textbf{Martino Andrea Scarpolini:} Conceptualization, Methodology, Software, Validation,  Formal analysis, Data curation, Writing -- original draft.
\textbf{Fabio Guglietta:} Conceptualization, Methodology, Writing -- review \& editing.
\textbf{Joshua Romero:} Software, Methodology, Writing -- review \& editing.
\textbf{Massimiliano Fatica:} Software, Methodology, Writing -- review \& editing.
\textbf{Roberto Verzicco:} Conceptualization, Methodology, Writing -- review \& editing, Supervision.
\textbf{Francesco Viola:} Conceptualization, Methodology, Software, Writing -- review \& editing, Supervision, Project administration, Funding acquisition.

\section*{Declaration of competing interest}

The authors declare that they have no known competing financial interests or personal relationships that could have appeared to
influence the work reported in this paper.

\section*{Data availability}
The source code, the input files and the reference data of all the test cases presented in this paper are available in the repository indicated in the Program summary.

\section*{Acknowledgments}
This project has received funding from the European Research Council (ERC) under the European Union’s Horizon Europe research and innovation program (Grant No. 101039657, CARDIOTRIALS to F.V.). Support/funding from Tor Vergata University project AI4HEART and INFN/FIELDTURB project are also acknowledged. The authors thank Prof. Luca Biferale for fruitful discussions.

\bibliographystyle{elsarticle-num-names}
\bibliography{refs}


\end{document}

%% file: setup.tex
\begin{tikzpicture}[font=\small, line cap=round, line join=round,
  >=Latex, scale=1]
\definecolor{ib}{HTML}{8ADBD2}
\definecolor{ibd}{HTML}{2A9D8F}
\definecolor{mk}{HTML}{F28C28}
\begin{scope}[shift={(0.3,-2.7)}, scale=1.35]
\begin{scope}[shift={(0,0)}]
  \def\W{2.6}\def\H{4.6}\def\dx{0.9}\def\dy{0.55}
  \fill[gray!22] (\dx,\dy) rectangle ++(\W,\H);
  \fill[gray!35] (0,0) -- (\dx,\dy) -- (\dx,\dy+\H) -- (0,\H) -- cycle;
  \fill[gray!30] (0,0) -- (\W,0) -- (\W+\dx,\dy) -- (\dx,\dy) -- cycle;
  \draw[gray!60, dashed, thin] (0,0) -- (\dx,\dy) -- (\W+\dx,\dy)  (\dx,\dy) -- (\dx,\dy+\H);
  \shade[inner color=ib!35!white, outer color=ibd] (\W/2+\dx/2,\H/2+\dy/2) ellipse (0.7 and 1.85);
  \fill[gray!45, opacity=0.6] (0,\H) -- (\W,\H) -- (\W+\dx,\H+\dy) -- (\dx,\H+\dy) -- cycle;
  \fill[gray!55, opacity=0.6] (\W,0) -- (\W+\dx,\dy) -- (\W+\dx,\H+\dy) -- (\W,\H) -- cycle;
  \fill[gray!20, opacity=0.25] (0,0) rectangle (\W,\H);
  \draw[black!70, thin] (0,0) rectangle (\W,\H);
  \draw[black!70, thin] (0,\H) -- (\dx,\H+\dy) -- (\W+\dx,\H+\dy) -- (\W,\H);
  \draw[black!70, thin] (\W+\dx,\H+\dy) -- (\W+\dx,\dy) -- (\W,0);
  \node[font=\small\bfseries] at (-0.35,\H+\dy+0.25) {a)};
  \node[above] at (\W/2+\dx/2,\H+\dy+0.05) {Eulerian box (fluid domain)};
  \draw[->] (1.2,-0.05) to[bend left=30] (\W/2+\dx/2-0.22,\H/2+\dy/2-1.78);
  \node[align=center, below] at (1.2,-0.05) {Immersed body\\(solid)};
  \begin{scope}[shift={(\W+\dx+0.35,0.15)}]
    \draw[->] (0,0) -- (0,0.45) node[above, inner sep=1pt] {$x_3$};
    \draw[->] (0,0) -- (0.45,0) node[right, inner sep=1pt] {$x_2$};
    \draw[->] (0,0) -- (-0.45*\dx/1.055,-0.45*\dy/1.055) node[below left, inner sep=0.5pt] {$x_1$};
  \end{scope}
\end{scope}
\draw[->] (3.9,2.2) -- (4.5,2.2);
\begin{scope}[shift={(4.9,0)}]
  \coordinate (c) at (1.7,2.25);
  \def\a{0.95}\def\b{2.05}
  \shade[inner color=ib!35!white, outer color=ibd] (c) ellipse [x radius=\a, y radius=\b];
  \begin{scope}
    \clip (c) ellipse [x radius=\a, y radius=\b];
    \clip plot[domain=-0.3:3.7, samples=60, variable=\x] ({\x},{1.9+0.09*sin(deg(3.5*\x))}) -- (3.7,-0.5) -- (-0.3,-0.5) -- cycle;
    \input{setup_mesh.tex}
  \end{scope}
  \draw[ibd!50!black, very thin] (c) ellipse [x radius=\a, y radius=\b];
  \begin{scope}
    \clip plot[domain=-0.3:3.7, samples=60, variable=\x] ({\x},{1.9+0.09*sin(deg(3.5*\x))}) -- (3.7,5.4) -- (-0.3,5.4) -- cycle;
    \fill[gray!25, opacity=0.55] (0,1.6) rectangle (3.4,5.15);
    \draw[gray!70, step=0.1cm, very thin, opacity=0.7] (0,1.6) grid (3.4,5.15);
    \draw[gray!70] (0,1.6) rectangle (3.4,5.15);
  \end{scope}
  \draw[thick] plot[domain=-0.3:3.7, samples=60, variable=\x]
     ({\x},{1.9+0.09*sin(deg(3.5*\x))});
  \coordinate (h) at (1.3,1.1);
  \draw[black, thin] (h) ++(0.16,0) -- ++(0.06,0.14) -- ++(-0.12,0.1) -- ++(-0.15,-0.02) -- ++(-0.09,-0.13) -- ++(0.06,-0.14) -- ++(0.15,-0.03) -- cycle;
  \coordinate (h1) at (1.46,1.1);
  \coordinate (h2) at (1.26,1.34);
  \node[font=\small\bfseries] at (-0.55,5.4) {b)};
  \node[above] at (2.6,5.2) {Cartesian grid};
  \draw[->] (2.85,5.2) to[bend left=40] (2.55,4.85);
  \node[below] at (0.75,-0.2) {Lagrangian grid};
  \draw[->] (0.95,-0.15) to[bend left=30] (1.2,0.3);
\end{scope}
\end{scope}
\begin{scope}[shift={(13.0,-1.4)}, scale=1.3]
\begin{scope}[shift={(1.5,3.5)}]
  \begin{scope}[scale=1.6, every node/.style={transform shape=false}]
    \input{setup_patch.tex}
  \end{scope}
  \node[font=\small\bfseries] at (-1.9,1.75) {c)};
  \node[above] at (0.4,1.6) {triangle vertices};
  \draw[->] (0.75,1.58) to[bend left=30] (ptop);
  \node[below] at (0.7,-1.4) {Lagrangian marker};
  \draw[->] (0.7,-1.36) to[bend right=30] (pmark);
\end{scope}
\draw[gray!70, thin] (h1) -- (z1);
\draw[gray!70, thin] (h2) -- (z2);
\begin{scope}[shift={(0,-2.05)}]
  \def\g{0.7}
  \coordinate (cc) at (2.95,-0.75);   
  \def\rr{2.667}                        
  \coordinate (Xl) at (1.60,1.55);
  \begin{scope}
    \clip (-0.15,-0.15) rectangle (2.95,2.95);
    \fill[ib!35] (cc) circle (\rr);
  \end{scope}
  \begin{scope}\clip (-0.15,-0.15) rectangle (2.95,2.95);\draw[thick, ib!80!black] (cc) ++(178:\rr) arc (178:65:\rr);\end{scope}
  \foreach \i in {0,...,4} { \draw (\i*\g,-0.15) -- (\i*\g,2.95); \draw (-0.15,\i*\g) -- (2.95,\i*\g); }
  \foreach \i in {0,...,4} \foreach \j in {0,...,4} { \fill (\i*\g,\j*\g) circle (1.4pt); }
  \draw[red, densely dashed, thick] ($(Xl)+(-1.5*\g,-1.5*\g)$) rectangle ($(Xl)+(1.5*\g,1.5*\g)$);
  \foreach \ang in {160.4,140.4,100.4} {
    \fill[mk, draw=black, very thin] (cc) ++(\ang:\rr) ++(-1.5pt,-1.5pt) rectangle ++(3pt,3pt);
  }
  \fill[red, draw=black, very thin] (Xl) ++(-1.8pt,-1.8pt) rectangle ++(3.6pt,3.6pt);
  \node[above right, inner sep=1pt] at ($(Xl)+(-0.03,0.12)$) {$\mathbf{X}_l$};
  \node[above right, inner sep=1pt] at (0.72,2.12) {$\mathbf{x}_k$};
  \draw[<->] (0.7,3.15) -- (1.4,3.15) node[midway, above] {$\Delta x_2$};
  \draw[<->] (3.15,1.4) -- (3.15,2.1) node[midway, right] {$\Delta x_1$};
  \node[font=\small\bfseries] at (-0.5,3.3) {d)};
  \node[above] at (2.65,3.05) {support domain};
  \draw[->] (2.35,3.05) to[bend left=15] (1.85,2.63);
  \node[below] at (1.4,-0.4) {selected Lagrangian marker};
  \draw[->] (1.4,-0.35) to[bend right=15] ($(Xl)+(0.0,-0.1)$);
\end{scope}
\end{scope}
\end{tikzpicture}

%% file: setup_mesh.tex
\draw[ibd!70!black, line width=0.3pt, opacity=0.9] (2.251,1.536)--(2.354,1.318) (1.437,0.777)--(1.360,0.590) (2.479,3.157)--(2.520,3.272) (1.966,4.115)--(1.946,4.230) (2.043,2.599)--(2.060,2.847) (1.928,2.733)--(2.043,2.599) (1.642,4.181)--(1.571,4.228) (1.823,0.439)--(1.793,0.291) (2.399,0.865)--(2.372,0.800) (1.200,3.171)--(1.229,3.412) (1.751,2.241)--(1.843,2.443) (1.718,4.093)--(1.837,4.130) (2.577,1.478)--(2.593,1.549) (1.780,1.726)--(1.534,1.753) (2.055,1.852)--(2.114,1.603) (2.623,2.002)--(2.646,2.071) (2.042,2.156)--(2.238,2.196) (1.846,0.678)--(1.756,0.551) (2.130,3.937)--(2.193,3.965) (1.342,3.809)--(1.185,3.657) (2.076,0.907)--(2.170,0.853) (1.223,1.251)--(1.106,1.105) (0.895,3.226)--(0.839,3.116) (2.550,2.362)--(2.633,2.445) (1.461,3.348)--(1.515,3.576) (1.471,0.583)--(1.408,0.442) (2.402,1.127)--(2.478,1.074) (1.780,2.010)--(1.650,1.861) (2.319,1.078)--(2.413,0.972) (1.408,0.442)--(1.472,0.328) (1.327,3.066)--(1.200,3.171) (0.989,1.797)--(0.856,1.642) (2.058,3.825)--(2.172,3.771) (0.852,1.468)--(0.812,1.535) (1.375,0.324)--(1.454,0.270) (1.781,3.963)--(1.718,4.093) (0.958,3.325)--(0.896,3.335) (1.542,0.825)--(1.437,0.777) (2.393,3.341)--(2.469,3.335) (0.751,2.235)--(0.750,2.250) (1.935,0.567)--(1.823,0.439) (1.781,0.855)--(1.846,0.678) (1.160,1.828)--(0.989,1.797) (0.798,1.716)--(0.782,1.719) (1.601,2.635)--(1.469,2.525) (2.458,1.632)--(2.568,1.587) (1.025,0.807)--(0.972,0.932) (1.704,3.210)--(1.777,3.403) (1.205,1.032)--(1.107,0.928) (1.304,0.387)--(1.299,0.392) (0.924,1.250)--(0.877,1.225) (0.943,1.083)--(0.972,0.932) (1.107,0.928)--(1.132,0.754) (1.364,3.524)--(1.432,3.724) (1.225,4.025)--(1.155,3.929) (1.039,2.470)--(0.846,2.448) (2.287,3.293)--(2.387,3.158) (0.801,2.295)--(0.764,2.408) (1.200,3.171)--(1.094,3.195) (2.381,3.517)--(2.428,3.568) (2.016,3.988)--(2.130,3.937) (2.080,3.625)--(2.058,3.825) (2.350,1.798)--(2.458,1.632) (1.257,3.844)--(1.187,3.951) (2.400,3.633)--(2.372,3.700) (2.193,3.965)--(2.175,4.025) (1.334,1.244)--(1.223,1.251) (1.718,4.093)--(1.642,4.181) (2.503,2.592)--(2.585,2.531) (2.055,1.852)--(2.220,1.763) (1.584,0.621)--(1.471,0.583) (1.946,0.270)--(2.025,0.324) (2.224,1.068)--(2.341,0.894) (1.808,2.711)--(1.924,2.998) (1.846,0.678)--(1.935,0.567) (2.527,1.715)--(2.610,1.730) (1.077,1.640)--(0.942,1.625) (0.942,1.625)--(0.924,1.422) (1.596,0.319)--(1.616,0.209) (1.350,2.485)--(1.249,2.344) (2.377,2.618)--(2.445,2.444) (0.921,3.083)--(0.895,3.226) (1.461,3.348)--(1.364,3.524) (2.025,0.324)--(2.101,0.392) (2.298,3.483)--(2.393,3.341) (2.261,3.064)--(2.287,3.293) (1.534,1.753)--(1.683,1.548) (1.745,4.297)--(1.783,4.292) (2.064,0.375)--(2.101,0.392) (1.273,1.788)--(1.160,1.828) (2.058,3.825)--(2.016,3.988) (1.175,3.833)--(1.123,3.877) (1.487,4.234)--(1.535,4.269) (2.446,1.863)--(2.527,1.715) (1.312,0.995)--(1.205,1.032) (0.839,3.116)--(0.807,2.951) (1.187,3.951)--(1.155,3.929) (2.096,0.493)--(2.064,0.375) (1.542,0.825)--(1.584,0.621) (1.098,2.262)--(0.942,2.342) (1.015,3.554)--(0.972,3.568) (0.942,2.342)--(0.874,2.207) (2.303,2.016)--(2.350,1.798) (0.924,1.422)--(0.924,1.250) (0.921,3.083)--(0.839,3.116) (1.185,3.657)--(1.175,3.833) (0.807,2.951)--(0.782,2.781) (2.428,0.932)--(2.478,1.074) (2.280,2.450)--(2.377,2.618) (2.103,0.670)--(2.014,0.505) (1.894,4.215)--(1.944,4.231) (2.577,1.478)--(2.561,1.384) (1.534,4.100)--(1.487,4.234) (2.287,3.293)--(2.298,3.483) (2.319,1.078)--(2.402,1.127) (0.812,1.535)--(0.839,1.384) (2.521,1.969)--(2.635,1.897) (1.175,1.535)--(1.077,1.640) (2.561,3.116)--(2.523,3.275) (1.717,0.357)--(1.663,0.247) (2.633,2.622)--(2.636,2.606) (1.334,1.244)--(1.312,0.995) (2.478,1.074)--(2.523,1.225) (1.092,2.968)--(0.990,2.921) (1.808,2.711)--(2.043,2.599) (0.874,2.207)--(0.829,2.064) (2.523,3.275)--(2.478,3.426) (1.098,3.778)--(1.033,3.694) (2.377,2.618)--(2.436,2.801) (2.520,3.272)--(2.523,3.275) (1.185,2.112)--(1.098,2.262) (2.402,1.127)--(2.476,1.222) (2.503,2.592)--(2.636,2.606) (1.273,1.788)--(1.175,1.535) (0.782,1.719)--(0.807,1.549) (1.386,2.215)--(1.350,2.485) (1.793,0.291)--(1.745,0.214) (1.129,3.429)--(1.185,3.657) (1.488,2.058)--(1.312,2.022) (0.990,2.921)--(0.907,2.830) (1.770,4.222)--(1.894,4.215) (1.432,3.724)--(1.274,3.622) (1.450,4.137)--(1.372,4.137) (1.410,4.202)--(1.375,4.176) (1.540,0.251)--(1.454,0.270) (2.303,2.016)--(2.413,2.099) (1.616,0.209)--(1.617,0.208) (1.132,0.754)--(1.089,0.680) (2.015,1.284)--(2.142,1.380) (1.249,2.344)--(1.039,2.470) (1.874,0.265)--(1.865,0.231) (2.103,0.670)--(2.185,0.658) (2.153,1.973)--(2.220,1.763) (2.101,4.108)--(2.025,4.176) (2.096,0.493)--(2.101,0.392) (2.341,0.894)--(2.428,0.932) (1.734,3.685)--(1.586,3.749) (2.440,1.371)--(2.561,1.384) (1.640,0.469)--(1.596,0.319) (1.884,1.563)--(1.772,1.362) (1.772,1.362)--(1.885,1.282) (0.943,1.083)--(0.900,1.148) (0.776,2.689)--(0.782,2.781) (1.931,1.821)--(2.003,1.533) (0.765,1.969)--(0.764,1.894) (2.278,3.672)--(2.353,3.673) (1.187,3.951)--(1.225,4.025) (1.966,3.343)--(1.869,3.538) (1.386,2.215)--(1.249,2.344) (2.256,1.308)--(2.319,1.078) (1.331,0.767)--(1.248,0.643) (2.549,2.944)--(2.577,3.039) (0.858,2.978)--(0.807,2.951) (1.229,3.412)--(1.274,3.622) (0.804,1.891)--(0.765,1.969) (0.846,2.448)--(0.754,2.429) (1.837,4.130)--(1.966,4.115) (2.114,1.603)--(2.142,1.380) (2.466,2.989)--(2.549,2.944) (2.153,1.973)--(2.303,2.016) (1.756,0.551)--(1.640,0.469) (2.245,3.829)--(2.305,3.814) (2.060,2.847)--(2.039,3.102) (1.711,2.502)--(1.469,2.525) (2.011,0.720)--(2.103,0.670) (2.577,2.106)--(2.645,2.181) (1.125,1.322)--(1.003,1.261) (2.353,3.673)--(2.372,3.700) (1.515,3.576)--(1.586,3.749) (1.949,0.276)--(2.025,0.324) (1.360,0.590)--(1.292,0.499) (1.772,1.362)--(1.614,1.261) (2.130,3.937)--(2.175,4.025) (0.923,1.988)--(0.804,1.891) (2.175,0.520)--(2.245,0.571) (2.172,3.771)--(2.278,3.672) (1.334,3.303)--(1.229,3.412) (2.520,3.272)--(2.561,3.116) (1.884,1.563)--(1.885,1.282) (0.798,1.716)--(0.772,1.811) (1.006,1.076)--(0.980,0.927) (2.494,2.227)--(2.577,2.106) (1.015,3.554)--(0.974,3.572) (2.251,1.536)--(2.256,1.308) (1.931,1.821)--(2.114,1.603) (2.175,4.025)--(2.101,4.108) (1.437,0.777)--(1.331,0.767) (2.479,3.157)--(2.541,3.120) (2.084,4.059)--(2.101,4.108) (1.966,3.343)--(2.148,3.118) (1.823,0.439)--(1.717,0.357) (1.069,1.975)--(0.923,1.988) (2.440,1.371)--(2.534,1.339) (1.685,0.708)--(1.756,0.551) (1.562,3.049)--(1.369,2.820) (1.777,3.403)--(1.869,3.538) (1.106,1.105)--(1.006,1.076) (2.614,2.270)--(2.649,2.327) (1.037,3.404)--(1.015,3.554) (2.522,1.224)--(2.523,1.225) (1.650,1.861)--(1.534,1.753) (1.432,3.724)--(1.513,3.889) (1.028,0.800)--(1.089,0.680) (1.002,2.685)--(0.843,2.720) (2.387,3.158)--(2.466,2.989) (1.410,4.202)--(1.454,4.230) (2.081,3.392)--(2.190,3.586) (0.876,1.831)--(0.798,1.716) (2.130,3.937)--(2.245,3.829) (1.123,3.877)--(1.089,3.820) (2.362,1.548)--(2.440,1.371) (1.344,4.006)--(1.296,4.098) (2.015,1.284)--(1.985,1.052) (1.229,3.412)--(1.129,3.429) (1.223,1.251)--(1.125,1.322) (1.837,4.130)--(1.770,4.222) (0.954,3.472)--(0.922,3.426) (2.550,2.362)--(2.614,2.270) (1.089,0.680)--(1.155,0.571) (1.471,0.583)--(1.360,0.590) (1.421,1.017)--(1.437,0.777) (1.711,2.502)--(1.601,2.635) (2.170,0.853)--(2.269,0.697) (2.336,0.760)--(2.312,0.682) (2.341,0.894)--(2.372,0.800) (1.924,2.998)--(2.039,3.102) (1.756,0.551)--(1.823,0.439) (0.989,1.797)--(0.876,1.831) (2.513,1.474)--(2.577,1.478) (1.342,3.809)--(1.344,4.006) (1.107,0.928)--(1.025,0.807) (1.617,4.292)--(1.535,4.269) (1.734,3.685)--(1.835,3.784) (2.183,2.839)--(2.352,2.965) (1.515,3.576)--(1.432,3.724) (0.912,2.587)--(0.793,2.559) (2.352,2.965)--(2.387,3.158) (1.160,1.828)--(1.069,1.975) (2.172,3.771)--(2.130,3.937) (2.458,1.632)--(2.513,1.474) (0.765,1.969)--(0.754,2.071) (1.535,4.269)--(1.454,4.230) (1.205,1.032)--(1.106,1.105) (1.535,0.231)--(1.617,0.208) (1.284,1.512)--(1.223,1.251) (1.991,0.360)--(1.949,0.276) (1.249,2.344)--(1.185,2.112) (1.843,2.443)--(2.043,2.599) (1.039,2.470)--(0.912,2.587) (2.262,0.854)--(2.336,0.760) (1.292,0.499)--(1.351,0.382) (1.437,0.777)--(1.471,0.583) (1.966,3.343)--(2.081,3.392) (0.942,1.625)--(0.812,1.535) (2.350,1.798)--(2.362,1.548) (1.257,3.844)--(1.266,4.000) (1.650,1.861)--(1.488,2.058) (1.617,0.208)--(1.700,0.200) (1.897,3.999)--(1.837,4.130) (1.094,3.195)--(0.958,3.325) (2.305,3.814)--(2.311,3.820) (2.358,2.283)--(2.445,2.444) (2.014,0.505)--(1.904,0.374) (2.541,3.120)--(2.520,3.272) (2.387,3.158)--(2.393,3.341) (2.224,1.068)--(2.262,0.854) (2.021,4.160)--(1.946,4.230) (1.642,4.181)--(1.620,4.290) (1.077,1.640)--(0.989,1.797) (1.312,2.022)--(1.160,1.828) (1.089,3.820)--(1.028,3.700) (1.596,0.319)--(1.540,0.251) (1.469,2.525)--(1.350,2.485) (1.006,1.076)--(1.028,0.907) (1.223,1.251)--(1.205,1.032) (1.966,4.115)--(2.084,4.059) (2.060,2.847)--(2.183,2.839) (0.846,2.448)--(0.801,2.295) (2.130,3.937)--(2.250,3.922) (2.190,3.586)--(2.172,3.771) (1.949,0.276)--(1.946,0.270) (1.175,3.833)--(1.187,3.951) (1.965,2.384)--(2.043,2.599) (1.620,4.290)--(1.700,4.300) (2.130,3.937)--(2.245,3.929) (2.445,2.444)--(2.503,2.592) (2.247,0.585)--(2.311,0.680) (2.175,0.520)--(2.175,0.475) (1.296,4.098)--(1.299,4.108) (1.098,2.262)--(1.039,2.470) (2.413,0.972)--(2.478,1.074) (2.341,0.894)--(2.413,0.972) (1.098,3.778)--(1.089,3.820) (0.876,1.831)--(0.856,1.642) (1.160,1.828)--(1.077,1.640) (2.513,1.474)--(2.593,1.549) (1.663,0.247)--(1.616,0.209) (1.185,3.657)--(1.257,3.844) (1.258,2.715)--(1.165,2.525) (2.175,0.475)--(2.245,0.571) (1.000,3.159)--(0.921,3.083) (1.513,3.889)--(1.342,3.809) (2.646,2.429)--(2.636,2.606) (2.522,1.224)--(2.478,1.074) (1.571,4.228)--(1.487,4.234) (1.683,1.548)--(1.772,1.362) (2.350,1.798)--(2.446,1.863) (1.165,2.525)--(1.002,2.685) (1.717,0.357)--(1.596,0.319) (2.513,1.474)--(2.568,1.587) (1.985,1.052)--(2.127,1.136) (0.772,1.811)--(0.782,1.719) (0.868,1.301)--(0.877,1.225) (2.220,1.763)--(2.251,1.536) (2.014,0.505)--(2.096,0.493) (0.912,2.587)--(0.846,2.448) (0.764,2.606)--(0.754,2.429) (1.469,2.525)--(1.485,2.800) (1.485,2.800)--(1.369,2.820) (1.539,2.308)--(1.350,2.485) (1.835,3.784)--(1.680,3.878) (0.900,1.148)--(0.922,1.074) (0.801,2.295)--(0.750,2.250) (2.577,3.039)--(2.593,2.951) (1.966,4.115)--(1.894,4.215) (1.528,0.429)--(1.472,0.328) (1.965,2.384)--(2.042,2.156) (2.262,0.854)--(2.341,0.894) (1.068,0.756)--(1.089,0.680) (1.885,1.282)--(1.714,1.114) (1.714,1.114)--(1.840,1.036) (2.003,1.533)--(2.015,1.284) (1.132,0.754)--(1.126,0.620) (1.094,3.195)--(1.000,3.159) (2.381,3.517)--(2.441,3.497) (2.127,2.330)--(2.358,2.283) (1.869,3.538)--(2.081,3.392) (2.081,3.392)--(1.973,3.618) (1.249,2.344)--(1.165,2.525) (1.652,3.524)--(1.734,3.685) (1.213,0.823)--(1.132,0.754) (1.683,1.548)--(1.547,1.483) (1.274,3.622)--(1.342,3.809) (1.745,0.214)--(1.700,0.200) (0.942,2.342)--(0.801,2.295) (1.547,1.483)--(1.614,1.261) (0.774,2.138)--(0.751,2.235) (2.305,3.814)--(2.372,3.700) (1.973,3.618)--(2.058,3.825) (0.924,1.422)--(0.868,1.301) (2.142,1.380)--(2.127,1.136) (1.612,4.010)--(1.642,4.181) (2.269,0.697)--(2.311,0.680) (2.538,2.757)--(2.600,2.706) (1.485,2.800)--(1.562,3.049) (1.770,4.222)--(1.783,4.292) (2.220,1.763)--(2.350,1.798) (1.640,0.469)--(1.528,0.429) (2.353,3.673)--(2.400,3.633) (2.183,2.839)--(2.148,3.118) (1.450,4.137)--(1.375,4.176) (1.863,3.228)--(1.966,3.343) (1.026,1.451)--(0.924,1.422) (2.585,1.865)--(2.635,1.897) (2.635,1.897)--(2.636,1.894) (1.586,3.749)--(1.680,3.878) (1.248,0.643)--(1.179,0.601) (1.092,2.968)--(0.921,3.083) (0.858,2.978)--(0.845,3.105) (1.714,1.114)--(1.523,1.104) (2.148,3.118)--(2.287,3.293) (0.874,2.207)--(0.774,2.138) (2.278,3.672)--(2.381,3.517) (1.885,1.282)--(1.840,1.036) (1.941,3.832)--(2.016,3.988) (0.924,1.250)--(0.900,1.148) (2.521,1.969)--(2.585,1.865) (2.003,1.533)--(2.142,1.380) (2.377,2.618)--(2.538,2.757) (1.331,0.767)--(1.213,0.823) (2.549,2.944)--(2.598,2.872) (0.782,2.781)--(0.764,2.606) (2.513,1.474)--(2.561,1.384) (1.928,2.733)--(2.060,2.847) (0.994,2.150)--(0.874,2.207) (2.402,1.127)--(2.477,1.096) (1.906,2.100)--(1.965,2.384) (1.869,3.538)--(1.973,3.618) (2.585,2.531)--(2.636,2.606) (0.843,2.720)--(0.782,2.781) (1.003,1.261)--(0.924,1.250) (1.094,3.623)--(1.098,3.778) (1.547,1.483)--(1.400,1.567) (1.652,3.524)--(1.586,3.749) (0.990,2.921)--(0.858,2.978) (2.174,2.565)--(2.377,2.618) (0.810,2.851)--(0.806,2.944) (0.772,1.811)--(0.764,1.894) (2.190,3.358)--(2.298,3.483) (0.829,2.064)--(0.765,1.969) (2.354,1.318)--(2.402,1.127) (1.450,4.137)--(1.410,4.202) (0.868,1.301)--(0.839,1.384) (2.478,3.426)--(2.428,3.568) (2.303,2.016)--(2.521,1.969) (1.125,1.322)--(1.026,1.451) (2.577,2.106)--(2.623,2.002) (1.360,0.590)--(1.248,0.643) (1.113,2.735)--(0.990,2.921) (1.779,3.010)--(1.562,3.049) (2.441,3.497)--(2.478,3.426) (1.819,0.216)--(1.865,0.231) (2.103,0.670)--(2.175,0.520) (1.292,0.499)--(1.299,0.392) (2.039,3.102)--(2.148,3.118) (0.923,1.988)--(0.829,2.064) (2.476,1.222)--(2.522,1.224) (2.245,3.829)--(2.311,3.820) (1.425,3.968)--(1.450,4.137) (1.634,2.106)--(1.488,2.058) (1.015,3.554)--(1.033,3.694) (1.068,0.756)--(1.028,0.800) (2.286,2.757)--(2.436,2.801) (0.907,2.830)--(0.810,2.851) (1.351,0.382)--(1.375,0.324) (1.664,2.872)--(1.485,2.800) (1.069,1.975)--(0.994,2.150) (2.440,1.371)--(2.476,1.222) (2.278,3.672)--(2.245,3.829) (1.935,0.567)--(2.014,0.505) (1.106,1.105)--(1.003,1.261) (1.874,0.265)--(1.819,0.216) (2.043,2.599)--(2.174,2.565) (1.002,2.685)--(0.907,2.830) (2.185,0.658)--(2.247,0.585) (1.179,0.601)--(1.233,0.481) (1.364,3.524)--(1.274,3.622) (1.751,2.241)--(1.539,2.308) (2.025,4.176)--(1.946,4.230) (2.362,1.548)--(2.354,1.318) (1.344,4.006)--(1.372,4.137) (1.751,2.241)--(1.965,2.384) (2.238,2.196)--(2.413,2.099) (1.129,3.429)--(1.015,3.554) (2.016,3.988)--(1.966,4.115) (1.904,0.374)--(1.793,0.291) (1.033,3.694)--(1.028,3.700) (2.127,1.136)--(2.256,1.308) (2.170,0.853)--(2.185,0.658) (2.466,2.989)--(2.479,3.157) (1.770,4.222)--(1.745,4.297) (0.845,3.105)--(0.807,2.951) (1.584,0.621)--(1.640,0.469) (0.989,1.797)--(0.923,1.988) (0.895,3.226)--(0.877,3.275) (1.342,3.809)--(1.425,3.968) (1.003,1.261)--(0.900,1.148) (1.369,2.820)--(1.258,2.715) (1.472,0.328)--(1.424,0.295) (0.793,2.559)--(0.754,2.429) (2.523,1.225)--(2.561,1.384) (1.680,3.878)--(1.513,3.889) (1.634,2.106)--(1.539,2.308) (2.298,3.483)--(2.278,3.672) (1.225,0.475)--(1.299,0.392) (2.127,2.330)--(2.174,2.565) (1.840,1.036)--(1.985,1.052) (2.494,2.227)--(2.550,2.362) (1.312,0.995)--(1.331,0.767) (1.039,2.470)--(1.002,2.685) (2.269,0.697)--(2.336,0.760) (0.923,1.988)--(0.765,1.969) (1.601,2.635)--(1.485,2.800) (0.943,1.083)--(0.922,1.074) (1.257,3.844)--(1.344,4.006) (1.207,2.942)--(1.113,2.735) (0.807,1.549)--(0.839,1.384) (1.973,3.618)--(1.835,3.784) (1.126,0.620)--(1.155,0.571) (1.037,3.404)--(0.958,3.325) (1.033,3.694)--(0.974,3.572) (2.400,3.633)--(2.428,3.568) (1.612,4.010)--(1.425,3.968) (2.600,2.706)--(2.618,2.781) (2.224,1.068)--(2.170,0.853) (2.436,2.801)--(2.466,2.989) (1.906,2.100)--(2.042,2.156) (1.751,2.241)--(1.906,2.100) (2.137,4.065)--(2.101,4.108) (1.614,1.261)--(1.714,1.114) (1.693,4.274)--(1.620,4.290) (2.193,3.965)--(2.245,3.929) (2.476,1.222)--(2.534,1.339) (2.362,1.548)--(2.458,1.632) (1.596,0.319)--(1.472,0.328) (1.175,1.535)--(1.125,1.322) (1.923,0.842)--(2.076,0.907) (1.894,4.215)--(1.865,4.269) (1.904,0.374)--(1.991,0.360) (0.907,2.830)--(0.843,2.720) (1.369,2.820)--(1.446,3.109) (0.876,1.831)--(0.772,1.811) (1.446,3.109)--(1.327,3.066) (2.148,3.118)--(2.081,3.392) (1.372,4.137)--(1.296,4.098) (1.342,3.809)--(1.257,3.844) (1.819,0.216)--(1.783,0.208) (1.350,2.485)--(1.258,2.715) (1.941,3.832)--(1.781,3.963) (2.413,2.099)--(2.494,2.227) (2.135,0.428)--(2.175,0.475) (2.064,0.375)--(2.025,0.324) (1.534,1.753)--(1.547,1.483) (2.084,4.059)--(2.021,4.160) (1.408,0.442)--(1.351,0.382) (2.185,0.658)--(2.269,0.697) (1.840,1.036)--(1.639,0.938) (2.336,0.760)--(2.372,0.800) (1.185,2.112)--(1.069,1.975) (1.639,0.938)--(1.781,0.855) (1.155,0.571)--(1.225,0.475) (1.781,3.963)--(1.837,4.130) (1.028,0.907)--(1.025,0.807) (2.476,1.222)--(2.523,1.225) (1.129,3.429)--(1.037,3.404) (2.650,2.250)--(2.646,2.071) (1.601,2.635)--(1.664,2.872) (1.664,2.872)--(1.562,3.049) (2.527,1.715)--(2.618,1.719) (2.174,2.565)--(2.183,2.839) (1.770,4.222)--(1.693,4.274) (1.165,2.525)--(1.113,2.735) (1.107,0.928)--(1.028,0.907) (1.751,2.241)--(1.780,2.010) (1.906,2.100)--(1.931,1.821) (1.614,1.261)--(1.444,1.322) (0.812,1.535)--(0.807,1.549) (1.444,1.322)--(1.523,1.104) (2.645,2.181)--(2.646,2.071) (2.080,3.625)--(2.172,3.771) (0.856,1.642)--(0.812,1.535) (2.127,1.136)--(2.076,0.907) (1.718,4.093)--(1.770,4.222) (2.358,2.283)--(2.550,2.362) (0.845,3.105)--(0.839,3.116) (1.446,3.109)--(1.598,3.315) (1.539,2.308)--(1.488,2.058) (1.528,0.429)--(1.408,0.442) (2.387,3.158)--(2.469,3.335) (0.793,2.559)--(0.764,2.606) (1.369,2.820)--(1.327,3.066) (0.980,0.927)--(0.972,0.932) (0.942,1.625)--(0.856,1.642) (0.829,2.064)--(0.754,2.071) (1.350,2.485)--(1.165,2.525) (1.534,1.753)--(1.410,1.821) (1.233,0.481)--(1.299,0.392) (1.132,0.754)--(1.068,0.756) (1.410,1.821)--(1.400,1.567) (1.680,3.878)--(1.781,3.963) (1.639,0.938)--(1.421,1.017) (2.538,2.757)--(2.618,2.781) (2.261,3.064)--(2.387,3.158) (0.846,2.448)--(0.764,2.408) (1.840,1.036)--(1.781,0.855) (2.058,3.825)--(2.130,3.937) (1.601,2.635)--(1.808,2.711) (1.664,2.872)--(1.779,3.010) (2.445,2.444)--(2.585,2.531) (1.213,0.823)--(1.107,0.928) (2.600,2.706)--(2.633,2.622) (1.386,2.215)--(1.185,2.112) (0.942,2.342)--(0.846,2.448) (2.341,0.894)--(2.399,0.865) (1.616,0.209)--(1.535,0.231) (1.132,0.754)--(1.028,0.800) (1.973,3.618)--(2.080,3.625) (0.924,1.422)--(0.852,1.468) (1.408,0.442)--(1.375,0.324) (1.444,1.322)--(1.284,1.512) (2.280,2.450)--(2.445,2.444) (1.780,1.726)--(1.931,1.821) (2.137,4.065)--(2.175,4.025) (1.000,3.159)--(0.895,3.226) (2.114,1.603)--(2.251,1.536) (1.614,1.261)--(1.523,1.104) (2.287,3.293)--(2.393,3.341) (1.894,4.215)--(1.946,4.230) (2.469,3.335)--(2.520,3.272) (2.319,1.078)--(2.341,0.894) (2.350,1.798)--(2.527,1.715) (1.571,4.228)--(1.535,4.269) (1.026,1.451)--(0.942,1.625) (2.585,1.865)--(2.610,1.730) (1.334,1.244)--(1.205,1.032) (1.092,2.968)--(1.000,3.159) (2.014,0.505)--(2.064,0.375) (1.780,2.010)--(1.906,2.100) (2.148,3.118)--(2.261,3.064) (0.874,2.207)--(0.801,2.295) (1.772,1.362)--(1.714,1.114) (0.974,3.572)--(0.972,3.568) (1.410,1.821)--(1.312,2.022) (0.806,2.944)--(0.782,2.781) (1.534,1.753)--(1.400,1.567) (2.377,2.618)--(2.503,2.592) (2.585,2.531)--(2.633,2.445) (2.251,1.536)--(2.362,1.548) (0.764,2.408)--(0.750,2.250) (2.534,1.339)--(2.561,1.384) (0.994,2.150)--(0.942,2.342) (2.402,1.127)--(2.413,0.972) (1.273,1.788)--(1.077,1.640) (1.823,0.439)--(1.904,0.374) (2.585,1.865)--(2.636,1.894) (2.633,2.445)--(2.636,2.606) (2.614,2.270)--(2.650,2.250) (2.174,2.565)--(2.280,2.450) (0.990,2.921)--(0.921,3.083) (1.432,3.724)--(1.342,3.809) (1.751,2.241)--(1.711,2.502) (1.450,4.137)--(1.487,4.234) (2.042,2.156)--(2.127,2.330) (2.303,2.016)--(2.446,1.863) (2.130,3.937)--(2.084,4.059) (2.527,1.715)--(2.568,1.587) (1.700,0.200)--(1.783,0.208) (2.015,1.284)--(2.127,1.136) (2.076,0.907)--(2.224,1.068) (1.808,2.711)--(1.664,2.872) (1.113,2.735)--(1.092,2.968) (2.103,0.670)--(2.096,0.493) (1.471,0.583)--(1.528,0.429) (1.179,0.601)--(1.155,0.571) (1.425,3.968)--(1.534,4.100) (1.351,0.382)--(1.304,0.387) (1.327,3.066)--(1.207,2.942) (1.734,3.685)--(1.680,3.878) (1.781,3.963)--(1.612,4.010) (1.823,4.274)--(1.783,4.292) (2.393,3.341)--(2.381,3.517) (1.233,0.481)--(1.225,0.475) (0.972,3.568)--(0.922,3.426) (1.781,0.855)--(1.923,0.842) (1.472,0.328)--(1.454,0.270) (1.248,0.643)--(1.126,0.620) (2.084,4.059)--(2.175,4.025) (1.205,1.032)--(1.213,0.823) (0.874,2.207)--(0.751,2.235) (1.562,3.049)--(1.446,3.109) (1.534,4.100)--(1.571,4.228) (1.344,4.006)--(1.450,4.137) (1.200,3.171)--(1.092,2.968) (2.245,0.571)--(2.311,0.680) (1.650,1.861)--(1.410,1.821) (2.399,0.865)--(2.428,0.932) (2.503,2.592)--(2.538,2.757) (1.123,3.877)--(1.155,3.929) (2.055,1.852)--(2.153,1.973) (2.636,2.606)--(2.618,2.781) (1.523,1.104)--(1.639,0.938) (0.896,3.335)--(0.877,3.275) (0.954,3.472)--(0.972,3.568) (1.846,0.678)--(2.011,0.720) (1.077,1.640)--(1.026,1.451) (0.810,2.851)--(0.782,2.781) (0.921,3.083)--(0.858,2.978) (2.153,1.973)--(2.350,1.798) (1.461,3.348)--(1.334,3.303) (0.754,2.429)--(0.750,2.250) (2.060,2.847)--(2.148,3.118) (2.446,1.863)--(2.521,1.969) (1.793,0.291)--(1.663,0.247) (2.436,2.801)--(2.598,2.872) (1.780,2.010)--(1.931,1.821) (1.400,1.567)--(1.444,1.322) (2.096,0.493)--(2.175,0.520) (1.375,4.176)--(1.299,4.108) (2.593,1.549)--(2.618,1.719) (1.542,0.825)--(1.685,0.708) (1.098,2.262)--(0.994,2.150) (2.477,1.096)--(2.522,1.224) (1.006,1.076)--(0.900,1.148) (1.185,3.657)--(1.094,3.623) (1.562,3.049)--(1.704,3.210) (0.907,2.830)--(0.806,2.944) (2.372,3.700)--(2.311,3.820) (1.704,3.210)--(1.598,3.315) (1.025,0.807)--(1.028,0.800) (2.633,2.445)--(2.646,2.429) (1.894,4.215)--(1.823,4.274) (1.304,0.387)--(1.375,0.324) (2.278,3.672)--(2.305,3.814) (2.250,3.922)--(2.245,3.929) (2.549,2.944)--(2.593,2.951) (2.021,4.160)--(2.025,4.176) (1.717,0.357)--(1.793,0.291) (2.393,3.341)--(2.478,3.426) (1.523,1.104)--(1.334,1.244) (1.334,1.244)--(1.421,1.017) (2.042,2.156)--(2.153,1.973) (0.980,0.927)--(1.025,0.807) (0.804,1.891)--(0.772,1.811) (1.711,2.502)--(1.808,2.711) (2.076,0.907)--(2.011,0.720) (1.808,2.711)--(1.779,3.010) (1.485,2.800)--(1.258,2.715) (1.837,4.130)--(1.894,4.215) (1.461,3.348)--(1.652,3.524) (2.466,2.989)--(2.541,3.120) (1.408,0.442)--(1.292,0.499) (1.327,3.066)--(1.334,3.303) (1.965,2.384)--(2.127,2.330) (2.247,0.585)--(2.245,0.571) (1.780,2.010)--(1.780,1.726) (1.400,1.567)--(1.273,1.788) (1.781,3.963)--(1.897,3.999) (1.028,0.907)--(0.980,0.927) (1.273,1.788)--(1.284,1.512) (2.413,0.972)--(2.428,0.932) (2.256,1.308)--(2.224,1.068) (1.793,0.291)--(1.874,0.265) (2.352,2.965)--(2.466,2.989) (0.843,2.720)--(0.776,2.689) (1.472,0.328)--(1.375,0.324) (2.568,1.587)--(2.618,1.719) (1.781,0.855)--(1.685,0.708) (2.172,3.771)--(2.245,3.829) (1.266,4.000)--(1.225,4.025) (1.258,2.715)--(1.207,2.942) (1.704,3.210)--(1.863,3.228) (1.284,1.512)--(1.125,1.322) (2.494,2.227)--(2.614,2.270) (1.107,0.928)--(1.006,1.076) (1.249,2.344)--(1.098,2.262) (1.562,3.049)--(1.598,3.315) (1.513,3.889)--(1.612,4.010) (1.292,0.499)--(1.233,0.481) (2.269,0.697)--(2.312,0.682) (1.274,3.622)--(1.185,3.657) (2.080,3.625)--(2.190,3.586) (0.856,1.642)--(0.798,1.716) (1.985,1.052)--(1.923,0.842) (1.897,3.999)--(1.966,4.115) (2.358,2.283)--(2.494,2.227) (1.037,3.404)--(0.954,3.472) (0.896,3.335)--(0.922,3.426) (2.561,1.384)--(2.593,1.549) (2.436,2.801)--(2.549,2.944) (1.523,1.104)--(1.421,1.017) (2.541,3.120)--(2.577,3.039) (2.387,3.158)--(2.479,3.157) (2.142,1.380)--(2.256,1.308) (1.711,2.502)--(1.843,2.443) (2.633,2.622)--(2.618,2.781) (1.808,2.711)--(1.928,2.733) (1.312,2.022)--(1.069,1.975) (2.362,1.548)--(2.513,1.474) (1.640,0.469)--(1.717,0.357) (0.900,1.148)--(0.877,1.225) (1.469,2.525)--(1.258,2.715) (0.774,2.138)--(0.754,2.071) (1.835,3.784)--(1.941,3.832) (1.006,1.076)--(0.943,1.083) (1.944,4.231)--(1.865,4.269) (1.904,0.374)--(1.949,0.276) (0.856,1.642)--(0.807,1.549) (1.586,3.749)--(1.513,3.889) (2.261,3.064)--(2.352,2.965) (0.846,2.448)--(0.793,2.559) (1.714,1.114)--(1.639,0.938) (2.190,3.586)--(2.278,3.672) (1.965,2.384)--(2.174,2.565) (0.839,1.384)--(0.877,1.225) (2.413,2.099)--(2.577,2.106) (1.400,1.567)--(1.284,1.512) (2.445,2.444)--(2.550,2.362) (2.614,2.270)--(2.645,2.181) (1.885,1.282)--(1.985,1.052) (2.256,1.308)--(2.354,1.318) (1.331,0.767)--(1.360,0.590) (1.488,2.058)--(1.410,1.821) (2.081,3.392)--(2.190,3.358) (0.876,1.831)--(0.804,1.891) (2.312,0.682)--(2.372,0.800) (1.258,2.715)--(1.113,2.735) (1.869,3.538)--(1.835,3.784) (2.280,2.450)--(2.358,2.283) (1.000,3.159)--(0.958,3.325) (2.250,3.922)--(2.311,3.820) (1.547,1.483)--(1.444,1.322) (1.513,3.889)--(1.425,3.968) (1.540,0.251)--(1.535,0.231) (1.571,4.228)--(1.620,4.290) (1.946,4.230)--(1.865,4.269) (2.153,1.973)--(2.238,2.196) (2.021,4.160)--(2.101,4.108) (1.793,0.291)--(1.783,0.208) (1.874,0.265)--(1.946,0.270) (1.985,1.052)--(2.076,0.907) (1.693,4.274)--(1.700,4.300) (2.011,0.720)--(2.170,0.853) (1.125,1.322)--(1.106,1.105) (1.092,2.968)--(1.094,3.195) (2.014,0.505)--(1.991,0.360) (2.096,0.493)--(2.175,0.475) (1.450,4.137)--(1.454,4.230) (2.183,2.839)--(2.286,2.757) (0.912,2.587)--(0.843,2.720) (1.360,0.590)--(1.408,0.442) (2.262,0.854)--(2.372,0.800) (1.865,4.269)--(1.783,4.292) (1.334,3.303)--(1.200,3.171) (1.266,4.000)--(1.296,4.098) (1.037,3.404)--(0.922,3.426) (1.835,3.784)--(1.781,3.963) (2.387,3.158)--(2.520,3.272) (1.966,4.115)--(2.021,4.160) (1.931,1.821)--(2.055,1.852) (2.262,0.854)--(2.269,0.697) (0.776,2.689)--(0.764,2.606) (1.069,1.975)--(0.989,1.797) (1.843,2.443)--(1.808,2.711) (1.685,0.708)--(1.846,0.678) (2.568,1.587)--(2.593,1.549) (2.645,2.181)--(2.650,2.250) (1.106,1.105)--(1.107,0.928) (1.094,3.195)--(1.037,3.404) (1.488,2.058)--(1.386,2.215) (1.386,2.215)--(1.312,2.022) (2.577,2.106)--(2.646,2.071) (2.043,2.599)--(2.183,2.839) (1.598,3.315)--(1.461,3.348) (1.642,4.181)--(1.693,4.274) (2.081,3.392)--(2.080,3.625) (1.650,1.861)--(1.780,1.726) (1.229,3.412)--(1.094,3.195) (2.550,2.362)--(2.585,2.531) (0.912,2.587)--(0.764,2.606) (1.299,4.108)--(1.225,4.025) (2.114,1.603)--(2.220,1.763) (1.421,1.017)--(1.542,0.825) (0.989,1.797)--(0.942,1.625) (1.756,0.551)--(1.935,0.567) (1.711,2.502)--(1.539,2.308) (1.779,3.010)--(1.704,3.210) (0.958,3.325)--(0.895,3.226) (2.220,1.763)--(2.362,1.548) (2.541,3.120)--(2.561,3.116) (2.183,2.839)--(2.261,3.064) (1.865,0.231)--(1.946,0.270) (1.515,3.576)--(1.364,3.524) (1.944,4.231)--(1.946,4.230) (0.856,1.642)--(0.782,1.719) (2.458,1.632)--(2.527,1.715) (1.931,1.821)--(1.884,1.563) (2.503,2.592)--(2.633,2.622) (1.884,1.563)--(2.003,1.533) (1.351,0.382)--(1.299,0.392) (1.284,1.512)--(1.334,1.244) (1.372,4.137)--(1.299,4.108) (1.991,0.360)--(2.064,0.375) (1.039,2.470)--(0.942,2.342) (1.437,0.777)--(1.584,0.621) (1.534,4.100)--(1.642,4.181) (1.175,3.833)--(1.089,3.820) (1.257,3.844)--(1.175,3.833) (1.598,3.315)--(1.777,3.403) (1.777,3.403)--(1.652,3.524) (0.877,3.275)--(0.839,3.116) (2.312,0.682)--(2.311,0.680) (2.381,3.517)--(2.400,3.633) (2.477,1.096)--(2.478,1.074) (2.224,1.068)--(2.319,1.078) (1.312,2.022)--(1.273,1.788) (2.610,1.730)--(2.636,1.894) (2.527,1.715)--(2.585,1.865) (1.596,0.319)--(1.663,0.247) (1.779,3.010)--(1.924,2.998) (2.011,0.720)--(1.935,0.567) (1.924,2.998)--(1.863,3.228) (1.175,3.833)--(1.098,3.778) (1.634,2.106)--(1.780,2.010) (2.538,2.757)--(2.598,2.872) (1.334,3.303)--(1.364,3.524) (2.127,2.330)--(2.238,2.196) (1.663,0.247)--(1.745,0.214) (0.858,2.978)--(0.806,2.944) (2.393,3.341)--(2.441,3.497) (1.685,0.708)--(1.584,0.621) (1.126,0.620)--(1.089,0.680) (1.534,4.100)--(1.450,4.137) (0.852,1.468)--(0.839,1.384) (1.777,3.403)--(1.966,3.343) (1.935,0.567)--(1.904,0.374) (1.207,2.942)--(1.200,3.171) (2.521,1.969)--(2.623,2.002) (1.424,0.295)--(1.454,0.270) (1.165,2.525)--(1.039,2.470) (1.006,1.076)--(0.972,0.932) (1.598,3.315)--(1.652,3.524) (1.179,0.601)--(1.126,0.620) (1.612,4.010)--(1.718,4.093) (1.906,2.100)--(2.055,1.852) (1.663,0.247)--(1.700,0.200) (0.994,2.150)--(0.829,2.064) (0.801,2.295)--(0.751,2.235) (1.923,0.842)--(1.846,0.678) (2.016,3.988)--(2.084,4.059) (1.094,3.623)--(1.033,3.694) (2.127,1.136)--(2.224,1.068) (2.503,2.592)--(2.600,2.706) (1.421,1.017)--(1.312,0.995) (1.248,0.643)--(1.132,0.754) (1.446,3.109)--(1.461,3.348) (1.584,0.621)--(1.528,0.429) (1.924,2.998)--(2.060,2.847) (1.528,0.429)--(1.596,0.319) (2.354,1.318)--(2.476,1.222) (0.922,3.426)--(0.877,3.275) (1.369,2.820)--(1.207,2.942) (2.135,0.428)--(2.101,0.392) (0.924,1.250)--(0.868,1.301) (1.941,3.832)--(2.058,3.825) (1.779,3.010)--(1.863,3.228) (1.680,3.878)--(1.612,4.010) (0.921,3.083)--(0.845,3.105) (1.113,2.735)--(0.907,2.830) (2.352,2.965)--(2.436,2.801) (1.620,4.290)--(1.617,4.292) (1.639,0.938)--(1.542,0.825) (2.298,3.483)--(2.381,3.517) (2.381,3.517)--(2.353,3.673) (2.127,2.330)--(2.280,2.450) (1.296,4.098)--(1.225,4.025) (1.840,1.036)--(1.923,0.842) (2.446,1.863)--(2.585,1.865) (1.284,1.512)--(1.175,1.535) (2.623,2.002)--(2.635,1.897) (2.311,0.680)--(2.372,0.800) (1.372,4.137)--(1.375,4.176) (1.003,1.261)--(0.924,1.422) (1.175,3.833)--(1.155,3.929) (1.843,2.443)--(1.965,2.384) (1.312,0.995)--(1.213,0.823) (1.213,0.823)--(1.248,0.643) (2.096,0.493)--(2.135,0.428) (1.098,3.778)--(1.028,3.700) (2.269,0.697)--(2.247,0.585) (2.618,2.781)--(2.593,2.951) (0.958,3.325)--(0.877,3.275) (2.190,3.358)--(2.287,3.293) (0.829,2.064)--(0.774,2.138) (1.292,0.499)--(1.179,0.601) (1.207,2.942)--(1.092,2.968) (2.372,0.800)--(2.428,0.932) (0.907,2.830)--(0.782,2.781) (1.897,3.999)--(2.016,3.988) (1.683,1.548)--(1.884,1.563) (2.358,2.283)--(2.413,2.099) (1.185,3.657)--(1.098,3.778) (1.973,3.618)--(1.941,3.832) (1.444,1.322)--(1.334,1.244) (1.612,4.010)--(1.534,4.100) (2.436,2.801)--(2.538,2.757) (1.906,2.100)--(2.153,1.973) (2.610,1.730)--(2.618,1.719) (1.693,4.274)--(1.745,4.297) (2.649,2.327)--(2.646,2.429) (0.750,2.250)--(0.754,2.071) (1.312,2.022)--(1.185,2.112) (2.476,1.222)--(2.477,1.096) (1.923,0.842)--(2.011,0.720) (2.598,2.872)--(2.593,2.951) (1.175,1.535)--(1.026,1.451) (1.026,1.451)--(1.003,1.261) (1.904,0.374)--(1.874,0.265) (1.539,2.308)--(1.386,2.215) (2.286,2.757)--(2.377,2.618) (0.907,2.830)--(0.858,2.978) (0.804,1.891)--(0.764,1.894) (2.469,3.335)--(2.478,3.426) (1.248,0.643)--(1.292,0.499) (0.754,2.071)--(0.764,1.894) (1.446,3.109)--(1.334,3.303) (2.190,3.586)--(2.298,3.483) (2.618,1.719)--(2.636,1.894) (2.148,3.118)--(2.190,3.358) (0.924,1.422)--(0.839,1.384) (2.577,3.039)--(2.561,3.116) (1.410,1.821)--(1.273,1.788) (2.413,2.099)--(2.521,1.969) (1.941,3.832)--(1.897,3.999) (2.521,1.969)--(2.577,2.106) (2.466,2.989)--(2.577,3.039) (2.084,4.059)--(2.137,4.065) (2.550,2.362)--(2.646,2.429) (2.003,1.533)--(2.114,1.603) (2.185,0.658)--(2.175,0.520) (1.185,2.112)--(0.994,2.150) (1.639,0.938)--(1.685,0.708) (2.636,1.894)--(2.646,2.071) (0.994,2.150)--(0.923,1.988) (2.238,2.196)--(2.303,2.016) (1.129,3.429)--(1.094,3.623) (1.003,1.261)--(1.006,1.076) (1.793,0.291)--(1.819,0.216) (1.094,3.623)--(1.015,3.554) (2.245,3.929)--(2.175,4.025) (1.652,3.524)--(1.515,3.576) (0.922,1.074)--(0.972,0.932) (2.170,0.853)--(2.103,0.670) (1.770,4.222)--(1.823,4.274) (2.174,2.565)--(2.286,2.757) (1.424,0.295)--(1.375,0.324) (1.274,3.622)--(1.129,3.429) (1.547,1.483)--(1.772,1.362) (2.190,3.358)--(2.190,3.586) (2.354,1.318)--(2.440,1.371) (1.663,0.247)--(1.617,0.208) (1.745,0.214)--(1.783,0.208) (1.780,1.726)--(1.884,1.563) (2.494,2.227)--(2.645,2.181) (2.278,3.672)--(2.372,3.700) (1.991,0.360)--(2.025,0.324) (2.142,1.380)--(2.251,1.536) (1.113,2.735)--(1.002,2.685) (1.571,4.228)--(1.617,4.292) (2.185,0.658)--(2.245,0.571) (1.640,0.469)--(1.823,0.439) (0.923,1.988)--(0.876,1.831) (1.539,2.308)--(1.469,2.525) (1.425,3.968)--(1.344,4.006) (1.863,3.228)--(1.777,3.403) (1.015,3.554)--(0.954,3.472) (0.974,3.572)--(1.028,3.700) (2.286,2.757)--(2.352,2.965) (1.586,3.749)--(1.432,3.724) (2.635,1.897)--(2.646,2.071) (1.634,2.106)--(1.650,1.861) (0.806,2.944)--(0.807,2.951) (2.440,1.371)--(2.513,1.474) (2.003,1.533)--(1.885,1.282) (2.550,2.362)--(2.649,2.327) (1.935,0.567)--(2.103,0.670) (0.751,2.235)--(0.754,2.071) (2.534,1.339)--(2.523,1.225) (1.094,3.195)--(1.129,3.429) (0.764,2.408)--(0.754,2.429) (1.700,4.300)--(1.617,4.292) (1.874,0.265)--(1.949,0.276) (1.885,1.282)--(2.015,1.284) (1.002,2.685)--(0.912,2.587) (1.642,4.181)--(1.770,4.222) (1.364,3.524)--(1.229,3.412) (1.928,2.733)--(1.924,2.998) (1.344,4.006)--(1.266,4.000) (1.652,3.524)--(1.869,3.538) (0.958,3.325)--(0.922,3.426) (1.454,0.270)--(1.535,0.231) (2.238,2.196)--(2.358,2.283) (1.751,2.241)--(1.634,2.106) (1.869,3.538)--(1.734,3.685) (2.256,1.308)--(2.402,1.127) (0.843,2.720)--(0.764,2.606) (2.170,0.853)--(2.262,0.854) (1.780,1.726)--(1.683,1.548) (1.584,0.621)--(1.756,0.551) (1.472,0.328)--(1.540,0.251) (2.649,2.327)--(2.650,2.250) (2.598,2.872)--(2.618,2.781) (2.245,3.829)--(2.250,3.922) (1.863,3.228)--(2.039,3.102) (2.469,3.335)--(2.523,3.275) (2.441,3.497)--(2.428,3.568) (2.039,3.102)--(1.966,3.343) (1.266,4.000)--(1.187,3.951) (1.823,4.274)--(1.865,4.269) (1.745,4.297)--(1.700,4.300) (1.233,0.481)--(1.155,0.571) (1.312,0.995)--(1.437,0.777) (1.487,4.234)--(1.454,4.230) (1.107,0.928)--(1.028,0.800) (0.942,1.625)--(0.852,1.468) (1.540,0.251)--(1.616,0.209);

%% file: setup_patch.tex
\fill[ib, draw=ibd!60!black, thin] (0.728,-0.106) -- (0.387,0.171) -- (0.797,0.199) -- cycle;
\fill[ib, draw=ibd!60!black, thin] (0.299,-0.335) -- (0.728,-0.106) -- (0.387,0.171) -- cycle;
\fill[ib, draw=ibd!60!black, thin] (-0.376,-0.133) -- (-0.740,0.202) -- (-0.265,0.286) -- cycle;
\fill[ib, draw=ibd!60!black, thin] (0.611,-0.531) -- (0.299,-0.335) -- (0.728,-0.106) -- cycle;
\fill[ib, draw=ibd!60!black, thin] (0.387,0.171) -- (0.090,0.410) -- (0.507,0.545) -- cycle;
\fill[ib, draw=ibd!60!black, thin] (0.090,0.410) -- (0.507,0.545) -- (0.139,0.815) -- cycle;
\fill[ib, draw=ibd!60!black, thin] (0.023,0.058) -- (-0.265,0.286) -- (0.090,0.410) -- cycle;
\fill[ib, draw=ibd!60!black, thin] (-0.265,0.286) -- (0.090,0.410) -- (-0.223,0.701) -- cycle;
\fill[ib, draw=ibd!60!black, thin] (-0.779,-0.286) -- (-0.376,-0.133) -- (-0.740,0.202) -- cycle;
\fill[ib, draw=ibd!60!black, thin] (0.023,0.058) -- (0.387,0.171) -- (0.090,0.410) -- cycle;
\fill[ib, draw=ibd!60!black, thin] (-0.065,-0.392) -- (-0.376,-0.133) -- (0.023,0.058) -- cycle;
\fill[ib, draw=ibd!60!black, thin] (-0.376,-0.133) -- (0.023,0.058) -- (-0.265,0.286) -- cycle;
\fill[ib, draw=ibd!60!black, thin] (0.090,0.410) -- (-0.223,0.701) -- (0.139,0.815) -- cycle;
\fill[ib, draw=ibd!60!black, thin] (-0.467,-0.554) -- (-0.779,-0.286) -- (-0.376,-0.133) -- cycle;
\fill[ib, draw=ibd!60!black, thin] (-0.740,0.202) -- (-0.265,0.286) -- (-0.595,0.527) -- cycle;
\fill[ib, draw=ibd!60!black, thin] (-0.265,0.286) -- (-0.595,0.527) -- (-0.223,0.701) -- cycle;
\fill[ib, draw=ibd!60!black, thin] (0.196,-0.692) -- (-0.467,-0.554) -- (-0.065,-0.392) -- cycle;
\fill[ib, draw=ibd!60!black, thin] (0.387,0.171) -- (0.797,0.199) -- (0.507,0.545) -- cycle;
\fill[ib, draw=ibd!60!black, thin] (0.299,-0.335) -- (0.023,0.058) -- (0.387,0.171) -- cycle;
\fill[ib, draw=ibd!60!black, thin] (0.196,-0.692) -- (0.611,-0.531) -- (0.299,-0.335) -- cycle;
\fill[ib, draw=ibd!60!black, thin] (0.196,-0.692) -- (-0.065,-0.392) -- (0.299,-0.335) -- cycle;
\fill[ib, draw=ibd!60!black, thin] (-0.065,-0.392) -- (0.299,-0.335) -- (0.023,0.058) -- cycle;
\fill[ib, draw=ibd!60!black, thin] (-0.467,-0.554) -- (-0.065,-0.392) -- (-0.376,-0.133) -- cycle;
\fill (0.196,-0.692) circle (1.3pt);
\fill (0.611,-0.531) circle (1.3pt);
\fill (-0.467,-0.554) circle (1.3pt);
\fill (-0.065,-0.392) circle (1.3pt);
\fill (0.299,-0.335) circle (1.3pt);
\fill (0.728,-0.106) circle (1.3pt);
\fill (-0.779,-0.286) circle (1.3pt);
\fill (-0.376,-0.133) circle (1.3pt);
\fill (0.023,0.058) circle (1.3pt);
\fill (0.387,0.171) circle (1.3pt);
\fill (0.797,0.199) circle (1.3pt);
\fill (-0.740,0.202) circle (1.3pt);
\fill (-0.265,0.286) circle (1.3pt);
\fill (0.090,0.410) circle (1.3pt);
\fill (0.507,0.545) circle (1.3pt);
\fill (-0.595,0.527) circle (1.3pt);
\fill (-0.223,0.701) circle (1.3pt);
\fill (0.139,0.815) circle (1.3pt);
\fill[mk, draw=black, very thin] (0.638,0.088) ++(-1.2pt,-1.2pt) rectangle ++(2.4pt,2.4pt);
\fill[mk, draw=black, very thin] (0.472,-0.090) ++(-1.2pt,-1.2pt) rectangle ++(2.4pt,2.4pt);
\fill[mk, draw=black, very thin] (-0.460,0.119) ++(-1.2pt,-1.2pt) rectangle ++(2.4pt,2.4pt);
\fill[mk, draw=black, very thin] (0.546,-0.324) ++(-1.2pt,-1.2pt) rectangle ++(2.4pt,2.4pt);
\fill[mk, draw=black, very thin] (0.328,0.375) ++(-1.2pt,-1.2pt) rectangle ++(2.4pt,2.4pt);
\fill[mk, draw=black, very thin] (0.246,0.590) ++(-1.2pt,-1.2pt) rectangle ++(2.4pt,2.4pt);
\fill[mk, draw=black, very thin] (-0.051,0.251) ++(-1.2pt,-1.2pt) rectangle ++(2.4pt,2.4pt);
\fill[mk, draw=black, very thin] (-0.133,0.466) ++(-1.2pt,-1.2pt) rectangle ++(2.4pt,2.4pt);
\fill[mk, draw=black, very thin] (-0.632,-0.072) ++(-1.2pt,-1.2pt) rectangle ++(2.4pt,2.4pt);
\fill[mk, draw=black, very thin] (0.167,0.213) ++(-1.2pt,-1.2pt) rectangle ++(2.4pt,2.4pt);
\fill[mk, draw=black, very thin] (-0.139,-0.156) ++(-1.2pt,-1.2pt) rectangle ++(2.4pt,2.4pt);
\fill[mk, draw=black, very thin] (-0.206,0.070) ++(-1.2pt,-1.2pt) rectangle ++(2.4pt,2.4pt);
\fill[mk, draw=black, very thin] (0.002,0.642) ++(-1.2pt,-1.2pt) rectangle ++(2.4pt,2.4pt);
\fill[mk, draw=black, very thin] (-0.540,-0.324) ++(-1.2pt,-1.2pt) rectangle ++(2.4pt,2.4pt);
\fill[mk, draw=black, very thin] (-0.534,0.339) ++(-1.2pt,-1.2pt) rectangle ++(2.4pt,2.4pt);
\fill[mk, draw=black, very thin] (-0.361,0.505) ++(-1.2pt,-1.2pt) rectangle ++(2.4pt,2.4pt);
\fill[mk, draw=black, very thin] (-0.112,-0.546) ++(-1.2pt,-1.2pt) rectangle ++(2.4pt,2.4pt);
\fill[mk, draw=black, very thin] (0.564,0.305) ++(-1.2pt,-1.2pt) rectangle ++(2.4pt,2.4pt);
\fill[mk, draw=black, very thin] (0.236,-0.035) ++(-1.2pt,-1.2pt) rectangle ++(2.4pt,2.4pt);
\fill[mk, draw=black, very thin] (0.369,-0.519) ++(-1.2pt,-1.2pt) rectangle ++(2.4pt,2.4pt);
\fill[mk, draw=black, very thin] (0.143,-0.473) ++(-1.2pt,-1.2pt) rectangle ++(2.4pt,2.4pt);
\fill[mk, draw=black, very thin] (0.086,-0.223) ++(-1.2pt,-1.2pt) rectangle ++(2.4pt,2.4pt);
\fill[mk, draw=black, very thin] (-0.302,-0.360) ++(-1.2pt,-1.2pt) rectangle ++(2.4pt,2.4pt);
\coordinate (ptop) at (0.139,0.815);
\coordinate (pmark) at (0.369,-0.519);
\coordinate (z1) at (-0.779,-0.286);
\coordinate (z2) at (-0.595,0.527);

%% file: loose_strong.tex
\begin{tikzpicture}[
  font=\small,
  box/.style={draw, align=center, inner sep=5pt, minimum width=40mm},
  tinybox/.style={draw, align=center, inner sep=4pt},
  diamond/.style={draw, diamond, aspect=2.2, align=center, inner sep=2pt},
  lab/.style={font=\small},
  node distance=7mm
]

\coordinate (o)  at (0,0) {};
\node [right=1mm of o] (vqn) {$\Fluid_{n}$};
\node [left =1mm of o] (vzn) {$\Solid_{n}$};

\node[box, thick, below=4mm of o] (predz) {%
$t_{n}\rightarrow t_{n+1}$\\
$\dot{\Solid}=\mathcal{Z}(\Solid,\Fluid_{n})\Rightarrow \Solid_{n+1}^p$
};

\coordinate [below=4mm of predz] (zpqn) {};
\node [right=1mm of zpqn] (vqn_) {$\Fluid_n$};
\node [left =1mm of zpqn] (vznp1_p) {$\Solid_{n+1}^p$};

\node[box, thick, below=4mm of zpqn] (predq) {%
$t_{n}\rightarrow t_{n+1}$\\
$\dot{\Fluid}=\mathcal{Q}(\Fluid,\Solid_{n+1}^p)\Rightarrow \Fluid_{n+1}^p$
};

\coordinate [below=4mm of predq] (zpqp) {};
\node [right=1mm of zpqp] (vqnp1_p)  {$\Fluid_{n+1}^p$};
\node [left =1mm of zpqp] (vznp1_p_) {$\Solid_{n+1}^p$};



\node[box, right=4cm of o, draw=black, dotted] (initc) {%
Initialize $k=1$\\
$\Solid_{n+1}^k = \Solid_{n+1}^p ;\quad \Fluid_{n+1}^k = \Fluid_{n+1}^p$
};

\coordinate [below=2mm of initc] (halfway_arrow_initc) {};
\node[box, right=2.5cm of halfway_arrow_initc, draw=black, dotted ,minimum width=2.5cm] (reinitc) {%
$k=k+1$\\
$\Solid_{n+1}^k = \Solid_{n+1}^{k+1}$\\
$\Fluid_{n+1}^k = \Fluid_{n+1}^{k+1}$
};

\node [box, thick, below=4mm of initc] (corrz) {%
$t_n \rightarrow t_{n+1}$\\
$\dot{\Solid}=\mathcal{Z}(\Solid,\Fluid_{n+1}^k)\Rightarrow \Solid_{n+1}^{k+1}$
};

\coordinate [below=4mm of corrz] (znp1kp1qnp1k) {};
\node [right=1mm of znp1kp1qnp1k] (vqnp1_k_)  {$\Fluid_{n+1}^k$};
\node [left =1mm of znp1kp1qnp1k] (vznp1_kp1) {$\Solid_{n+1}^{k+1}$};

\node [shape=trapezium, below=4mm of znp1kp1qnp1k, rounded corners, draw=black, dotted] (checkz) {%
check $|| \Solid_{n+1}^{k+1} - \Solid_{n+1}^k ||_\infty < \epsilon $
};

\node [box, thick, below=4mm of checkz] (corrq) {%
$t_n \rightarrow t_{n+1}$\\
$\dot{\Fluid}=\mathcal{Q}(\Fluid,\Solid_{n+1}^{k+1})\Rightarrow \Fluid_{n+1}^{k+1}$
};
\draw [->, thick] (checkz.south) -- (corrq.north) node [midway,left] {no};

\coordinate [below=4mm of corrq] (znp1kp1qnp1kp1) {};
\node [right=1mm of znp1kp1qnp1kp1] (vqnp1_kp1)  {$\Fluid_{n+1}^{k+1}$};
\node [left =1mm of znp1kp1qnp1kp1] (vznp1_kp1_) {$\Solid_{n+1}^{k+1}$};


\draw [->] (initc.south) -- (corrz.north);

\draw [->, thick]  (reinitc.west) -| (corrz.north);

\draw [->, thick] ($(znp1kp1qnp1kp1)+(0,-3mm)$) -- ++(0,-2mm) -| (reinitc.south);

\node[draw,
    thick,
    rounded corners,
    inner ysep=4mm,
    inner xsep=4mm,
    fit=(vqn)(vzn)(predz)(vqn_)(vznp1_p)(predq)(vqnp1_p)(vznp1_p_),
    label=above:{\textbf{Predictor}}] (predictor) {};

\node[draw,
    thick,
    rounded corners,
    inner ysep=6mm,
    inner xsep=5.5mm,
    fit=(initc)(reinitc)(corrz)(vqnp1_k_)(vznp1_kp1)(checkz)(corrq)(vqnp1_kp1)(vznp1_kp1_),
    label=above:{\textbf{Corrector}}] (corrector) {};

\draw [->, dotted, thick] (predictor.south) -- ++(0,-6mm) -- ++(2.9cm,0) node [midway,below] (sc) {(strong coupling)} |- (initc.west) ;

\coordinate [below=1.5cm of corrector.south] (out_corr) {};
\coordinate (out_pred) at (out_corr -| predictor.south) {};
\node at (out_pred) (out_pred)  {$\Solid_{n+1}=\Solid_{n+1}^{p}; \quad \Fluid_{n+1}=\Fluid_{n+1}^{p}$};
\node at (out_corr) (out_corr)  {$\Solid_{n+1}=\Solid_{n+1}^{k+1}; \quad \Fluid_{n+1}=\Fluid_{n+1}^{k}$};

\coordinate [above=2mm of corrector.south] (corrector_south_inside) {};
\draw [->] (checkz.west) -- ++(-4mm,0) |- (corrector_south_inside) -- (out_corr.north);
\node [anchor=south, inner sep=1pt] (yes) at ($(checkz.west)+(-3mm,0.6mm)$) {yes};
\draw [->] (predictor.south) -- (out_pred.north) node[midway,left] (lc) {(loose coupling)};
\draw [->] (predictor.south) -- (out_pred.north) node[midway,left] {(loose coupling)};

\end{tikzpicture}

%% file: decomposition.tex
\begin{tikzpicture}[font=\small, line cap=round, line join=round, >=Latex]
\definecolor{ib}{HTML}{8ADBD2}
\definecolor{ibd}{HTML}{2A9D8F}
\def\W{2.6}\def\H{4.6}\def\dx{0.9}\def\dy{0.55}
\def\ea{0.7}\def\eb{1.85}  
\def\nsl{6} 
\colorlet{c0}{blue!60!black}
\colorlet{c1}{blue!70!cyan}
\colorlet{c2}{cyan!70!teal}
\colorlet{c3}{yellow!75!orange}
\colorlet{c4}{orange!90!red}
\colorlet{c5}{red!75!black}
\begin{scope}
  \coordinate (e) at (\W/2+\dx/2,\H/2+\dy/2);
  \fill[gray!22] (\dx,\dy) rectangle ++(\W,\H);
  \fill[gray!35] (0,0) -- (\dx,\dy) -- (\dx,\dy+\H) -- (0,\H) -- cycle;
  \fill[gray!30] (0,0) -- (\W,0) -- (\W+\dx,\dy) -- (\dx,\dy) -- cycle;
  \draw[gray!60, dashed, thin] (0,0) -- (\dx,\dy) -- (\W+\dx,\dy)  (\dx,\dy) -- (\dx,\dy+\H);
  \shade[inner color=gray!5, outer color=gray!60] (e) ellipse [x radius=\ea, y radius=\eb];
  \begin{scope}[shift={(e)}, xscale=\ea/0.8, yscale=\eb/2.05]
    \clip (0,0) ellipse [x radius=0.8, y radius=2.05];
    \begin{scope}[gray!40!black]\input{decomp_mesh.tex}\end{scope}
  \end{scope}
  \draw[gray!50!black, very thin] (e) ellipse [x radius=\ea, y radius=\eb];
  \fill[gray!45, opacity=0.6] (0,\H) -- (\W,\H) -- (\W+\dx,\H+\dy) -- (\dx,\H+\dy) -- cycle;
  \fill[gray!55, opacity=0.6] (\W,0) -- (\W+\dx,\dy) -- (\W+\dx,\H+\dy) -- (\W,\H) -- cycle;
  \fill[gray!20, opacity=0.25] (0,0) rectangle (\W,\H);
  \draw[black!70, thin] (0,0) rectangle (\W,\H);
  \draw[black!70, thin] (0,\H) -- (\dx,\H+\dy) -- (\W+\dx,\H+\dy) -- (\W,\H);
  \draw[black!70, thin] (\W+\dx,\H+\dy) -- (\W+\dx,\dy) -- (\W,0);
  \node[above] at (\W/2+\dx/2,\H+\dy+0.15) {Single GPU};
  \begin{scope}[shift={(\W+\dx+0.35,0.15)}]
    \draw[->] (0,0) -- (0,0.5) node[above, inner sep=1pt] {$x_3$};
    \draw[->] (0,0) -- (0.5,0) node[right, inner sep=1pt] {$x_2$};
    \draw[->] (0,0) -- (-0.5*\dx/1.055,-0.5*\dy/1.055) node[below left, inner sep=0.5pt] {$x_1$};
  \end{scope}
\end{scope}
\draw[->, thick] (4.3,\H/2+\dy/2) -- (5.3,\H/2+\dy/2);
\begin{scope}[shift={(6.2,0)}]
  \coordinate (e) at (\W/2+\dx/2,\H/2+\dy/2);
  \pgfmathsetmacro{\hs}{\H/\nsl}
  \foreach \k in {0,...,5} {
    \pgfmathsetmacro{\zl}{\k*\hs}\pgfmathsetmacro{\zh}{(\k+1)*\hs}
    \fill[c\k!18!white] (\dx,\dy+\zl) rectangle (\W+\dx,\dy+\zh);
    \fill[c\k!30!white] (0,\zl) -- (\dx,\dy+\zl) -- (\dx,\dy+\zh) -- (0,\zh) -- cycle;
  }
  \fill[gray!30] (0,0) -- (\W,0) -- (\W+\dx,\dy) -- (\dx,\dy) -- cycle;
  \draw[gray!60, dashed, thin] (0,0) -- (\dx,\dy) -- (\W+\dx,\dy)  (\dx,\dy) -- (\dx,\dy+\H);
  \shade[inner color=gray!5, outer color=gray!60] (e) ellipse [x radius=\ea, y radius=\eb];
  \begin{scope}
    \clip (e) ellipse [x radius=\ea, y radius=\eb];
    \foreach \k in {0,...,5} {
      \pgfmathsetmacro{\zl}{\k*\hs}\pgfmathsetmacro{\zh}{(\k+1)*\hs}
      \fill[c\k, opacity=0.8] (-1,\zl+\dy/2-0.03*\k) -- (\W+\dx+1,\zl+\dy/2-0.03*\k+0.0) -- (\W+\dx+1,\zh+\dy/2-0.03*\k) -- (-1,\zh+\dy/2-0.03*\k) -- cycle;
    }
    \begin{scope}[shift={(e)}, xscale=\ea/0.8, yscale=\eb/2.05]\begin{scope}[black!55]\input{decomp_mesh.tex}\end{scope}\end{scope}
  \end{scope}
  \draw[gray!50!black, very thin] (e) ellipse [x radius=\ea, y radius=\eb];
  \foreach \k in {0,...,5} {
    \pgfmathsetmacro{\zl}{\k*\hs}\pgfmathsetmacro{\zh}{(\k+1)*\hs}
    \fill[c\k, opacity=0.07] (0,\zl) rectangle (\W,\zh);
    \fill[c\k, opacity=0.2] (0,\zh) -- (\W,\zh) -- (\W+\dx,\zh+\dy) -- (\dx,\zh+\dy) -- cycle;
    \fill[c\k, opacity=0.35] (\W,\zl) -- (\W+\dx,\zl+\dy) -- (\W+\dx,\zh+\dy) -- (\W,\zh) -- cycle;
    \draw[c\k!70!black, thin] (0,\zl) rectangle (\W,\zh);
    \draw[c\k!70!black, thin] (0,\zh) -- (\dx,\zh+\dy) -- (\W+\dx,\zh+\dy) -- (\W,\zh);
    \draw[c\k!70!black, thin] (\W+\dx,\zh+\dy) -- (\W+\dx,\zl+\dy) -- (\W,\zl);
  }
  \node[above] at (\W/2+\dx/2,\H+\dy+0.15) {Multi-GPU};
  \node[right] at (\W+\dx+0.15,0.5*\hs+\dy) {MPI domain $0$};
  \node[right] at (\W+\dx+0.15,1.5*\hs+\dy) {MPI domain $1$};
  \node[right] at (\W+\dx+0.55,3.5*\hs+\dy) {$\vdots$};
  \node[right] at (\W+\dx+0.15,5.5*\hs+\dy) {MPI domain $N-1$};
\end{scope}
\end{tikzpicture}

%% file: decomp_mesh.tex
\draw[line width=0.25pt, opacity=0.8] (0.450,-0.675)--(0.533,-0.901) (-0.213,-1.429)--(-0.276,-1.612) (0.642,0.880)--(0.683,1.014) (0.186,0.475)--(0.281,0.327) (0.281,0.327)--(0.292,0.567) (-0.041,1.889)--(-0.112,1.941) (0.105,-1.763)--(0.072,-1.920) (-0.403,0.894)--(-0.383,1.122) (-0.564,-1.446)--(-0.566,-1.450) (0.656,1.167)--(0.693,1.025) (0.042,0.001)--(0.115,0.190) (0.019,1.790)--(0.117,1.833) (0.773,0.200)--(0.797,0.179) (0.061,-0.507)--(-0.132,-0.470) (0.287,-0.388)--(0.336,-0.616) (0.459,1.669)--(0.459,1.679) (0.697,0.692)--(0.752,0.701) (-0.747,0.313)--(-0.791,0.268) (0.272,-0.105)--(0.430,-0.064) (0.117,-1.528)--(0.040,-1.648) (0.356,1.639)--(0.415,1.681) (-0.287,1.522)--(-0.421,1.360) (0.313,-1.280)--(0.387,-1.349) (0.696,0.120)--(0.773,0.200) (-0.383,-0.964)--(-0.486,-1.097) (-0.467,-1.462)--(-0.549,-1.331) (-0.191,1.069)--(-0.148,1.279) (-0.186,-1.613)--(-0.241,-1.761) (0.635,1.062)--(0.656,1.167) (0.292,0.567)--(0.474,0.502) (0.066,-0.229)--(-0.044,-0.372) (0.506,-1.133)--(0.593,-1.244) (-0.241,-1.761)--(-0.184,-1.886) (-0.004,-2.049)--(-0.070,-2.042) (-0.297,0.791)--(-0.403,0.894) (-0.433,-1.616)--(-0.514,-1.570) (0.296,1.522)--(0.387,1.475) (-0.699,-0.771)--(-0.741,-0.707) (-0.070,2.042)--(-0.139,2.019) (-0.540,1.512)--(-0.566,1.450) (0.272,-1.926)--(0.274,-1.926) (0.074,1.657)--(0.019,1.790) (-0.608,1.053)--(-0.671,1.080) (0.454,-1.654)--(0.459,-1.679) (-0.124,-1.369)--(-0.213,-1.429) (0.568,1.064)--(0.635,1.062) (0.097,2.004)--(0.139,2.019) (0.183,-1.638)--(0.105,-1.763) (0.067,-1.341)--(0.117,-1.528) (-0.439,-0.390)--(-0.578,-0.431) (-0.074,0.364)--(-0.186,0.266) (-0.139,-2.019)--(-0.070,-2.042) (-0.177,1.960)--(-0.207,1.980) (0.617,-0.605)--(0.719,-0.639) (-0.001,0.916)--(0.058,1.123) (-0.403,-1.178)--(-0.484,-1.291) (0.727,0.279)--(0.780,0.359) (0.239,-1.855)--(0.272,-1.926) (-0.359,1.718)--(-0.400,1.775) (-0.269,1.238)--(-0.223,1.422) (-0.536,0.210)--(-0.697,0.212) (0.461,-1.355)--(0.587,-1.366) (0.480,0.993)--(0.560,0.877) (-0.739,0.062)--(-0.777,0.159) (-0.403,0.894)--(-0.491,0.908) (0.264,1.687)--(0.356,1.639) (0.315,1.324)--(0.296,1.522) (0.524,-0.440)--(0.617,-0.605) (-0.361,1.557)--(-0.423,1.683) (-0.062,-0.936)--(0.146,-0.941) (-0.298,-0.968)--(-0.383,-0.964) (0.019,1.790)--(-0.041,1.889) (0.656,0.324)--(0.727,0.279) (-0.665,0.961)--(-0.710,0.940) (0.287,-0.388)--(0.423,-0.464) (-0.091,-1.575)--(-0.186,-1.613) (0.426,-1.143)--(0.531,-1.319) (-0.671,1.080)--(-0.655,1.176) (0.117,-1.528)--(0.183,-1.638) (-0.510,-0.581)--(-0.618,-0.606) (-0.618,-0.606)--(-0.638,-0.792) (0.678,-0.517)--(0.757,-0.502) (-0.762,-0.620)--(-0.773,-0.531) (-0.290,0.226)--(-0.366,0.084) (0.548,0.345)--(0.607,0.180) (-0.191,1.069)--(-0.269,1.238) (-0.566,1.450)--(-0.613,1.318) (0.487,1.200)--(0.568,1.064) (0.338,-1.858)--(0.400,-1.775) (-0.132,-0.470)--(-0.009,-0.669) (-0.297,0.791)--(-0.495,0.694) (-0.341,-0.449)--(-0.439,-0.390) (0.296,1.522)--(0.264,1.687) (-0.439,1.534)--(-0.486,1.610) (0.609,-0.362)--(0.678,-0.517) (0.686,0.499)--(0.772,0.527) (-0.314,-1.212)--(-0.403,-1.178) (0.800,0.000)--(0.797,0.179) (0.329,-1.715)--(0.302,-1.851) (-0.124,-1.369)--(-0.091,-1.575) (-0.487,0.003)--(-0.615,0.090) (-0.615,0.090)--(-0.675,-0.046) (-0.205,1.847)--(-0.112,1.941) (0.488,-0.247)--(0.524,-0.440) (-0.421,1.360)--(-0.439,1.534) (-0.025,-1.968)--(-0.004,-2.049) (0.472,0.193)--(0.548,0.345) (-0.773,0.531)--(-0.788,0.356) (0.330,-1.536)--(0.256,-1.702) (0.165,1.929)--(0.205,1.969) (0.477,1.378)--(0.584,1.377) (0.506,-1.133)--(0.574,-1.096) (-0.290,-1.838)--(-0.387,-1.749) (0.673,-0.275)--(0.782,-0.360) (0.459,1.669)--(0.514,1.570) (-0.430,-0.669)--(-0.510,-0.581) (-0.131,-1.972)--(-0.139,-2.019) (0.011,-1.845)--(-0.025,-1.968) (-0.298,-0.968)--(-0.314,-1.212) (-0.788,0.356)--(-0.797,0.179) (0.560,1.243)--(0.613,1.318) (-0.495,0.694)--(-0.579,0.641) (0.262,1.889)--(0.338,1.858) (0.693,-1.025)--(0.725,-0.866) (-0.601,-1.307)--(-0.637,-1.238) (0.143,-1.956)--(0.207,-1.980) (-0.009,1.995)--(0.000,2.050) (0.079,0.440)--(0.281,0.327) (-0.675,-0.046)--(-0.714,-0.199) (0.329,-1.715)--(0.400,-1.775) (0.655,1.176)--(0.613,1.318) (-0.500,1.497)--(-0.555,1.429) (0.461,-1.355)--(0.566,-1.450) (0.725,-0.866)--(0.752,-0.701) (0.548,0.345)--(0.599,0.525) (0.097,2.004)--(0.144,2.016) (-0.540,1.512)--(-0.514,1.570) (-0.411,-0.133)--(-0.487,0.003) (-0.539,1.124)--(-0.655,1.176) (0.574,-1.096)--(0.638,-1.013) (-0.710,0.940)--(-0.725,0.866) (0.613,1.318)--(0.566,1.450) (-0.341,-0.449)--(-0.430,-0.669) (-0.466,1.150)--(-0.421,1.360) (0.072,-1.920)--(0.040,-2.013) (0.105,-1.763)--(0.256,-1.702) (-0.771,0.433)--(-0.788,0.356) (0.040,-2.013)--(0.106,-2.024) (-0.170,-0.178)--(-0.319,-0.217) (-0.579,0.641)--(-0.648,0.570) (0.061,1.932)--(0.165,1.929) (-0.223,1.422)--(-0.347,1.327) (0.719,-0.639)--(0.752,-0.701) (0.787,-0.088)--(0.800,0.000) (-0.205,1.847)--(-0.272,1.858) (0.720,-0.128)--(0.797,-0.179) (-0.725,-0.866)--(-0.693,-1.025) (0.488,-0.247)--(0.582,-0.155) (-0.366,0.084)--(-0.536,0.210) (0.415,1.681)--(0.459,1.669) (0.255,-0.931)--(0.351,-0.852) (0.330,-1.536)--(0.401,-1.551) (0.366,-0.262)--(0.423,-0.464) (-0.186,-1.613)--(-0.041,-1.730) (0.031,1.381)--(-0.094,1.438) (-0.041,-1.730)--(-0.089,-1.888) (0.539,1.400)--(0.542,1.507) (0.151,-0.645)--(0.053,-0.856) (0.402,1.772)--(0.400,1.775) (0.053,-0.856)--(0.146,-0.941) (0.698,0.850)--(0.725,0.866) (-0.766,-0.094)--(-0.800,0.000) (-0.627,-1.145)--(-0.672,-1.090) (0.205,1.969)--(0.207,1.980) (0.151,-0.645)--(0.336,-0.616) (0.403,1.287)--(0.560,1.243) (0.477,1.378)--(0.539,1.400) (-0.290,-1.838)--(-0.338,-1.858) (-0.272,1.858)--(-0.338,1.858) (0.215,1.051)--(0.137,1.253) (-0.250,-0.032)--(-0.366,0.084) (0.317,1.778)--(0.402,1.772) (0.452,-0.904)--(0.506,-1.133) (-0.302,-1.432)--(-0.375,-1.559) (0.697,0.692)--(0.734,0.787) (-0.383,1.122)--(-0.347,1.327) (-0.737,-0.370)--(-0.778,-0.281) (0.508,-1.576)--(0.514,-1.570) (0.117,1.833)--(0.212,1.829) (0.336,-0.616)--(0.351,-0.852) (0.645,-1.147)--(0.655,-1.176) (0.629,0.705)--(0.697,0.692) (-0.739,0.575)--(-0.772,0.541) (0.387,1.475)--(0.542,1.507) (0.366,-0.262)--(0.488,-0.247) (0.757,-0.502)--(0.788,-0.356) (0.040,-1.648)--(-0.041,-1.730) (0.292,0.567)--(0.273,0.830) (0.450,1.545)--(0.501,1.554) (0.008,0.246)--(-0.186,0.266) (0.031,1.381)--(0.221,1.311) (-0.222,1.675)--(-0.361,1.557) (0.252,-1.479)--(0.330,-1.536) (-0.469,-0.900)--(-0.568,-0.966) (0.720,-0.128)--(0.787,-0.088) (-0.148,1.279)--(-0.094,1.438) (-0.276,-1.612)--(-0.338,-1.710) (0.698,0.850)--(0.713,0.932) (-0.704,0.461)--(-0.739,0.575) (0.053,-0.856)--(-0.062,-0.936) (-0.296,1.019)--(-0.383,1.122) (-0.631,-0.265)--(-0.737,-0.370) (0.387,1.475)--(0.477,1.378) (-0.752,-0.701)--(-0.725,-0.866) (-0.747,-0.531)--(-0.777,-0.448) (0.797,-0.187)--(0.788,-0.356) (-0.566,-1.153)--(-0.601,-1.307) (0.647,-0.031)--(0.720,-0.128) (-0.561,1.288)--(-0.611,1.312) (0.450,-0.675)--(0.452,-0.904) (0.186,-0.397)--(0.336,-0.616) (-0.213,-1.429)--(-0.302,-1.432) (0.642,0.880)--(0.698,0.850) (0.215,1.051)--(0.363,0.849) (0.105,-1.763)--(0.011,-1.845) (-0.511,-0.266)--(-0.631,-0.265) (0.607,-0.850)--(0.690,-0.902) (-0.011,-1.484)--(0.040,-1.648) (0.734,-0.755)--(0.725,-0.866) (-0.771,0.433)--(-0.773,0.531) (-0.637,-1.238)--(-0.655,-1.176) (0.058,1.123)--(0.137,1.253) (-0.778,-0.281)--(-0.788,-0.356) (-0.486,-1.097)--(-0.566,-1.153) (0.754,0.015)--(0.797,0.099) (-0.539,1.124)--(-0.561,1.288) (0.143,-1.956)--(0.165,-2.006) (-0.044,-0.372)--(-0.132,-0.470) (-0.481,-1.614)--(-0.514,-1.570) (0.274,1.926)--(0.207,1.980) (-0.568,0.426)--(-0.704,0.461) (0.401,-1.551)--(0.508,-1.576) (-0.747,0.313)--(-0.771,0.433) (0.560,0.877)--(0.629,0.705) (0.281,0.327)--(0.474,0.502) (0.745,0.432)--(0.788,0.356) (-0.234,-1.933)--(-0.207,-1.980) (0.305,1.100)--(0.403,1.287) (0.356,1.639)--(0.450,1.545) (0.540,-0.671)--(0.607,-0.850) (-0.291,1.718)--(-0.334,1.835) (-0.672,-0.426)--(-0.747,-0.531) (0.255,-0.931)--(0.226,-1.160) (-0.741,-0.707)--(-0.762,-0.620) (-0.383,1.122)--(-0.466,1.150) (-0.383,-0.964)--(-0.469,-0.900) (0.117,1.833)--(0.061,1.932) (-0.618,1.205)--(-0.651,1.192) (0.696,0.120)--(0.754,0.015) (-0.186,-1.613)--(-0.276,-1.612) (-0.227,-1.189)--(-0.213,-1.429) (0.008,0.246)--(-0.074,0.364) (0.387,-1.349)--(0.470,-1.518) (0.183,0.720)--(0.273,0.830) (0.040,-1.648)--(0.105,-1.763) (-0.578,-0.431)--(-0.672,-0.426) (0.667,-0.770)--(0.734,-0.755) (0.031,1.381)--(0.107,1.480) (-0.338,-1.710)--(-0.459,-1.679) (0.387,0.580)--(0.530,0.676) (0.402,1.772)--(0.459,1.679) (-0.148,1.279)--(-0.223,1.422) (-0.641,0.332)--(-0.747,0.313) (-0.400,-1.775)--(-0.338,-1.858) (0.205,1.969)--(0.274,1.926) (-0.439,-0.390)--(-0.511,-0.266) (0.144,2.016)--(0.139,2.019) (-0.359,1.718)--(-0.400,1.771) (0.617,-0.605)--(0.667,-0.770) (-0.069,2.024)--(-0.070,2.042) (-0.403,-1.178)--(-0.486,-1.097) (-0.207,1.980)--(-0.274,1.926) (-0.335,-0.703)--(-0.383,-0.964) (0.239,-1.855)--(0.210,-1.956) (-0.366,0.084)--(-0.411,-0.133) (0.115,0.190)--(0.281,0.327) (-0.536,0.210)--(-0.641,0.332) (0.461,-1.355)--(0.529,-1.466) (-0.334,1.835)--(-0.400,1.775) (-0.272,1.858)--(-0.274,1.926) (-0.338,-1.710)--(-0.290,-1.838) (-0.112,1.941)--(-0.009,1.995) (0.215,1.051)--(0.305,1.100) (-0.618,-0.606)--(-0.741,-0.707) (-0.500,1.497)--(-0.566,1.450) (0.524,-0.440)--(0.540,-0.671) (-0.361,1.557)--(-0.359,1.718) (-0.044,-0.372)--(-0.170,-0.178) (0.160,1.701)--(0.117,1.833) (-0.491,0.908)--(-0.608,1.053) (-0.608,1.053)--(-0.693,1.025) (0.536,0.045)--(0.607,0.180) (0.256,-1.702)--(0.173,-1.828) (0.262,1.889)--(0.306,1.891) (0.508,-1.576)--(0.459,-1.679) (0.560,0.877)--(0.568,1.064) (0.426,-1.143)--(0.461,-1.355) (-0.510,-0.581)--(-0.578,-0.431) (-0.319,-0.217)--(-0.439,-0.390) (-0.089,-1.888)--(-0.131,-1.972) (-0.186,0.266)--(-0.290,0.226) (0.187,-1.357)--(0.387,-1.349) (0.757,-0.502)--(0.773,-0.531) (-0.383,-0.964)--(-0.403,-1.178) (0.797,0.099)--(0.797,0.179) (0.070,-2.042)--(0.139,-2.019) (0.302,-1.851)--(0.272,-1.926) (0.212,1.829)--(0.317,1.778) (0.292,0.567)--(0.387,0.580) (0.748,0.603)--(0.752,0.701) (-0.697,0.212)--(-0.739,0.062) (0.356,1.639)--(0.459,1.669) (0.211,0.120)--(0.281,0.327) (-0.737,-0.370)--(-0.788,-0.356) (-0.423,1.683)--(-0.486,1.610) (0.607,0.180)--(0.656,0.324) (0.734,0.787)--(0.725,0.866) (-0.487,0.003)--(-0.536,0.210) (0.531,-1.319)--(0.593,-1.244) (0.296,1.522)--(0.450,1.545) (-0.655,1.176)--(-0.693,1.025) (-0.672,-0.426)--(-0.696,-0.585) (0.696,0.120)--(0.797,0.179) (-0.421,1.360)--(-0.361,1.557) (-0.439,-0.390)--(-0.510,-0.581) (-0.025,-1.968)--(-0.070,-2.027) (-0.070,-2.027)--(-0.004,-2.049) (-0.358,0.434)--(-0.436,0.247) (-0.569,0.891)--(-0.639,0.805) (0.568,1.064)--(0.683,1.014) (0.165,1.929)--(0.262,1.889) (-0.152,1.588)--(-0.287,1.522) (-0.112,1.941)--(-0.177,1.960) (-0.009,-0.669)--(0.053,-0.856) (-0.430,-1.727)--(-0.459,-1.679) (0.524,-0.440)--(0.609,-0.362) (-0.436,0.247)--(-0.568,0.426) (0.011,-1.845)--(-0.089,-1.888) (0.727,0.279)--(0.793,0.263) (0.501,1.554)--(0.542,1.507) (0.226,-1.160)--(0.340,-1.074) (-0.234,-1.933)--(-0.274,-1.926) (0.514,-1.570)--(0.566,-1.450) (-0.566,-1.153)--(-0.655,-1.176) (0.667,-0.770)--(0.719,-0.639) (0.423,-0.464)--(0.450,-0.675) (0.256,-1.702)--(0.329,-1.715) (0.040,-2.013)--(0.070,-2.042) (-0.641,0.332)--(-0.697,0.212) (0.773,0.531)--(0.752,0.701) (-0.186,0.266)--(-0.175,0.524) (-0.175,0.524)--(-0.274,0.525) (-0.387,-1.749)--(-0.338,-1.847) (-0.125,0.064)--(-0.290,0.226) (-0.000,-2.050)--(0.070,-2.042) (0.107,1.480)--(-0.021,1.568) (0.212,1.829)--(0.165,1.929) (-0.143,-1.770)--(-0.184,-1.886) (0.211,0.120)--(0.272,-0.105) (0.461,-1.355)--(0.531,-1.319) (0.146,-0.941)--(0.012,-1.095) (0.725,-0.866)--(0.752,-0.701) (0.012,-1.095)--(0.118,-1.158) (-0.651,1.192)--(-0.655,1.176) (0.239,-0.682)--(0.255,-0.931) (-0.749,0.689)--(-0.752,0.701) (-0.491,0.908)--(-0.569,0.891) (0.560,1.243)--(0.615,1.229) (0.348,0.065)--(0.536,0.045) (-0.777,0.159)--(-0.797,0.179) (0.690,-0.902)--(0.693,-1.025) (0.137,1.253)--(0.305,1.100) (0.305,1.100)--(0.221,1.311) (-0.366,0.084)--(-0.436,0.247) (-0.314,-1.212)--(-0.484,-1.291) (0.793,0.263)--(0.788,0.356) (-0.038,1.223)--(0.031,1.381) (-0.392,-1.393)--(-0.467,-1.462) (0.745,0.432)--(0.772,0.527) (-0.009,-0.669)--(-0.115,-0.738) (-0.347,1.327)--(-0.287,1.522) (-0.115,-0.738)--(-0.062,-0.936) (-0.615,0.090)--(-0.739,0.062) (-0.766,-0.094)--(-0.793,0.005) (-0.608,1.053)--(-0.655,1.176) (0.221,1.311)--(0.296,1.522) (-0.638,-0.792)--(-0.690,-0.944) (-0.569,0.891)--(-0.711,0.845) (0.686,0.499)--(0.745,0.432) (-0.175,0.524)--(-0.111,0.765) (0.423,-0.464)--(0.524,-0.440) (0.273,0.830)--(0.387,0.580) (-0.041,-1.730)--(-0.143,-1.770) (0.539,1.400)--(0.584,1.377) (0.797,0.099)--(0.800,0.000) (-0.186,0.266)--(-0.274,0.525) (0.132,0.948)--(0.215,1.051) (-0.547,-0.786)--(-0.638,-0.792) (0.728,-0.386)--(0.782,-0.360) (0.748,0.603)--(0.773,0.531) (-0.125,0.064)--(-0.366,0.084) (-0.170,-1.999)--(-0.139,-2.019) (-0.094,1.438)--(-0.021,1.568) (-0.375,-1.559)--(-0.433,-1.616) (-0.495,0.694)--(-0.639,0.805) (0.306,1.891)--(0.338,1.858) (-0.694,0.703)--(-0.711,0.845) (0.615,1.229)--(0.613,1.318) (0.012,-1.095)--(-0.142,-1.100) (0.635,1.062)--(0.693,1.025) (0.363,0.849)--(0.480,0.993) (-0.675,-0.046)--(-0.766,-0.094) (0.477,1.378)--(0.560,1.243) (0.031,2.043)--(0.000,2.050) (0.146,-0.941)--(0.118,-1.158) (0.200,1.524)--(0.264,1.687) (-0.638,-0.983)--(-0.672,-1.090) (0.673,-0.275)--(0.728,-0.386) (-0.500,1.497)--(-0.540,1.512) (0.788,0.356)--(0.773,0.531) (0.548,0.345)--(0.686,0.499) (0.239,-0.682)--(0.351,-0.852) (0.697,0.692)--(0.748,0.603) (-0.302,-1.432)--(-0.392,-1.393) (0.186,0.475)--(0.292,0.567) (-0.574,-0.098)--(-0.675,-0.046) (0.574,-1.096)--(0.645,-1.147) (0.169,-0.142)--(0.211,0.120) (0.137,1.253)--(0.221,1.311) (-0.568,-0.966)--(-0.638,-0.983) (0.765,-0.235)--(0.797,-0.187) (0.430,-1.728)--(0.400,-1.775) (-0.115,-0.738)--(-0.238,-0.662) (0.383,0.294)--(0.548,0.345) (-0.579,0.641)--(-0.694,0.703) (0.633,-1.250)--(0.655,-1.176) (-0.739,0.575)--(-0.749,0.689) (0.587,-1.366)--(0.566,-1.450) (0.394,1.079)--(0.487,1.200) (-0.714,-0.199)--(-0.778,-0.281) (0.734,-0.755)--(0.752,-0.701) (-0.205,1.847)--(-0.237,1.948) (0.533,-0.901)--(0.574,-1.096) (0.488,-0.247)--(0.673,-0.275) (-0.421,1.360)--(-0.555,1.429) (0.720,-0.128)--(0.765,-0.235) (-0.469,-0.900)--(-0.547,-0.786) (-0.276,-1.612)--(-0.375,-1.559) (-0.478,0.466)--(-0.579,0.641) (0.067,0.718)--(-0.111,0.765) (0.330,-1.536)--(0.393,-1.700) (0.273,0.830)--(0.363,0.849) (-0.631,-0.265)--(-0.714,-0.199) (0.638,-1.013)--(0.690,-1.022) (0.470,-1.518)--(0.514,-1.570) (-0.290,-1.838)--(-0.280,-1.920) (-0.050,-0.129)--(-0.170,-0.178) (0.474,0.502)--(0.599,0.525) (-0.648,0.570)--(-0.739,0.575) (0.773,-0.531)--(0.788,-0.356) (-0.651,1.192)--(-0.613,1.318) (-0.030,0.601)--(-0.175,0.524) (-0.511,-0.266)--(-0.574,-0.098) (0.607,-0.850)--(0.638,-1.013) (-0.272,1.858)--(-0.303,1.898) (-0.140,1.800)--(-0.291,1.718) (0.183,-1.638)--(0.256,-1.702) (-0.486,-1.097)--(-0.568,-0.966) (-0.793,0.005)--(-0.800,0.000) (0.143,-1.956)--(0.106,-2.024) (0.281,0.327)--(0.383,0.294) (-0.568,0.426)--(-0.648,0.570) (-0.564,-1.446)--(-0.613,-1.318) (0.401,-1.551)--(0.454,-1.654) (-0.747,-0.531)--(-0.773,-0.531) (-0.269,1.238)--(-0.347,1.327) (-0.009,1.995)--(0.097,2.004) (0.042,0.001)--(-0.125,0.064) (0.459,-1.679)--(0.514,-1.570) (0.540,-0.671)--(0.533,-0.901) (-0.291,1.718)--(-0.272,1.858) (-0.638,-0.983)--(-0.693,-1.025) (-0.338,-1.847)--(-0.338,-1.858) (0.042,0.001)--(0.211,0.120) (0.430,-0.064)--(0.582,-0.155) (-0.466,1.150)--(-0.561,1.288) (0.264,1.687)--(0.212,1.829) (-0.009,-0.669)--(-0.238,-0.662) (0.173,-1.828)--(0.072,-1.920) (0.340,-1.074)--(0.452,-0.904) (0.584,1.377)--(0.566,1.450) (0.387,-1.349)--(0.401,-1.551) (0.629,0.705)--(0.642,0.880) (-0.091,-1.575)--(-0.041,-1.730) (-0.578,-0.431)--(-0.631,-0.265) (-0.287,1.522)--(-0.222,1.675) (-0.184,-1.886)--(-0.234,-1.933) (-0.274,0.525)--(-0.358,0.434) (-0.568,-0.966)--(-0.672,-1.090) (-0.797,0.179)--(-0.800,0.000) (0.317,1.778)--(0.415,1.681) (-0.021,1.568)--(-0.152,1.588) (-0.050,-0.129)--(-0.125,0.064) (-0.749,0.689)--(-0.772,0.541) (-0.191,1.069)--(-0.383,1.122) (-0.704,0.461)--(-0.747,0.313) (0.450,1.545)--(0.542,1.507) (0.070,2.042)--(0.000,2.050) (0.348,0.065)--(0.383,0.294) (0.720,-0.128)--(0.797,-0.187) (0.118,-1.158)--(0.226,-1.160) (0.647,-0.031)--(0.696,0.120) (0.031,2.043)--(0.070,2.042) (0.450,1.545)--(0.514,1.570) (-0.314,-1.212)--(-0.302,-1.432) (0.302,-1.851)--(0.338,-1.858) (-0.536,0.210)--(-0.568,0.426) (0.752,-0.701)--(0.773,-0.531) (-0.177,1.960)--(-0.139,2.019) (0.470,-1.518)--(0.529,-1.466) (-0.631,-0.265)--(-0.778,-0.281) (-0.074,0.364)--(-0.175,0.524) (-0.274,-1.926)--(-0.207,-1.980) (-0.361,1.557)--(-0.291,1.718) (-0.566,-1.153)--(-0.637,-1.238) (-0.396,0.655)--(-0.478,0.466) (-0.423,1.683)--(-0.459,1.679) (0.221,1.311)--(0.107,1.480) (-0.539,1.124)--(-0.608,1.053) (-0.561,1.288)--(-0.613,1.318) (-0.071,1.708)--(-0.222,1.675) (0.599,0.525)--(0.629,0.705) (0.426,-1.143)--(0.387,-1.349) (0.042,0.001)--(0.169,-0.142) (0.169,-0.142)--(0.272,-0.105) (-0.009,1.995)--(-0.069,2.024) (-0.062,-0.936)--(0.012,-1.095) (-0.772,0.541)--(-0.773,0.531) (0.638,-1.013)--(0.690,-0.902) (0.540,-0.671)--(0.617,-0.605) (-0.089,-1.888)--(-0.184,-1.886) (0.633,-1.250)--(0.613,-1.318) (-0.430,-0.669)--(-0.469,-0.900) (0.187,-1.357)--(0.313,-1.280) (-0.693,-1.025)--(-0.655,-1.176) (0.173,-1.828)--(0.239,-1.855) (-0.648,0.570)--(-0.704,0.461) (-0.400,1.775)--(-0.459,1.679) (0.765,-0.235)--(0.788,-0.356) (-0.672,-0.426)--(-0.777,-0.448) (-0.203,0.827)--(-0.297,0.791) (0.780,0.359)--(0.788,0.356) (0.363,0.849)--(0.305,1.100) (-0.287,1.522)--(-0.361,1.557) (-0.383,-0.964)--(-0.568,-0.966) (-0.272,1.858)--(-0.334,1.835) (-0.797,-0.176)--(-0.797,-0.179) (0.200,1.524)--(0.074,1.657) (0.582,-0.155)--(0.647,-0.031) (-0.290,0.226)--(-0.358,0.434) (-0.611,1.312)--(-0.651,1.192) (0.317,1.778)--(0.262,1.889) (-0.241,-1.761)--(-0.290,-1.838) (0.401,-1.551)--(0.470,-1.518) (0.118,-1.158)--(-0.043,-1.256) (0.667,-0.770)--(0.725,-0.866) (-0.411,-0.133)--(-0.511,-0.266) (-0.043,-1.256)--(0.067,-1.341) (0.139,2.019)--(0.070,2.042) (0.683,1.014)--(0.693,1.025) (-0.549,-1.331)--(-0.564,-1.446) (-0.466,1.150)--(-0.539,1.124) (-0.074,0.364)--(-0.030,0.601) (-0.030,0.601)--(-0.111,0.765) (0.061,1.932)--(-0.009,1.995) (-0.436,0.247)--(-0.478,0.466) (-0.403,-1.178)--(-0.566,-1.153) (0.160,1.701)--(0.019,1.790) (-0.484,-1.291)--(-0.549,-1.331) (0.169,-0.142)--(0.186,-0.397) (-0.338,-1.847)--(-0.400,-1.775) (0.773,0.200)--(0.793,0.263) (-0.062,-0.936)--(-0.209,-0.895) (-0.209,-0.895)--(-0.142,-1.100) (-0.237,1.948)--(-0.274,1.926) (-0.070,-2.027)--(-0.070,-2.042) (-0.131,-1.972)--(-0.207,-1.980) (-0.467,-1.462)--(-0.514,-1.570) (0.315,1.324)--(0.387,1.475) (-0.696,-0.585)--(-0.741,-0.707) (0.143,-1.956)--(0.139,-2.019) (0.340,-1.074)--(0.313,-1.280) (0.019,1.790)--(0.061,1.932) (0.536,0.045)--(0.696,0.120) (-0.203,0.827)--(-0.088,1.029) (-0.143,-1.770)--(-0.241,-1.761) (0.615,1.229)--(0.656,1.167) (-0.274,0.525)--(-0.297,0.791) (0.531,-1.319)--(0.613,-1.318) (-0.618,-0.606)--(-0.696,-0.585) (0.719,-0.639)--(0.761,-0.625) (-0.290,0.226)--(-0.436,0.247) (-0.132,-0.470)--(-0.229,-0.416) (-0.021,1.568)--(0.074,1.657) (-0.229,-0.416)--(-0.238,-0.662) (-0.467,-1.462)--(-0.523,-1.476) (0.761,-0.625)--(0.773,-0.531) (-0.043,-1.256)--(-0.227,-1.189) (-0.710,0.940)--(-0.693,1.025) (0.454,0.785)--(0.560,0.877) (-0.697,0.212)--(-0.777,0.159) (0.118,-1.158)--(0.067,-1.341) (-0.791,0.268)--(-0.797,0.179) (0.296,1.522)--(0.356,1.639) (-0.074,0.364)--(0.079,0.440) (-0.030,0.601)--(0.067,0.718) (0.607,0.180)--(0.727,0.279) (-0.392,-1.393)--(-0.484,-1.291) (0.745,0.432)--(0.780,0.359) (-0.250,-0.032)--(-0.411,-0.133) (-0.615,0.090)--(-0.697,0.212) (0.772,0.527)--(0.773,0.531) (0.531,-1.319)--(0.587,-1.366) (0.221,1.311)--(0.315,1.324) (-0.638,-0.792)--(-0.699,-0.771) (0.678,-0.517)--(0.773,-0.531) (-0.209,-0.895)--(-0.335,-0.703) (-0.694,0.703)--(-0.752,0.701) (0.472,0.193)--(0.607,0.180) (0.061,-0.507)--(0.186,-0.397) (-0.569,0.891)--(-0.665,0.961) (-0.711,0.845)--(-0.710,0.940) (0.336,-0.616)--(0.450,-0.675) (-0.062,-0.936)--(-0.142,-1.100) (0.635,1.062)--(0.683,1.014) (0.480,0.993)--(0.568,1.064) (-0.697,0.212)--(-0.797,0.179) (0.506,-1.133)--(0.531,-1.319) (-0.112,1.941)--(-0.133,2.020) (0.524,-0.440)--(0.678,-0.517) (-0.361,1.557)--(-0.486,1.610) (-0.547,-0.786)--(-0.618,-0.606) (-0.741,-0.707)--(-0.752,-0.701) (0.728,-0.386)--(0.757,-0.502) (-0.298,-0.968)--(-0.403,-1.178) (-0.495,0.694)--(-0.569,0.891) (0.256,-1.702)--(0.302,-1.851) (-0.280,-1.920)--(-0.274,-1.926) (0.066,-0.229)--(0.169,-0.142) (0.363,0.849)--(0.454,0.785) (-0.675,-0.046)--(-0.739,0.062) (0.593,-1.244)--(0.633,-1.250) (-0.229,-0.416)--(-0.319,-0.217) (0.539,1.400)--(0.566,1.450) (0.210,-1.956)--(0.274,-1.926) (-0.132,-0.470)--(-0.238,-0.662) (0.548,0.345)--(0.656,0.324) (0.727,0.279)--(0.773,0.200) (0.393,-1.700)--(0.459,-1.679) (0.450,-0.675)--(0.540,-0.671) (-0.001,0.916)--(-0.203,0.827) (-0.574,-0.098)--(-0.615,0.090) (0.574,-1.096)--(0.593,-1.244) (-0.030,0.601)--(0.186,0.475) (-0.341,-0.449)--(-0.510,-0.581) (0.105,-1.763)--(0.173,-1.828) (0.186,0.475)--(0.067,0.718) (0.040,-2.013)--(-0.004,-2.049) (0.317,1.778)--(0.338,1.858) (0.383,0.294)--(0.472,0.193) (-0.579,0.641)--(-0.639,0.805) (-0.223,1.422)--(-0.287,1.522) (0.042,0.001)--(0.008,0.246) (-0.205,1.847)--(-0.177,1.960) (0.272,-0.105)--(0.348,0.065) (0.690,-1.022)--(0.693,-1.025) (0.488,-0.247)--(0.609,-0.362) (0.356,1.639)--(0.317,1.778) (0.678,-0.517)--(0.719,-0.639) (-0.237,1.948)--(-0.207,1.980) (-0.338,1.858)--(-0.400,1.775) (-0.070,-2.027)--(-0.139,-2.019) (0.255,-0.931)--(0.340,-1.074) (0.313,-1.280)--(0.426,-1.143) (0.079,0.440)--(-0.030,0.601) (-0.478,0.466)--(-0.495,0.694) (0.330,-1.536)--(0.329,-1.715) (0.165,1.929)--(0.144,2.016) (-0.486,1.610)--(-0.514,1.570) (-0.186,-1.613)--(-0.143,-1.770) (-0.618,1.205)--(-0.655,1.176) (-0.290,-1.838)--(-0.338,-1.847) (-0.638,-0.792)--(-0.725,-0.861) (-0.222,1.675)--(-0.140,1.800) (-0.297,0.791)--(-0.396,0.655) (0.031,1.381)--(-0.021,1.568) (0.074,1.657)--(-0.071,1.708) (-0.694,0.703)--(-0.739,0.575) (0.568,1.064)--(0.560,1.243) (0.683,1.014)--(0.713,0.932) (0.067,-1.341)--(0.187,-1.357) (0.757,-0.502)--(0.782,-0.360) (-0.791,0.268)--(-0.788,0.356) (-0.675,-0.046)--(-0.793,0.005) (-0.111,0.765)--(-0.203,0.827) (-0.778,-0.281)--(-0.797,-0.179) (-0.140,1.800)--(-0.112,1.941) (-0.291,1.718)--(-0.205,1.847) (-0.403,0.894)--(-0.495,0.694) (-0.693,1.025)--(-0.725,0.866) (-0.699,-0.771)--(-0.752,-0.701) (-0.044,-0.372)--(-0.229,-0.416) (0.772,0.527)--(0.788,0.356) (0.019,1.790)--(-0.140,1.800) (0.656,0.324)--(0.686,0.499) (0.656,1.167)--(0.655,1.176) (0.287,-0.388)--(0.366,-0.262) (-0.142,-1.100)--(-0.043,-1.256) (0.593,-1.244)--(0.645,-1.147) (0.501,1.554)--(0.514,1.570) (-0.568,-0.966)--(-0.690,-0.944) (-0.184,-1.886)--(-0.290,-1.838) (0.117,-1.528)--(0.252,-1.479) (-0.510,-0.581)--(-0.547,-0.786) (0.566,-1.450)--(0.613,-1.318) (-0.338,-1.710)--(-0.430,-1.727) (-0.639,0.805)--(-0.694,0.703) (-0.714,-0.199)--(-0.797,-0.176) (-0.191,1.069)--(-0.296,1.019) (0.752,0.701)--(0.725,0.866) (-0.280,-1.920)--(-0.338,-1.858) (-0.177,1.960)--(-0.237,1.948) (0.609,-0.362)--(0.673,-0.275) (0.072,-1.920)--(-0.025,-1.968) (0.599,0.525)--(0.748,0.603) (0.066,-0.229)--(0.186,-0.397) (-0.004,-2.049)--(-0.000,-2.050) (0.144,2.016)--(0.207,1.980) (-0.238,-0.662)--(-0.209,-0.895) (0.329,-1.715)--(0.393,-1.700) (0.210,-1.956)--(0.207,-1.980) (0.067,-1.341)--(-0.124,-1.369) (-0.797,-0.179)--(-0.788,-0.356) (-0.069,2.024)--(0.000,2.050) (-0.124,-1.369)--(-0.011,-1.484) (-0.487,0.003)--(-0.574,-0.098) (0.393,-1.700)--(0.400,-1.775) (-0.566,-1.153)--(-0.672,-1.090) (-0.334,1.835)--(-0.338,1.858) (-0.421,1.360)--(-0.497,1.337) (0.593,-1.244)--(0.655,-1.176) (-0.111,0.765)--(-0.001,0.916) (-0.001,0.916)--(-0.088,1.029) (-0.500,1.497)--(-0.514,1.570) (0.165,1.929)--(0.097,2.004) (0.667,-0.770)--(0.752,-0.701) (-0.111,0.765)--(-0.297,0.791) (0.317,1.778)--(0.400,1.775) (0.011,-1.845)--(0.072,-1.920) (-0.142,-1.100)--(-0.298,-0.968) (-0.298,-0.968)--(-0.227,-1.189) (0.393,-1.700)--(0.454,-1.654) (0.272,-0.105)--(0.366,-0.262) (-0.777,-0.448)--(-0.773,-0.531) (-0.737,-0.370)--(-0.777,-0.448) (0.008,0.246)--(0.079,0.440) (-0.690,-0.944)--(-0.693,-1.025) (0.117,1.833)--(0.165,1.929) (-0.191,1.069)--(-0.038,1.223) (0.629,0.705)--(0.698,0.850) (-0.241,-1.761)--(-0.338,-1.710) (-0.762,-0.620)--(-0.752,-0.701) (-0.297,0.791)--(-0.296,1.019) (0.097,2.004)--(0.031,2.043) (0.211,0.120)--(0.348,0.065) (0.690,-0.902)--(0.725,-0.866) (-0.184,-1.886)--(-0.280,-1.920) (-0.672,-1.090)--(-0.655,-1.176) (-0.238,-0.662)--(-0.341,-0.449) (0.074,1.657)--(0.160,1.701) (0.734,0.787)--(0.752,0.701) (-0.341,-0.449)--(-0.335,-0.703) (-0.549,-1.331)--(-0.601,-1.307) (0.452,-0.904)--(0.426,-1.143) (-0.523,-1.476)--(-0.514,-1.570) (0.072,-1.920)--(0.143,-1.956) (0.530,0.676)--(0.629,0.705) (-0.704,0.461)--(-0.771,0.433) (0.067,-1.341)--(-0.011,-1.484) (0.387,1.475)--(0.450,1.545) (-0.358,0.434)--(-0.396,0.655) (-0.001,0.916)--(0.132,0.948) (-0.335,-0.703)--(-0.469,-0.900) (0.647,-0.031)--(0.754,0.015) (-0.484,-1.291)--(-0.566,-1.153) (-0.366,0.084)--(-0.487,0.003) (-0.152,1.588)--(-0.071,1.708) (-0.338,-1.710)--(-0.387,-1.749) (0.470,-1.518)--(0.508,-1.576) (-0.347,1.327)--(-0.421,1.360) (-0.133,2.020)--(-0.139,2.019) (0.315,1.324)--(0.403,1.287) (-0.696,-0.585)--(-0.747,-0.531) (-0.549,-1.331)--(-0.613,-1.318) (0.226,-1.160)--(0.187,-1.357) (0.160,1.701)--(0.212,1.829) (-0.400,1.771)--(-0.459,1.679) (0.536,0.045)--(0.647,-0.031) (-0.539,1.124)--(-0.618,1.205) (0.040,-2.013)--(-0.000,-2.050) (0.351,-0.852)--(0.452,-0.904) (0.599,0.525)--(0.697,0.692) (-0.555,1.429)--(-0.613,1.318) (0.698,0.850)--(0.734,0.787) (0.560,0.877)--(0.642,0.880) (-0.142,-1.100)--(-0.227,-1.189) (0.008,0.246)--(0.115,0.190) (-0.800,0.000)--(-0.797,-0.179) (0.079,0.440)--(0.186,0.475) (0.262,1.889)--(0.274,1.926) (-0.319,-0.217)--(-0.511,-0.266) (0.540,-0.671)--(0.667,-0.770) (-0.618,-0.606)--(-0.672,-0.426) (-0.041,-1.730)--(0.011,-1.845) (-0.186,0.266)--(-0.358,0.434) (0.107,1.480)--(0.200,1.524) (-0.566,-1.153)--(-0.627,-1.145) (0.173,-1.828)--(0.210,-1.956) (0.725,-0.866)--(0.725,-0.866) (-0.094,1.438)--(-0.152,1.588) (-0.205,1.847)--(-0.274,1.926) (0.454,0.785)--(0.530,0.676) (-0.697,0.212)--(-0.747,0.313) (0.012,-1.095)--(-0.043,-1.256) (0.782,-0.360)--(0.788,-0.356) (0.403,1.287)--(0.477,1.378) (0.211,0.120)--(0.383,0.294) (-0.423,1.683)--(-0.400,1.771) (0.514,1.570)--(0.459,1.679) (0.582,-0.155)--(0.720,-0.128) (-0.238,-0.662)--(-0.335,-0.703) (0.754,0.015)--(0.787,-0.088) (0.607,0.180)--(0.696,0.120) (0.146,-0.941)--(0.226,-1.160) (0.452,-0.904)--(0.533,-0.901) (-0.302,-1.432)--(-0.276,-1.612) (-0.170,-0.178)--(-0.229,-0.416) (-0.213,-1.429)--(-0.186,-1.613) (0.305,1.100)--(0.394,1.079) (-0.672,-0.426)--(-0.737,-0.370) (0.593,-1.244)--(0.613,-1.318) (0.793,0.263)--(0.797,0.179) (-0.358,0.434)--(-0.478,0.466) (0.713,0.932)--(0.693,1.025) (0.719,-0.639)--(0.773,-0.531) (0.472,0.193)--(0.536,0.045) (-0.044,-0.372)--(0.186,-0.397) (-0.618,1.205)--(-0.611,1.312) (-0.569,0.891)--(-0.608,1.053) (-0.115,-0.738)--(-0.209,-0.895) (-0.152,1.588)--(-0.222,1.675) (0.727,0.279)--(0.788,0.356) (-0.704,0.461)--(-0.773,0.531) (-0.112,1.941)--(-0.069,2.024) (0.366,-0.262)--(0.430,-0.064) (-0.613,-1.318)--(-0.566,-1.450) (0.226,-1.160)--(0.313,-1.280) (-0.777,-0.448)--(-0.788,-0.356) (-0.484,-1.291)--(-0.601,-1.307) (0.252,-1.479)--(0.387,-1.349) (-0.495,0.694)--(-0.491,0.908) (0.256,-1.702)--(0.239,-1.855) (0.387,0.580)--(0.474,0.502) (-0.690,-0.944)--(-0.725,-0.866) (-0.276,-1.612)--(-0.241,-1.761) (-0.641,0.332)--(-0.704,0.461) (-0.175,0.524)--(-0.297,0.791) (-0.696,-0.585)--(-0.762,-0.620) (-0.296,1.019)--(-0.403,0.894) (-0.387,-1.749)--(-0.430,-1.727) (-0.170,-1.999)--(-0.207,-1.980) (-0.359,1.718)--(-0.334,1.835) (0.615,1.229)--(0.655,1.176) (0.107,1.480)--(0.074,1.657) (0.053,-0.856)--(0.239,-0.682) (0.106,-2.024)--(0.139,-2.019) (0.560,0.877)--(0.683,1.014) (-0.672,-1.090)--(-0.693,-1.025) (0.212,1.829)--(0.262,1.889) (0.186,-0.397)--(0.287,-0.388) (0.461,-1.355)--(0.470,-1.518) (0.000,2.050)--(-0.070,2.042) (-0.523,-1.476)--(-0.566,-1.450) (-0.511,-0.266)--(-0.578,-0.431) (0.115,0.190)--(0.079,0.440) (-0.011,-1.484)--(0.117,-1.528) (-0.131,-1.972)--(-0.170,-1.999) (-0.467,-1.462)--(-0.564,-1.446) (-0.491,0.908)--(-0.539,1.124) (-0.170,-0.178)--(-0.250,-0.032) (-0.250,-0.032)--(-0.319,-0.217) (-0.697,0.212)--(-0.791,0.268) (-0.088,1.029)--(-0.191,1.069) (-0.041,1.889)--(-0.009,1.995) (0.305,1.100)--(0.315,1.324) (-0.044,-0.372)--(0.061,-0.507) (-0.383,1.122)--(-0.491,0.908) (-0.207,-1.980)--(-0.139,-2.019) (-0.392,-1.393)--(-0.549,-1.331) (-0.133,2.020)--(-0.070,2.042) (0.696,0.120)--(0.727,0.279) (0.568,1.064)--(0.656,1.167) (0.336,-0.616)--(0.423,-0.464) (-0.400,1.771)--(-0.400,1.775) (-0.227,-1.189)--(-0.124,-1.369) (0.645,-1.147)--(0.693,-1.025) (-0.375,-1.559)--(-0.514,-1.570) (-0.578,-0.431)--(-0.618,-0.606) (-0.555,1.429)--(-0.566,1.450) (0.040,-1.648)--(0.183,-1.638) (0.008,0.246)--(-0.125,0.064) (0.067,0.718)--(-0.001,0.916) (-0.665,0.961)--(-0.693,1.025) (-0.711,0.845)--(-0.752,0.701) (-0.608,1.053)--(-0.665,0.961) (0.423,-0.464)--(0.540,-0.671) (-0.050,-0.129)--(0.169,-0.142) (0.387,0.580)--(0.454,0.785) (-0.148,1.279)--(-0.269,1.238) (-0.459,1.679)--(-0.514,1.570) (-0.777,0.159)--(-0.793,0.005) (-0.050,-0.129)--(-0.250,-0.032) (-0.069,2.024)--(-0.133,2.020) (0.617,-0.605)--(0.678,-0.517) (0.186,-0.397)--(0.151,-0.645) (0.656,0.324)--(0.780,0.359) (0.151,-0.645)--(0.239,-0.682) (0.797,-0.187)--(0.797,-0.179) (-0.025,-1.968)--(-0.131,-1.972) (-0.335,-0.703)--(-0.298,-0.968) (0.239,-1.855)--(0.302,-1.851) (0.363,0.849)--(0.560,0.877) (0.274,-1.926)--(0.338,-1.858) (-0.536,0.210)--(-0.615,0.090) (-0.140,1.800)--(-0.041,1.889) (-0.213,-1.429)--(-0.091,-1.575) (-0.223,1.422)--(-0.152,1.588) (-0.361,1.557)--(-0.439,1.534) (-0.088,1.029)--(0.058,1.123) (0.058,1.123)--(-0.038,1.223) (-0.665,0.961)--(-0.711,0.845) (-0.637,-1.238)--(-0.613,-1.318) (-0.627,-1.145)--(-0.655,-1.176) (0.262,1.889)--(0.205,1.969) (0.713,0.932)--(0.725,0.866) (0.560,1.243)--(0.584,1.377) (0.426,-1.143)--(0.506,-1.133) (-0.481,-1.614)--(-0.459,-1.679) (0.584,1.377)--(0.613,1.318) (-0.319,-0.217)--(-0.341,-0.449) (0.678,-0.517)--(0.728,-0.386) (0.542,1.507)--(0.514,1.570) (-0.089,-1.888)--(-0.025,-1.968) (0.415,1.681)--(0.459,1.679) (0.359,1.808)--(0.338,1.858) (0.302,-1.851)--(0.367,-1.808) (0.394,1.079)--(0.568,1.064) (-0.766,-0.094)--(-0.797,-0.176) (0.067,0.718)--(0.183,0.720) (0.252,-1.479)--(0.183,-1.638) (0.183,0.720)--(0.132,0.948) (-0.439,1.534)--(-0.500,1.497) (-0.050,-0.129)--(0.066,-0.229) (-0.148,1.279)--(0.031,1.381) (0.686,0.499)--(0.748,0.603) (-0.747,-0.531)--(-0.762,-0.620) (0.205,1.969)--(0.144,2.016) (0.273,0.830)--(0.454,0.785) (0.106,-2.024)--(0.070,-2.042) (0.348,0.065)--(0.430,-0.064) (-0.699,-0.771)--(-0.725,-0.861) (0.367,-1.808)--(0.400,-1.775) (0.302,-1.851)--(0.274,-1.926) (-0.025,-1.968)--(0.040,-2.013) (0.529,-1.466)--(0.566,-1.450) (-0.694,0.703)--(-0.749,0.689) (0.568,1.064)--(0.615,1.229) (-0.011,-1.484)--(-0.091,-1.575) (0.638,-1.013)--(0.693,-1.025) (-0.140,1.800)--(-0.205,1.847) (0.058,1.123)--(0.215,1.051) (0.673,-0.275)--(0.765,-0.235) (-0.436,0.247)--(-0.536,0.210) (-0.088,1.029)--(-0.038,1.223) (-0.071,1.708)--(0.019,1.790) (-0.433,-1.616)--(-0.481,-1.614) (-0.439,1.534)--(-0.514,1.570) (0.169,-0.142)--(0.287,-0.388) (-0.772,0.541)--(-0.752,0.701) (-0.574,-0.098)--(-0.714,-0.199) (-0.739,0.062)--(-0.793,0.005) (0.187,-1.357)--(0.117,-1.528) (0.264,1.687)--(0.317,1.778) (-0.497,1.337)--(-0.555,1.429) (0.787,-0.088)--(0.797,-0.179) (0.340,-1.074)--(0.426,-1.143) (0.656,0.324)--(0.745,0.432) (-0.227,-1.189)--(-0.314,-1.212) (-0.375,-1.559)--(-0.467,-1.462) (0.207,-1.980)--(0.274,-1.926) (-0.203,0.827)--(-0.191,1.069) (0.183,0.720)--(0.292,0.567) (-0.797,-0.176)--(-0.800,0.000) (0.533,-0.901)--(0.638,-1.013) (-0.143,-1.770)--(-0.089,-1.888) (0.690,-0.902)--(0.690,-1.022) (-0.274,0.525)--(-0.396,0.655) (-0.711,0.845)--(-0.725,0.866) (0.200,1.524)--(0.296,1.522) (-0.638,-0.983)--(-0.690,-0.944) (-0.725,-0.861)--(-0.725,-0.866) (0.067,0.718)--(0.132,0.948) (-0.747,0.313)--(-0.788,0.356) (-0.021,1.568)--(-0.071,1.708) (-0.639,0.805)--(-0.711,0.845) (-0.478,0.466)--(-0.648,0.570) (0.530,0.676)--(0.599,0.525) (-0.714,-0.199)--(-0.797,-0.179) (-0.601,-1.307)--(-0.613,-1.318) (-0.043,-1.256)--(-0.124,-1.369) (0.487,1.200)--(0.560,1.243) (0.348,0.065)--(0.472,0.193) (0.118,-1.158)--(0.187,-1.357) (0.609,-0.362)--(0.728,-0.386) (-0.335,-0.703)--(-0.430,-0.669) (0.686,0.499)--(0.773,0.531) (0.765,-0.235)--(0.782,-0.360) (-0.568,-0.966)--(-0.638,-0.792) (-0.725,0.866)--(-0.752,0.701) (-0.314,-1.212)--(-0.392,-1.393) (0.115,0.190)--(0.211,0.120) (-0.338,-1.710)--(-0.433,-1.616) (-0.392,-1.393)--(-0.375,-1.559) (0.329,-1.715)--(0.367,-1.808) (0.165,-2.006)--(0.207,-1.980) (0.394,1.079)--(0.480,0.993) (-0.714,-0.199)--(-0.766,-0.094) (-0.396,0.655)--(-0.495,0.694) (0.160,1.701)--(0.264,1.687) (0.221,1.311)--(0.200,1.524) (-0.009,-0.669)--(0.151,-0.645) (0.536,0.045)--(0.582,-0.155) (-0.303,1.898)--(-0.338,1.858) (0.800,0.000)--(0.797,-0.179) (-0.421,1.360)--(-0.500,1.497) (-0.209,-0.895)--(-0.298,-0.968) (-0.071,1.708)--(-0.140,1.800) (0.599,0.525)--(0.686,0.499) (0.542,1.507)--(0.566,1.450) (0.169,-0.142)--(0.366,-0.262) (0.477,1.378)--(0.542,1.507) (0.423,-0.464)--(0.488,-0.247) (-0.009,1.995)--(0.031,2.043) (-0.319,-0.217)--(-0.411,-0.133) (0.638,-1.013)--(0.645,-1.147) (0.359,1.808)--(0.400,1.775) (0.655,-1.176)--(0.693,-1.025) (0.187,-1.357)--(0.252,-1.479) (-0.430,-0.669)--(-0.547,-0.786) (-0.547,-0.786)--(-0.568,-0.966) (0.173,-1.828)--(0.143,-1.956) (-0.125,0.064)--(-0.250,-0.032) (0.474,0.502)--(0.548,0.345) (-0.648,0.570)--(-0.694,0.703) (0.393,-1.700)--(0.430,-1.728) (-0.375,-1.559)--(-0.338,-1.710) (-0.112,1.941)--(-0.139,2.019) (0.693,1.025)--(0.655,1.176) (-0.203,0.827)--(-0.296,1.019) (0.403,1.287)--(0.487,1.200) (0.363,0.849)--(0.394,1.079) (-0.788,-0.356)--(-0.773,-0.531) (0.761,-0.625)--(0.752,-0.701) (-0.229,-0.416)--(-0.341,-0.449) (0.582,-0.155)--(0.673,-0.275) (0.200,1.524)--(0.160,1.701) (0.673,-0.275)--(0.720,-0.128) (-0.497,1.337)--(-0.613,1.318) (0.629,0.705)--(0.734,0.787) (0.239,-0.682)--(0.336,-0.616) (0.317,1.778)--(0.359,1.808) (-0.611,1.312)--(-0.613,1.318) (0.367,-1.808)--(0.338,-1.858) (0.401,-1.551)--(0.393,-1.700) (-0.411,-0.133)--(-0.574,-0.098) (-0.574,-0.098)--(-0.631,-0.265) (0.529,-1.466)--(0.514,-1.570) (-0.234,-1.933)--(-0.280,-1.920) (0.430,-0.064)--(0.488,-0.247) (-0.466,1.150)--(-0.497,1.337) (-0.777,0.159)--(-0.800,0.000) (-0.568,-0.966)--(-0.566,-1.153) (-0.497,1.337)--(-0.561,1.288) (0.072,-1.920)--(0.106,-2.024) (0.690,-0.902)--(0.725,-0.866) (-0.430,-1.727)--(-0.400,-1.775) (-0.038,1.223)--(-0.148,1.279) (0.383,0.294)--(0.474,0.502) (0.387,-1.349)--(0.330,-1.536) (0.061,1.932)--(0.097,2.004) (0.400,1.775)--(0.338,1.858) (0.754,0.015)--(0.800,0.000) (-0.347,1.327)--(-0.466,1.150) (-0.115,-0.738)--(0.053,-0.856) (0.394,1.079)--(0.403,1.287) (0.533,-0.901)--(0.607,-0.850) (0.061,-0.507)--(0.151,-0.645) (0.720,-0.128)--(0.754,0.015) (0.415,1.681)--(0.402,1.772) (0.255,-0.931)--(0.452,-0.904) (0.642,0.880)--(0.713,0.932) (0.351,-0.852)--(0.450,-0.675) (0.338,1.858)--(0.274,1.926) (-0.478,0.466)--(-0.568,0.426) (-0.566,-1.450)--(-0.514,-1.570) (-0.523,-1.476)--(-0.564,-1.446) (0.315,1.324)--(0.477,1.378) (0.607,-0.850)--(0.725,-0.866) (0.454,-1.654)--(0.508,-1.576) (-0.041,-1.730)--(0.105,-1.763) (-0.631,-0.265)--(-0.672,-0.426) (-0.125,0.064)--(-0.186,0.266) (-0.222,1.675)--(-0.291,1.718) (0.132,0.948)--(0.058,1.123) (-0.561,1.288)--(-0.618,1.205) (-0.433,-1.616)--(-0.459,-1.679) (0.474,0.502)--(0.530,0.676) (-0.514,-1.570)--(-0.459,-1.679) (-0.725,-0.861)--(-0.752,-0.701) (-0.094,1.438)--(-0.223,1.422) (-0.050,-0.129)--(-0.044,-0.372) (0.306,1.891)--(0.274,1.926) (0.607,-0.850)--(0.667,-0.770) (0.239,-0.682)--(0.146,-0.941) (0.696,0.120)--(0.797,0.099) (0.183,-1.638)--(0.330,-1.536) (0.146,-0.941)--(0.255,-0.931) (-0.491,0.908)--(-0.466,1.150) (0.097,2.004)--(0.070,2.042) (-0.387,-1.749)--(-0.400,-1.775) (-0.638,-0.792)--(-0.725,-0.866) (0.454,0.785)--(0.629,0.705) (0.143,-1.956)--(0.210,-1.956) (-0.568,0.426)--(-0.641,0.332) (-0.041,1.889)--(0.061,1.932) (-0.184,-1.886)--(-0.207,-1.980) (-0.269,1.238)--(-0.383,1.122) (0.186,0.475)--(0.183,0.720) (-0.291,1.718)--(-0.359,1.718) (-0.038,1.223)--(0.137,1.253) (0.165,-2.006)--(0.139,-2.019) (-0.139,2.019)--(-0.207,1.980) (0.430,-0.064)--(0.536,0.045) (0.042,0.001)--(-0.050,-0.129) (0.137,1.253)--(0.031,1.381) (0.452,-0.904)--(0.574,-1.096) (0.430,-1.728)--(0.459,-1.679) (0.387,-1.349)--(0.461,-1.355) (-0.303,1.898)--(-0.274,1.926) (0.061,-0.507)--(-0.009,-0.669) (-0.091,-1.575)--(0.040,-1.648) (0.587,-1.366)--(0.613,-1.318) (-0.486,1.610)--(-0.459,1.679) (-0.184,-1.886)--(-0.131,-1.972) (-0.070,-2.042)--(-0.000,-2.050) (-0.671,1.080)--(-0.693,1.025) (0.210,-1.956)--(0.272,-1.926) (-0.739,0.575)--(-0.773,0.531) (0.132,0.948)--(0.273,0.830) (0.450,1.545)--(0.459,1.669) (0.273,0.830)--(0.215,1.051) (-0.359,1.718)--(-0.423,1.683) (-0.314,-1.212)--(-0.213,-1.429) (-0.618,-0.606)--(-0.699,-0.771) (-0.131,-1.972)--(-0.070,-2.027);

%% file: refs.bib
@article{lupi2025impact,
  title={Impact of blood rheology on left heart haemodynamics: {Newtonian} vs. non-{Newtonian} modelling},
  author={Lupi, Valerio and Lombardi, Filippo Caruso and Scarpolini, Martino Andrea and Verzicco, Roberto and Viola, Francesco},
  journal={European Journal of Mechanics-B/Fluids},
  pages={204445},
  year={2025},
  publisher={Elsevier}
}

@article{carreau1972rheological,
  title={Rheological equations from molecular network theories},
  author={Carreau, Pierre J},
  journal={Transactions of the Society of Rheology},
  volume={16},
  number={1},
  pages={99--127},
  year={1972},
  publisher={The Society of Rheology}
}

@article{yasuda1981shear,
  title={Shear flow properties of concentrated solutions of linear and star branched polystyrenes},
  author={Yasuda, KY and Armstrong, RC and Cohen, RE},
  journal={Rheologica Acta},
  volume={20},
  number={2},
  pages={163--178},
  year={1981},
  publisher={Springer}
}

@article{hamming1959stable,
  title={Stable predictor-corrector methods for ordinary differential equations},
  author={Hamming, Richard Wesley},
  journal={Journal of the ACM (JACM)},
  volume={6},
  number={1},
  pages={37--47},
  year={1959},
  publisher={ACM New York, NY, USA}
}

@article{yang_and_stern2015,
  title={A non-iterative direct forcing immersed boundary method for strongly-coupled fluid--solid interactions},
  author={Yang, Jianming and Stern, Frederick},
  journal={Journal of Computational Physics},
  volume={295},
  pages={779--804},
  year={2015},
  publisher={Elsevier}
}

@article{gulan2018valve_axial_orientation,
  title={The influence of bileaflet prosthetic aortic valve orientation on the blood flow patterns in the ascending aorta},
  author={G{\"u}lan, Utku and Holzner, Markus},
  journal={Medical engineering \& physics},
  volume={60},
  number={1},
  pages={61--69},
  year={2018},
  publisher={IOP Publishing}
}

@article{vanellabalaras2009,
  title = {A Moving-Least-Squares Reconstruction for Embedded-Boundary Formulations},
  author = {Vanella, Marcos and Balaras, Elias},
  year = 2009,
  journal = {Journal of Computational Physics},
  volume = {228},
  number = {18},
  pages = {6617--6628}
}

@article{vagnoli2026taycorr,
title = {A moving-least-squares reconstruction of the hydrodynamic loads on deformable bodies: application to immersed boundary methods},
journal = {Journal of Computational Physics},
pages = {114804},
year = {2026},
issn = {0021-9991},
author = {Giovanni Vagnoli and Martino A. Scarpolini and Roberto Verzicco and Francesco Viola},
}

@article{Verzicco_Orlandi_Eisenga_Heijst_Carnevale_1996,
    title={Dynamics of a vortex ring in a rotating fluid},
    volume={317},
    pages={215--239},
    journal={Journal of Fluid Mechanics},
    author={Verzicco, R. and Orlandi, P. and Eisenga, A. H. M. and Heijst, G. J. F. Van and Carnevale, G. F.},
    year={1996}
}

@article{kimApplicationFractionalstepMethod1985,
	title = {Application of a Fractional-Step Method to Incompressible {{Navier-Stokes}} Equations},
	author = {Kim, J. and Moin, P.},
	year = {1985},
	journal = {Journal of Computational Physics},
	volume = {59},
	number = {2},
	pages = {308--323},
	langid = {english}
}

@article{mittal2011sharp_IB_method,
  title={A sharp-interface immersed boundary method with improved mass conservation and reduced spurious pressure oscillations},
  author={Seo, Jung Hee and Mittal, Rajat},
  journal={Journal of computational physics},
  volume={230},
  number={19},
  pages={7347--7363},
  year={2011},
  publisher={Elsevier}
}

@article{mittal2008sharp_IB_method,
  title={A versatile sharp interface immersed boundary method for incompressible flows with complex boundaries},
  author={Mittal, Rajat and Dong, Haibo and Bozkurttas, Meliha and Najjar, FM and Vargas, Abel and Von Loebbecke, Alfred},
  journal={Journal of computational physics},
  volume={227},
  number={10},
  pages={4825--4852},
  year={2008},
  publisher={Elsevier}
}

@inproceedings{mittal2026gpu,
  title={A {GPU-Accelerated} Sharp Interface Immersed Boundary Solver for Large Scale Flow Simulations},
  author={Kumar, Sushrut and Romero, Joshua and Seo, Jung-Hee and Fatica, Massimiliano and Mittal, Rajat},
  booktitle={AIAA SCITECH 2026 Forum},
  pages={0705},
  year={2026}
}

@book{ruetsch2024cuda,
  title={{CUDA Fortran} for scientists and engineers: best practices for efficient {CUDA Fortran} programming},
  author={Ruetsch, Gregory and Fatica, Massimiliano},
  year={2024},
  publisher={Elsevier}
}

@book{wieSpaceVehicleDynamics2008,
  title = {Space Vehicle Dynamics and Control},
  editor = {Wie, Bong},
  year = {2008},
  series = {{{AIAA}} Education Series},
  edition = {2nd ed},
  publisher = {{American Institute of Aeronautics and Astronautics}},
  address = {Reston, VA},
  langid = {english}
}

@article{moricheSingleOblateSpheroid2021,
  title = {A Single Oblate Spheroid Settling in Unbounded Ambient Fluid: {{A}} Benchmark for Simulations in Steady and Unsteady Wake Regimes},
  author = {Moriche, Manuel and Uhlmann, Markus and Du{\v s}ek, Jan},
  year = {2021},
  journal = {International Journal of Multiphase Flow},
  volume = {136}
}

@article{zhouPathInstabilitiesOblate2017,
  title = {Path Instabilities of Oblate Spheroids},
  author = {Zhou, W. and Chrust, M. and Du{\v s}ek, J.},
  year = {2017},
  journal = {Journal of Fluid Mechanics},
  volume = {833},
  pages = {445--468},
  langid = {english}
}

@article{quadrio2016does_forcing_affects,
  title={Does the choice of the forcing term affect flow statistics in {DNS} of turbulent channel flow?},
  author={Quadrio, Maurizio and Frohnapfel, Bettina and Hasegawa, Yosuke},
  journal={European Journal of Mechanics-B/Fluids},
  volume={55},
  pages={286--293},
  year={2016},
  publisher={Elsevier}
}

@article{weymouth2025waterlily,
  title={{WaterLily.jl}: A differentiable and backend-agnostic {Julia} solver for incompressible viscous flow around dynamic bodies},
  author={Weymouth, Gabriel D and Font, Bernat},
  journal={Computer Physics Communications},
  volume={315},
  pages={109748},
  year={2025},
  publisher={Elsevier}
}

@inproceedings{romero2022distributed,
  title={Distributed-memory simulations of turbulent flows on modern {GPU} systems using an adaptive pencil decomposition library},
  author={Romero, Joshua and Costa, Pedro and Fatica, Massimiliano},
  booktitle={Proceedings of the Platform for Advanced Scientific Computing Conference},
  pages={1--11},
  year={2022}
}

@article{huang2010flag,
  title={Three-dimensional simulation of a flapping flag in a uniform flow},
  author={Huang, Wei-Xi and Sung, Hyung Jin},
  journal={Journal of Fluid Mechanics},
  volume={653},
  pages={301--336},
  year={2010},
  publisher={Cambridge University Press}
}

@article{updegrove2017simvascular,
  title={{SimVascular}: an open source pipeline for cardiovascular simulation},
  author={Updegrove, Adam and Wilson, Nathan M and Merkow, Jameson and Lan, Hongzhi and Marsden, Alison L and Shadden, Shawn C},
  journal={Annals of biomedical engineering},
  volume={45},
  number={3},
  pages={525--541},
  year={2017},
  publisher={Springer}
}

@article{arthurs2021crimson,
  title={{CRIMSON}: An open-source software framework for cardiovascular integrated modelling and simulation},
  author={Arthurs, Christopher J and Khlebnikov, Rostislav and Melville, Alex and Mar{\v{c}}an, Marija and Gomez, Alberto and Dillon-Murphy, Desmond and Cuomo, Federica and Silva Vieira, Miguel and Schollenberger, Jonas and Lynch, Sabrina R and others},
  journal={PLoS computational biology},
  volume={17},
  number={5},
  pages={e1008881},
  year={2021},
  publisher={Public Library of Science San Francisco, CA USA}
}

@article{layton2011cuibm,
  title={{cuIBM}--a {GPU-accelerated} immersed boundary method},
  author={Layton, Simon K and Krishnan, Anush and Barba, Lorena A},
  journal={arXiv preprint arXiv:1109.3524},
  year={2011}
}

@article{chuang2018petibm,
  title={{PetIBM}: toolbox and applications of the immersed-boundary method on distributed-memory architectures},
  author={Chuang, Pi-Yueh and Mesnard, Olivier and Krishnan, Anush and A Barba, Lorena},
  journal={Journal of Open Source Software},
  volume={3},
  number={25},
  year={2018}
}

@article{cheng2019openifem,
  title={{OpenIFEM}: A high performance modular open-source software of the immersed finite element method for fluid-structure interactions},
  author={Cheng, Jie and Yu, Feimi and Zhang, Lucy T},
  journal={Computer Modeling in Engineering \& Sciences},
  volume={119},
  number={1},
  pages={91--124},
  year={2019},
  publisher={Tech Science Press}
}

@article{weller1998tensorial,
  title={A tensorial approach to computational continuum mechanics using object-oriented techniques},
  author={Weller, Henry G. and Tabor, Gavin and Jasak, Hrvoje and Fureby, Christer},
  journal={Computers in Physics},
  volume={12},
  number={6},
  pages={620--631},
  year={1998},
  publisher={American Institute of Physics}
}

@article{jansson2021neko,
  title={{Neko}: A modern, portable, and scalable framework for high-fidelity computational fluid dynamics},
  author={Jansson, Niclas and Karp, Martin and Podobas, Artur and Markidis, Stefano and Schlatter, Philipp},
  journal={Computers \& Fluids},
  volume={275},
  pages={106235},
  year={2024}
}

@article{cantwell2015nektar,
  title={Nektar++: An open-source spectral/hp element framework},
  author={Cantwell, Chris D. and Moxey, David and Comerford, Andrew and Bolis, Alessandro and Rocco, Gianmarco and Mengaldo, Gianmarco and De Grazia, Daniele and Yakovlev, Sergey and Lombard, Jean-Eudes and Ekelschot, David and Jordi, Benjamin and Xu, Hui and Mohamied, Y. and Eskilsson, Claes and Nelson, B. and Vos, Peter and Biotto, C. and Kirby, Robert M. and Sherwin, Spencer J.},
  journal={Computer Physics Communications},
  volume={192},
  pages={205--219},
  year={2015},
  publisher={Elsevier}
}

@article{anderson2019mfem,
  title={{MFEM}: A modular finite element methods library},
  author={Anderson, Robert and Andrej, Julian and Barker, Andrew and Bramwell, Jamie and Camier, Jean-Sylvain and Cerveny, Jakub and Dobrev, Veselin and Dudouit, Yohann and Fisher, Aaron and Kolev, Tzanio and Pazner, Will and Stowell, Mark and Tomov, Vladimir and Dahm, Johann and Medina, David and Zampini, Stefano},
  journal={Computers \& Mathematics with Applications},
  volume={81},
  pages={42--74},
  year={2021}
}

@article{arndt2019dealii,
  title={The {deal.II} finite element library: Design, features, and insights},
  author={Arndt, Daniel and Bangerth, Wolfgang and Davydov, Denis and Heister, Timo and Heltai, Luca and Kronbichler, Martin and Maier, Matthias and Pelteret, Jean-Paul and Turcksin, Bruno and Wells, David},
  journal={Computers \& Mathematics with Applications},
  volume={81},
  pages={407--422},
  year={2021}
}

@book{logg2012automated,
  title={Automated Solution of Differential Equations by the Finite Element Method: The FEniCS Book},
  editor={Logg, Anders and Mardal, Kent-Andre and Wells, Garth N.},
  year={2012},
  publisher={Springer}
}

@article{hecht2012freefem,
  title={New development in {FreeFem++}},
  author={Hecht, Frederic},
  journal={Journal of Numerical Mathematics},
  volume={20},
  number={3-4},
  pages={251--266},
  year={2012}
}

@article{bernardini2021streams,
  title={{STREAmS}: A high-fidelity accelerated solver for direct numerical simulation of compressible turbulent flows},
  author={Bernardini, Matteo and Modesti, Davide and Salvadore, Francesco and Pirozzoli, Sergio},
  journal={Computer Physics Communications},
  volume={263},
  pages={107906},
  year={2021},
  publisher={Elsevier}
}

@manual{lstc2024lsdyna,
  title={{LS-DYNA Keyword User's Manual}},
  organization={Ansys/Livermore Software Technology Corporation},
  year={2024}
}

@manual{siemens2025starccm,
  title={{Simcenter STAR-CCM+ Documentation}},
  organization={Siemens Digital Industries Software},
  year={2025}
}

@manual{dassault2024abaqus,
  title={{Abaqus Analysis User's Guide}},
  organization={Dassault Systemes Simulia Corp.},
  year={2024}
}

@manual{comsol2025manual,
  title={{COMSOL Multiphysics Reference Manual}},
  organization={COMSOL AB},
  year={2025}
}

@article{chourdakis2021precice,
  title={{preCICE} v2: A sustainable and user-friendly coupling library},
  author={Chourdakis, Gerasimos and Davis, Kyle and Rodenberg, Benjamin and Schulte, Miriam and Simonis, Frederic and Uekermann, Benjamin and Abrams, Georg and Bungartz, Hans-Joachim and Cheung Yau, Lucia and Desai, Ishaan and others},
  journal={Open Research Europe},
  volume={2},
  pages={51},
  year={2022}
}

@article{cardiff2018solids4foam,
  title={An open-source finite volume toolbox for solid mechanics and fluid-solid interaction simulations},
  author={Cardiff, Philip and Karac, Aleksandar and De Jaeger, Philippe and Jasak, Hrvoje and Nagy, Jozsef and Ivankovic, Alojz and Tukovic, Zeljko},
  journal={arXiv preprint arXiv:1808.10736},
  year={2018}
}

@article{africa2023lifexcfd,
  title={{lifex-cfd}: an open-source computational fluid dynamics solver for cardiovascular applications},
  author={Africa, Pasquale Claudio and Fumagalli, Ivan and Bucelli, Michele and Zingaro, Alberto and Fedele, Marco and Dede', Luca and Quarteroni, Alfio},
  journal={Computer Physics Communications},
  volume={296},
  pages={109039},
  year={2024}
}

@article{economon2016su2,
  title={{SU2}: An open-source suite for multiphysics simulation and design},
  author={Economon, Thomas D. and Palacios, Francisco and Copeland, Sean R. and Lukaczyk, Trent W. and Alonso, Juan J.},
  journal={AIAA Journal},
  volume={54},
  number={3},
  pages={828--846},
  year={2016},
  doi={10.2514/1.J053813}
}

@article{bhalla2013unified,
  title={A unified mathematical framework and an adaptive numerical method for fluid-structure interaction with rigid, deforming, and elastic bodies},
  author={Bhalla, Amneet Pal Singh and Bale, Rahul and Griffith, Boyce E. and Patankar, Neelesh A.},
  journal={Journal of Computational Physics},
  volume={250},
  pages={446--476},
  year={2013},
  publisher={Elsevier}
}

@article{zhang2021sphinxsys,
  title={{SPHinXsys}: An open-source multi-physics and multi-resolution library based on smoothed particle hydrodynamics},
  author={Zhang, Chi and Rezavand, Massoud and Zhu, Yujie and Yu, Yongchuan and Wu, Dong and Zhang, Wenbin and Wang, Jianhang and Hu, Xiangyu},
  journal={Computer Physics Communications},
  volume={267},
  pages={108066},
  year={2021},
  publisher={Elsevier}
}

@article{battista2017ib2d,
  title={{IB2d}: a {Python} and {MATLAB} implementation of the immersed boundary method},
  author={Battista, Nicholas A. and Strickland, W. Christopher and Miller, Laura A.},
  journal={Bioinspiration \& Biomimetics},
  volume={12},
  number={3},
  pages={036003},
  year={2017},
  publisher={IOP Publishing}
}

@article{ten2002sphere,
  title={Particle imaging velocimetry experiments and lattice-{Boltzmann} simulations on a single sphere settling under gravity},
  author={Ten Cate, A and Nieuwstad, CH and Derksen, Jacobus J and Van den Akker, HEA},
  journal={Physics of Fluids},
  volume={14},
  number={11},
  pages={4012--4025},
  year={2002},
  publisher={American Institute of Physics}
}

@article{viola2020fsei,
  title={Fluid--structure-electrophysiology interaction ({FSEI}) in the left-heart: a multi-way coupled computational model},
  author={Viola, Francesco and Meschini, Valentina and Verzicco, Roberto},
  journal={European Journal of Mechanics-B/Fluids},
  volume={79},
  pages={212--232},
  year={2020},
  publisher={Elsevier}
}

@article{viola2022fsei_gpu,
  title={{FSEI-GPU}: {GPU} accelerated simulations of the fluid--structure--electrophysiology interaction in the left heart},
  author={Viola, Francesco and Spandan, Vamsi and Meschini, Valentina and Romero, Joshua and Fatica, Massimiliano and de Tullio, Marco D and Verzicco, Roberto},
  journal={Computer physics communications},
  volume={273},
  pages={108248},
  year={2022},
  publisher={Elsevier}
}

@article{deTullio2016MLS,
  title={A moving-least-squares immersed boundary method for simulating the fluid--structure interaction of elastic bodies with arbitrary thickness},
  author={de Tullio, Marco D and Pascazio, Giuseppe},
  journal={Journal of Computational Physics},
  volume={325},
  pages={201--225},
  year={2016},
  publisher={Elsevier}
}

@article{verzicco1996finite,
  title={A finite-difference scheme for three-dimensional incompressible flows in cylindrical coordinates},
  author={Verzicco, R and Orlandi, Paolo},
  journal={Journal of Computational Physics},
  volume={123},
  number={2},
  pages={402--414},
  year={1996},
  publisher={Elsevier}
}

@article{van2015pencil,
  title={A pencil distributed finite difference code for strongly turbulent wall-bounded flows},
  author={Van Der Poel, Erwin P and Ostilla-M{\'o}nico, Rodolfo and Donners, John and Verzicco, Roberto},
  journal={Computers \& Fluids},
  volume={116},
  pages={10--16},
  year={2015},
  publisher={Elsevier}
}

@article{hammer2011mass,
  title={Mass-spring model for simulation of heart valve tissue mechanical behavior},
  author={Hammer, Peter E and Sacks, Michael S and Del Nido, Pedro J and Howe, Robert D},
  journal={Annals of biomedical engineering},
  volume={39},
  pages={1668--1679},
  year={2011},
  publisher={Springer}
}

@article{gelder1998elastic_membranes,
  title={Approximate simulation of elastic membranes by triangulated spring meshes},
  author={Gelder, Allen Van},
  journal={Journal of graphics tools},
  volume={3},
  number={2},
  pages={21--41},
  year={1998},
  publisher={Taylor \& Francis}
}

@article{fedosov2010systematic,
  title={Systematic coarse-graining of spectrin-level red blood cell models},
  author={Fedosov, Dmitry A and Caswell, Bruce and Karniadakis, George Em},
  journal={Computer Methods in Applied Mechanics and Engineering},
  volume={199},
  number={29-32},
  pages={1937--1948},
  year={2010},
  publisher={Elsevier}
}

@article{viola2023high,
  title={High-fidelity model of the human heart: an immersed boundary implementation},
  author={Viola, Francesco and Del Corso, Giulio and Verzicco, Roberto},
  journal={Physical Review Fluids},
  volume={8},
  number={10},
  pages={100502},
  year={2023},
  publisher={APS}
}

@article{verzicco_2022_EFM, title={Electro-fluid-mechanics of the heart}, volume={941}, journal={Journal of Fluid Mechanics}, publisher={Cambridge University Press}, author={Verzicco, R.}, year={2022}, pages={P1}}

@article{spandan_2017_potential,
  title={A parallel interaction potential approach coupled with the immersed boundary method for fully resolved simulations of deformable interfaces and membranes},
  author={Spandan, Vamsi and Meschini, Valentina and Ostilla-M{\'o}nico, Rodolfo and Lohse, Detlef and Querzoli, Giorgio and de Tullio, Marco D and Verzicco, Roberto},
  journal={Journal of computational physics},
  volume={348},
  pages={567--590},
  year={2017},
  publisher={Elsevier}
}

@article{spandan2018fast,
  title={A fast moving least squares approximation with adaptive {Lagrangian} mesh refinement for large scale immersed boundary simulations},
  author={Spandan, Vamsi and Lohse, Detlef and de Tullio, Marco D and Verzicco, Roberto},
  journal={Journal of computational physics},
  volume={375},
  pages={228--239},
  year={2018},
  publisher={Elsevier}
}

@article{deVita2016non-newtonian,
  title={Numerical simulation of the non-{Newtonian} blood flow through a mechanical aortic valve: Non-{Newtonian} blood flow in the aortic root},
  author={De Vita, F and De Tullio, MD and Verzicco, R},
  journal={Theoretical and computational fluid dynamics},
  volume={30},
  pages={129--138},
  year={2016},
  publisher={Springer}
}

@article{de2009dns_mech_valve,
  title={Direct numerical simulation of the pulsatile flow through an aortic bileaflet mechanical heart valve},
  author={De Tullio, MD and Cristallo, A and Balaras, E and Verzicco, R},
  journal={Journal of Fluid Mechanics},
  volume={622},
  pages={259--290},
  year={2009},
  publisher={Cambridge University Press}
}

@article{deTullio2012mechHemolAorticValveProstheses,
  title={Computational prediction of mechanical hemolysis in aortic valved prostheses},
  author={De Tullio, MD and Nam, J and Pascazio, G and Balaras, E and Verzicco, Roberto},
  journal={European Journal of Mechanics-B/Fluids},
  volume={35},
  pages={47--53},
  year={2012},
  publisher={Elsevier}
}

@article{vremanEddyviscositySubgridscaleModel2004,
  title = {An Eddy-Viscosity Subgrid-Scale Model for Turbulent Shear Flow: {{Algebraic}} Theory and Applications},
  author = {Vreman, A. W.},
  year = 2004,
  journal = {Physics of Fluids},
  volume = {16},
  number = {10},
  pages = {3670--3681},
  publisher = {AIP Publishing},
  langid = {english}
}

@book{fung1993biomechanics,
  author    = {Fung, Y. C.},
  title     = {Biomechanics: Mechanical Properties of Living Tissues},
  edition   = {2},
  year      = {1993},
  publisher = {Springer},
  address   = {New York},
  doi       = {10.1007/978-1-4757-2257-4}
}

@article{vagnoli2025kappa,
  title   = {A local and explicit forcing correction for {L}agrangian immersed boundary methods},
  author  = {Vagnoli, Giovanni and Scarpolini, Martino Andrea and Verzicco, Roberto and Viola, Francesco},
  journal = {Computer Physics Communications},
  volume  = {315},
  pages   = {109741},
  year    = {2025},
  doi     = {10.1016/j.cpc.2025.109741}
}

@article{yildiran2024pressure,
  title   = {Pressure boundary conditions for immersed-boundary methods},
  author  = {Yildiran, Ibrahim Nasuh and Beratlis, Nikolaos and Capuano, Francesco and Loke, Yue-Hin and Squires, Kyle and Balaras, Elias},
  journal = {Journal of Computational Physics},
  volume  = {510},
  pages   = {113057},
  year    = {2024},
  doi     = {10.1016/j.jcp.2024.113057}
}

@article{nicoud1999wale,
  title   = {Subgrid-scale stress modelling based on the square of the velocity gradient tensor},
  author  = {Nicoud, Franck and Ducros, Fr{\'e}d{\'e}ric},
  journal = {Flow, Turbulence and Combustion},
  volume  = {62},
  number  = {3},
  pages   = {183--200},
  year    = {1999},
  doi     = {10.1023/A:1009995426001}
}

@misc{vagnoli2026fastconsistentsharpinterfaceimmersed,
      title={A fast and consistent sharp-interface immersed boundary method for moving bodies of arbitrary thickness}, 
      author={Giovanni Vagnoli and Martino Andrea Scarpolini and Roberto Verzicco and Francesco Viola},
      year={2026},
      eprint={2606.09799},
      archivePrefix={arXiv},
      primaryClass={physics.flu-dyn},
      url={https://arxiv.org/abs/2606.09799}, 
}
